\documentclass[longauth]{aa}  

\usepackage{amssymb}
\usepackage{graphicx}
\usepackage{array, tabularx, multirow, longtable, makecell}  
\usepackage{txfonts}
\usepackage{comment}
\usepackage{caption, subcaption}
\usepackage{lscape}
\usepackage{titlesec} 
\usepackage[]{hyperref}
\hypersetup{
   colorlinks=true,
   linkcolor= blue,
   filecolor= magenta,
   urlcolor= blue,
   citecolor= blue}
\usepackage{float,amsmath}
\usepackage{color}
\usepackage[version=4]{mhchem}
\usepackage{booktabs}
\usepackage{soul} 
\usepackage{orcidlink} 

\usepackage{xcolor}
\usepackage{xifthen}
\usepackage[normalem]{ulem}
\newcommand{\stkout}[1]{\ifmmode\text{\sout{\ensuremath{#1}}}\else\sout{#1}\fi}
\newcommand{\edited}[2]{\ifthenelse{\isempty{#1}}{\textbf{#2}}{\ifthenelse{\isempty{#2}}{\textcolor{gray}{\stkout{#1}}}{\textcolor{gray}{\stkout{#1}} \textbf{#2}}}}

\usepackage{xstring,xifthen}
\newcommand{\replacedecimal}[2]{\ifthenelse{\isin{.}{#1}}{\text{\StrBefore{#1}{.}}\ensuremath{\overset{#2}{.}}\text{\StrBehind{#1}{.}}}{#1\ensuremath{^{#2}}}}

\begin{document}

   \title{JOYS+: link between ice and gas of complex organic molecules}

   \subtitle{Comparing JWST and ALMA data of two low-mass protostars}

   \author{Y. Chen\inst{1}\orcidlink{0000-0002-3395-5634}
          \and W. R. M. Rocha\inst{2}\orcidlink{0000-0001-6144-4113}
          \and E. F. van Dishoeck\inst{1,3}\orcidlink{0000-0001-7591-1907} 
          \and M. L. van Gelder\inst{1}\orcidlink{0000-0002-6312-8525}
          \and P. Nazari\inst{1,4}\orcidlink{0000-0002-4448-3871}
          \and K. Slavicinska\inst{1,2}\orcidlink{0000-0002-7433-1035} 
          \and L. Francis\inst{1}\orcidlink{0000-0001-8822-6327}
          \and B. Tabone\inst{5}\orcidlink{0000-0002-1103-3225}
          \and M. E. Ressler\inst{6}\orcidlink{0000-0001-5644-8830}
          \and P. D. Klaassen\inst{7}\orcidlink{0000-0001-9443-0463}
          \and H. Beuther\inst{8}\orcidlink{0000-0002-1700-090X}
          \and A. C. A. Boogert\inst{9}\orcidlink{0000-0001-9344-0096}
          \and C. Gieser\inst{3}\orcidlink{0000-0002-8120-1765}
          \and P. J. Kavanagh\inst{10}\orcidlink{0000-0001-6872-2358}
          \and G. Perotti\inst{8}\orcidlink{0000-0002-8545-6175}
          \and V. J. M. Le Gouellec\inst{11}\orcidlink{0000-0002-5714-799X}
          \and L. Majumdar\inst{12,13}\orcidlink{0000-0001-7031-8039}
          \and M. G\"{u}del\inst{14,15}\orcidlink{0000-0001-9818-0588}
          \and Th. Henning\inst{8}\orcidlink{0000-0002-1493-300X}
          }

   \institute{Leiden Observatory, Leiden University, P.O. Box 9513, 2300RA Leiden, The Netherlands \\
              \email{ychen@strw.leidenuniv.nl}
         \and Laboratory for Astrophysics, Leiden Observatory, Leiden University, PO Box 9513, 2300 RA Leiden, The Netherlands
         \and Max Planck Institut für Extraterrestrische Physik (MPE), Giessenbachstrasse 1, 85748 Garching, Germany
         \and European Southern Observatory (ESO), Karl-Schwarzschild-Strasse 2, 1780 85748 Garching, Germany
         \and Universite Paris-Saclay, CNRS, Institut d’Astrophysique Spatiale, 91405 Orsay, France
         \and Jet Propulsion Laboratory, California Institute of Technology, 4800 Oak Grove Drive, Pasadena, CA 91109, USA
         \and UK Astronomy Technology Centre, Royal Observatory Edinburgh, Blackford Hill, Edinburgh EH9 3HJ, UK
         \and Max Planck Institute for Astronomy (MPIA), K\"{o}nigstuhl 17, 69117 Heidelberg, Germany
         \and Institute for Astronomy, University of Hawaii at Manoa, 2680 Woodlawn Drive, Honolulu, HI 96822, USA
         \and Department of Experimental Physics, Maynooth University, Maynooth, Co. Kildare, Ireland
         \and NASA Ames Research Center, PO Box 1 Moffett Field, CA 94035-1000, USA
         \and School of Earth and Planetary Sciences, National Institute of Science Education and Research, Jatni 752050, Odisha, India
         \and Homi Bhabha National Institute, Training School Complex, Anushaktinagar, Mumbai 400094, India
         \and Department of Astrophysics, University of Vienna, T\"urkenschanzstrasse 17, A-1180 Vienna, Austria
         \and ETH Z\"urich, Institute for Particle Physics and Astrophysics, Wolfgang-Pauli-Strasse 27, 8093 Z\"urich, Switzerland
         }

   \date{Received 14 May 2024 / Accepted 29 Jul 2024}

 
  \abstract
   {A rich inventory of complex organic molecules (COMs) has been observed in high abundances in the gas phase toward Class 0 protostars. These molecules are suggested to be formed in ices and sublimate in the warm inner envelope close to the protostar. However, only the most abundant COM, methanol (\ce{CH3OH}), has been firmly detected in ices before the era of \textit{James Webb} Space Telescope (JWST). Now it is possible to detect the interstellar ices of other COMs and constrain their ice column densities quantitatively.}
   {We aim to determine the column densities of several oxygen-bearing COMs (O-COMs) in both gas and ice for two low-mass protostellar sources, NGC 1333 IRAS~2A (hereafter IRAS~2A) and B1-c, as case studies in our JWST Observations of Young protoStars (JOYS+) program. By comparing the column density ratios with respect to \ce{CH3OH} between both phases measured in the same sources, we can probe into the evolution of COMs from ice to gas in the early stages of star formation.}
   {The column densities of COMs in gas and ice are derived by fitting the spectra observed by the Atacama Large Millimeter/submillimeter Array (ALMA) and the JWST/Mid-InfraRed Instrument-Medium Resolution Spectroscopy (MIRI-MRS), respectively. The gas-phase emission lines are fit using local thermal equilibrium (LTE) models, and the ice absorption bands are fit by matching the infrared spectra measured in laboratories. The column density ratios of four O-COMs (\ce{CH3CHO}, \ce{C2H5OH}, \ce{CH3OCH3}, and \ce{CH3OCHO}) with respect to \ce{CH3OH} are compared between ice and gas in IRAS~2A and B1-c.} 
   {We are able to fit the fingerprints range of COM ices between 6.8 and 8.8 $\mu m$ in the JWST/MIRI-MRS spectra of B1-c using similar components as recently used for NGC 1333 IRAS~2A. We claim detection of \ce{CH4}, \ce{OCN^-}, \ce{HCOO^-}, \ce{HCOOH}, \ce{CH3CHO}, \ce{C2H5OH}, \ce{CH3OCH3}, \ce{CH3OCHO}, and \ce{CH3COCH3} in B1-c, and upper limits are estimated for \ce{SO2}, \ce{CH3COOH}, and \ce{CH3CN}. The total abundance of O-COM ices is constrained to be 15\% with respect to \ce{H2O} ice, 80\% of which is dominated by \ce{CH3OH}. The comparison of O-COM ratios with respect to \ce{CH3OH} between ice and gas shows two different cases. On one hand, the column density ratios of \ce{CH3OCHO} and \ce{CH3OCH3} match well between the two phases, which may be attributed to a direct inheritance from ice to gas or strong chemical links with \ce{CH3OH}. On the other hand, the ice ratios of \ce{CH3CHO} and \ce{C2H5OH} with respect to \ce{CH3OH} are higher than the gas ratios by 1--2 orders of magnitudes. This difference can be explained by the gas-phase reprocessing following sublimation, or different spatial distributions of COMs in the envelope, which is an observational effect since ALMA and JWST are tracing different components in a protostellar system.}
   {The firm detection of COM ices other than \ce{CH3OH} is reported in another well-studied low-mass protostar, B1-c, following the recent detection in NGC 1333 IRAS~2A. The column density ratios of four O-COMs with respect to \ce{CH3OH} show both similarities and differences between gas and ice. Although the straightforward explanations would be the direct inheritance from ice to gas and the gas-phase reprocessing, respectively, other possibilities such as different spatial distributions of molecules cannot be excluded.}

   \keywords{astrochemistry – stars: formation – stars: protostars – stars: low-mass – ISM: molecules – ISM: abundances
               }

   \maketitle
%
\begin{figure*}[!ht]
    \centering
    \includegraphics[width=\textwidth]{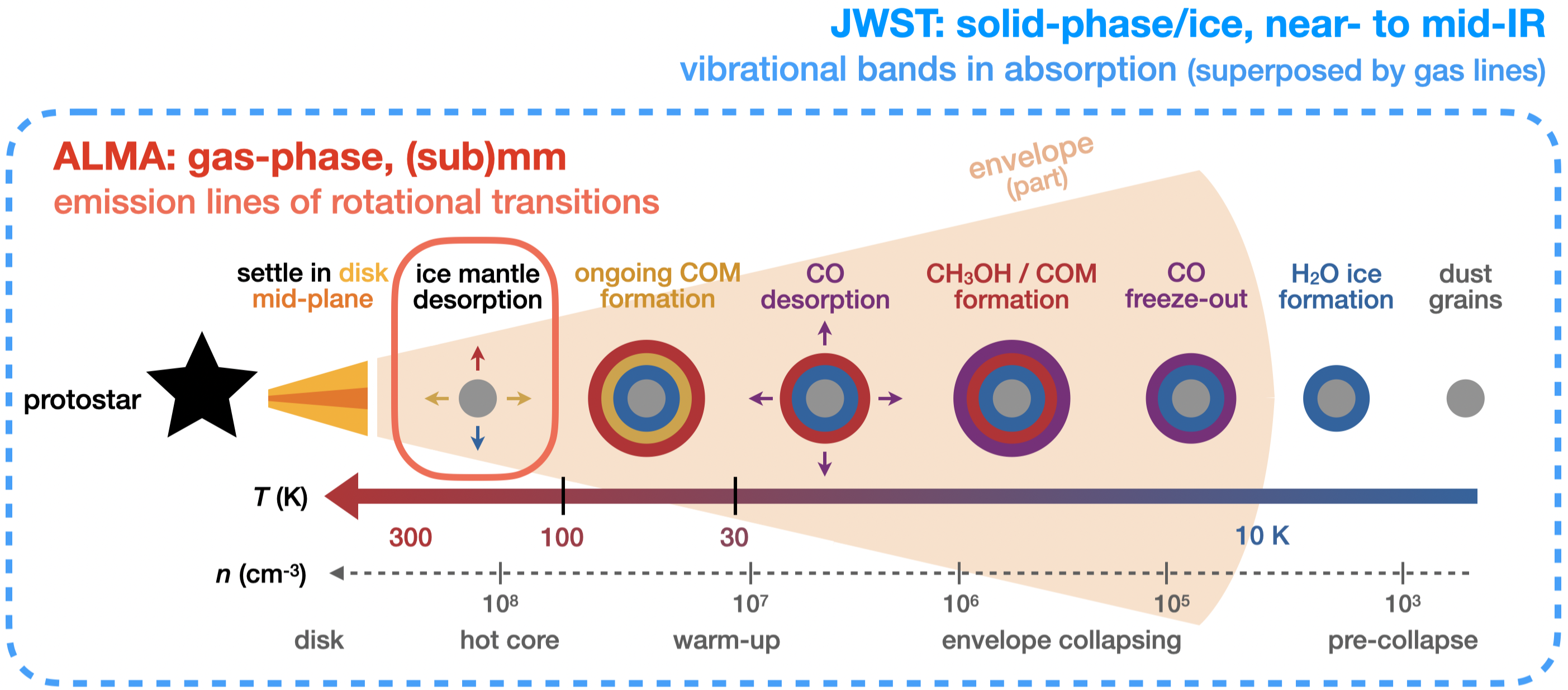}
    \caption{Schematic of chemical evolution on dust grains in protostellar stages, modified from Fig.~1.4 of \cite{MvH2019PhDT} and Fig.~14 of \cite{2009ARAA}. Ice layers are dominated by different species (e.g., \ce{H2O}, CO, and COMs) that are denoted in different colors. The typical temperature and density in different evolutionary stages are labeled on arrows in the bottom. The small red box indicates the hot core region where the temperature is high enough (>100 K) to sublimate most of the volatile materials and therefore the gas-phase molecules can be traced by radio telescopes at (sub-)millimeter wavelengths. The big blue box indicates that infrared telescopes such as JWST are tracing everything along the line of sight, including ice in the vast envelope and gas in the hot core.}
    \label{fig:COM_schematic}
\end{figure*}

\section{Introduction}\label{sect:intro}
How does the chemistry in the universe evolve from simple atoms in the diffuse interstellar or intergalactic medium to a prosperous biosphere on our Earth? The early phase of star formation, from molecular clouds to protostars, is probably the first key stage in this long journey. In particular, complex organic molecules \citep[COMs, typically defined as carbon-bearing molecules with at least six atoms;][]{2009ARAA}, have gained in popularity over the past several decades due to their importance of linking simple species with prebiotic molecules \citep{Jorgensen2020ARAA, Ceccarelli2022}. 

The formation and evolution history of COMs in star-forming regions is still a topic of active study. COMs are suggested to be abundantly formed during cold ($T\lesssim$10 K) early stages of star formation in the ice mantles of dust grains, and 
indirect evidence from gas-phase observations shows that the abundance ratios of commonly detected COMs (with respect to methanol, \ce{CH3OH}, the simplest and most abundant COM) remain consistent among sources with different luminosities \citep[e.g.,][]{Coletta2020, vanGelder2020, Nazari2022_NCOM, Chen2023}. 
Figure \ref{fig:COM_schematic} shows a schematic of the proposed formation history of COMs. Prior to star formation, the growth of ice mantles starts with the formation of \ce{H2O} ice and the subsequent freeze-out of CO gas as temperature and UV radiation intensity decrease toward the center of dense molecular clouds. In the icy mantles, CO is gradually hydrogenated to form formaldehyde (\ce{H2CO}), \ce{CH3OH}, and even larger COMs that contain two carbon atoms \citep[e.g.,][]{Watanabe2002, Fuchs2009, Cuppen2009, Fedoseev2015, Fedoseev2022, Simons2020}. 
As the central protostars gradually warm up their envelopes, volatile molecules such as CO in the ice mantles start to sublimate into the gas phase. In the meantime, COMs, especially those larger than \ce{CH3OH}, can keep forming via solid-phase chemistry \citep[e.g.,][]{Garrod2022}. COMs have high binding energies and sublimate at high temperatures \citep[$\lesssim$100 K;][]{Fedoseev2015, Fedoseev2022} toward the end of the warm-up stage.
In hot cores, where $T>$100 K, all the volatile ice mantles are expected to sublimate into the gas phase, and the chemistry also evolves fully in the gas phase. However, it is not certain how important the gas-phase chemistry is to the observed chemical composition of hot cores \citep{Balucani2015}. 

COMs in different phases are observed by different facilities at different wavelengths. Gas-phase COMs are usually observed at (sub-)millimeter wavelengths via their rotational transitions in emission using radio telescopes such as the Atacama Large Millimeter/submillimeter Array (ALMA). With its powerful sensitivity and resolution (both spatial and spectral), ALMA has detected a rich inventory of COMs in various star-forming regions \citep[e.g.,][]{Bacmann2012, Jimenez-Serra2016}, from starless cores to circumstellar disks \citep[e.g.,][]{Brunken2022, Yamato2024, Booth2024b}, but most commonly in protostars \citep[e.g.,][]{Jorgensen2020ARAA, vanGelder2020, QinSL2022, Nazari2022_NCOM, Baek2022, Chen2023}. These hot ($\gtrsim$100 K) inner regions around protostars are often referred to as hot cores, or hot corinos specifically for low-mass sources.

On the other hand, solid-phase COMs, also known as COM ices, are observed at near- and mid-infrared wavelengths via their vibrational bands in absorption. These infrared observations are better to be conducted in space to avoid absorption by species such as \ce{H2O}, \ce{CO2}, and \ce{CH4} in the Earth's atmosphere. Due to the limited sensitivity and spectral resolution of the previous Infrared Space Observatory (ISO) and \textit{Spitzer} Space Telescope, detection of COM ices have only been confirmed for \ce{CH3OH} (the simplest and most abundant COM) before the era of the \textit{James Webb} Space Telescope (JWST). Tentative identifications were made for two bands at 7.24 and 7.41~$\mu$m with ISO and \textit{Spitzer}, for which the possible contributors are \ce{HCOO^-}, \ce{C2H5OH}, and \ce{CH3CHO} \citep{Schutte1999, Oberg2011_Spitzer}.
With the unprecedented power of JWST, it is now promising to detect more COMs other than \ce{CH3OH} using the Medium Resolution Spectroscopy (MRS) mode of the Mid-InfraRed Instrument (MIRI). 
The wavelength range covered by MIRI-MRS (4.9--27.9 $\mu$m) contains the methanol band at 9.74~$\mu$m and a characteristic fingerprint range around 6.8--8.8~$\mu$m for COM ices, where multiple absorption bands of their vibrational modes are located \citep{Boogert2015ARAA}.

Recently, robust detection of COMs other than \ce{CH3OH}, including acetaldehyde (\ce{CH3CHO}), ethanol (\ce{C2H5OH}), and methyl formate (\ce{CH3OCHO}), has been reported in the MIRI-MRS spectra of two protostars, NGC 1333 IRAS~2A and IRAS 23385+6053 \citep[hereafter IRAS~2A and IRAS 23385;][]{Rocha2024}. Now, for the first time, we are able to directly compare the gas- and solid-phase abundances of these complex organics using a combination of ALMA and JWST. In fact, some gas-to-ice comparison has been made in previous studies \citep[e.g.,][]{Noble2017, Perotti2020, Perotti2021,Perotti2023}, but they are targeting simpler molecules (CO and \ce{H2O}, along with \ce{CH3OH}) in the cold envelopes in star-forming regions where the gas-phase material is non-thermally desorbed from dust grains. In this work, we are tracing the hot core regions where most COMs have sublimated into the gas phase.
As shown in Fig.~\ref{fig:COM_schematic}, 
ALMA traces the hot core region where most of the volatile species have been thermally sublimated into the gas phase.
On the other hand, JWST looks through the vast envelope and traces ice mantles along the line of sight, providing so-far the best mid-infrared spectra (i.e., with the highest sensitivity and spectral resolution) for COM ices. 
Using high-quality spectra from ALMA and JWST, we can trace the main reservoir of COMs in both phases and make direct comparison of their abundances between gas and ice. This will provide us with valuable observational evidence of how these molecules evolve from the earlier solid-phase stage to the subsequent gas-phase stage. 

\begin{figure*}[!ht]
    \centering
    \includegraphics[width=\textwidth]{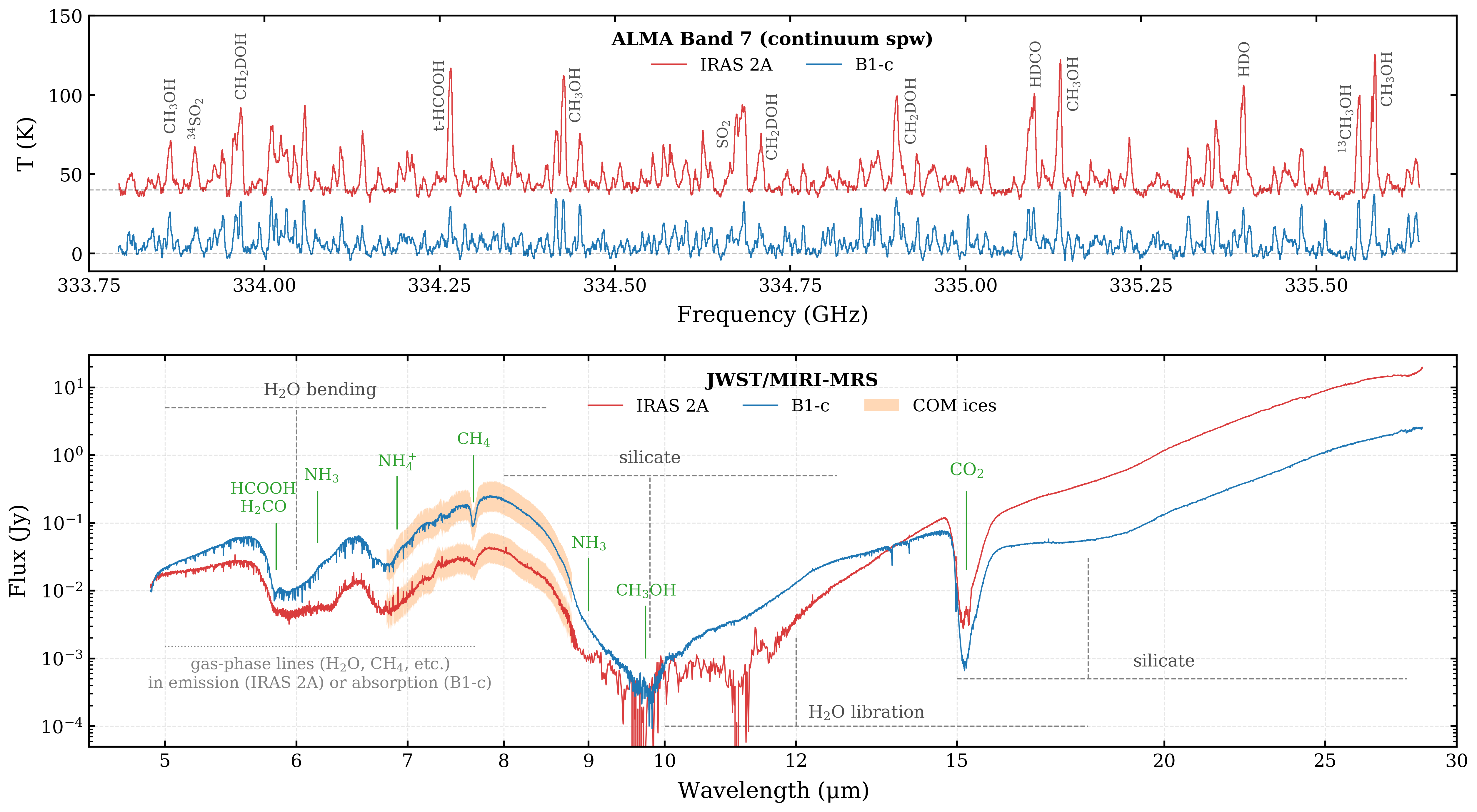}
    \caption{Overview of the ALMA spectra (the continuum spectral window from 333.78 to 335.65 GHz; top panel) and JWST/MIRI-MRS spectra (bottom panel) of NGC 1333 IRAS~2A and B1-c. The JWST spectrum of IRAS~2A between 8.85 and 11.7~$\mu$m was binned by a factor of 10 for better visualization. In the bottom panel, the fingerprint ranges of COM ices between 6.8 and 8.8 $\mu$m are highlight in orange. Important emission lines in the ALMA spectra and absorption bands in the JWST spectra are labelled in each panel.}
    \label{fig:spec_overview}
\end{figure*}

Ideally, the comparison should be made for the same sources that are rich in both gas and ice COMs. However, the sample of protostars for which gas-phase COMs have been detected using ALMA and other telescopes \citep[e.g.,][]{Jorgensen2016_PILS, Mininni2020_GUAPOS, Yang2021_PEACHES, Hsu2022_ALMASOP, Nazari2022_NCOM, Chen2023} is much larger than that for ice detections with IR telescopes. 
In this paper, we select two low-mass protostars, IRAS~2A and B1-c, for a case study as part of our JWST Observations of Young protoStars (JOYS+) program. These two sources are famous hot corinos that have been observed to have rich emission lines of gas-phase COMs \citep[e.g.,][]{Taquet2015, vanGelder2020}. They also have noticeable absorption features in the fingerprint range of COM ices between 6.8 and 8.8~$\mu$m, as revealed by previous \textit{Spitzer} observations \citep{Boogert2008} and our JWST/MIRI-MRS spectra. 

IRAS~2A and B1-c also have a relatively strong \ce{CH3OH} ice band at 9.74~$\mu$m, which is often severely extincted by silicate features but necessary to determine the ice column density of \ce{CH3OH} as a reference species. A higher SNR of this band will lead to better constraints on the column density of \ce{CH3OH} ice, and hence more accurate estimation on the COM ratios with respect to \ce{CH3OH}. Overall, IRAS~2A and B1-c are among the best candidates for the first direct comparison of COM abundances between gas and ice.

We emphasize that this observational study cannot be realized without the support by experimentalists. Over the past years, many laboratory infrared spectra (hereafter lab spectra) of COM ices have been measured under different conditions \citep[e.g.,][]{TvS2018, TvS2021, Hudson2018, Hudson2020, Hudson2021, Hudson2022, Rachid2020, Rachid2021, Rachid2022,  Slavicinska2023}. Most of these lab spectra are public available on the Leiden Ice Database for Astrochemistry \citep[LIDA; ][]{Rocha2022_LIDA}. The references for other lab spectra that are involved in this paper can be found at Table C.1 in \cite{Rocha2024}.

The paper is structured as follows: Sect.~\ref{sect:observations} provides information on the ALMA and JWST observations and the data reduction procedures. Sect.~\ref{sect:methods} describes the analyzing methods of the ALMA and JWST data, and the results are displayed in Sect.~\ref{sect:results}. Discussion is elaborated in Sect.~\ref{sect:discussion}, with a focus on the gas-to-ice ratios of four selected oxygen-bearing COMs. Conclusions are summarized in Sect.~\ref{sect:conclusions}. A considerable amount of content is reserved in the appendices, including (i) the images of ALMA and JWST observations (Appendix~\ref{appendix:maps}); (ii) details and supplements to the fitting methodology of the JWST spectrum of B1-c (Appendices~\ref{appendix:silicates}--\ref{appendix:NIRSpec_CH3OH}; (iii) additional figures and table for the fitting results of ALMA spectra (Appendices~\ref{appendix:ALMA_fitting} and \ref{appendix:table}); (iv) additional figures and table of the fitting results of the JWST spectrum of B1-c (Appendix~\ref{appendix:JWST_fitting}).

\section{Observations and data reduction}\label{sect:observations}
\subsection{ALMA}\label{sect:obs_ALMA}
The ALMA data of IRAS~2A and B1-c were taken as part of program 2021.1.01578.S (PI: B. Tabone), which targeted magnetic disk winds of five Class 0 protostars in Perseus and also observed a large number of emission lines of gas-phase COMs. The observations were taken with a combination of an extended configuration C-6 ($\theta_\mathrm{beam}$ = $0.12\arcsec\times0.09\arcsec$) and a more compact configuration C-3 ($\theta_\mathrm{beam}$ = $0.58\arcsec\times0.34\arcsec$). The maximum recoverable scales ($\theta_\mathrm{MRS}$) for the C-6 and C-3 datasets are 1.6$\arcsec$ and 6.2$\arcsec$, respectively. The data cover nine spectral windows in Band 7 between 333.8 and 347.6 GHz, including one 1.875-GHz wide continuum window with a spectral resolution of 0.87 km s$^{-1}$. Among the eight line windows, six windows have a spectral resolution of 0.22 km s$^{-1}$ and a frequency coverage of 0.12 GHz or 0.24 GHz; the remaining two windows are wider (0.48 GHz) but have a lower spectral resolution of 0.44 km s$^{-1}$. 

The data were pipeline calibrated using CASA versions 6.2.1.7 and 6.4.1.12 \citep{McMullin2007}. The measurement sets from the C-3 and C-6 configurations were combined via concatenation and subsequently continuum subtracted and imaged using the \texttt{concat}, \texttt{uvcontsub} and \texttt{tclean} tasks in CASA version 6.4.1.12.
For B1-c, a \texttt{briggs} weighting of 0.5 was used for all the spectral windows to improve the angular resolution, yielding a synthesized beam of $\theta_\mathrm{beam}$ = $0.08\arcsec\times0.11\arcsec$ for the 0.88 mm continuum and the spectral windows, which corresponds to a spatial resolution of $\sim$30 au at $\sim$320 pc (i.e., the distance of the Perseus star-forming region). For IRAS~2A, a higher \texttt{briggs} weighting of 2.0 was used to increase the SNR of the \ce{H^{13}CO^+} line at 348.998 GHz \citep[customized for][]{Nazari2024_diskwinds}, resulting in a larger beam size of $0.25\arcsec\times0.4\arcsec$ for that spectral window (336.93--337.05 GHz).
The rms is about 1.5--2.0 mJy beam$^{-1}$ for all the spectral windows except that the one with larger beam has a higher rms of 5.0 mJy beam$^{-1}$. The rms for the continuum is about 0.2 mJy beam$^{-1}$. 



\subsection{JWST}\label{sect:obs_JWST}
The JWST/MIRI-MRS data of IRAS~2A and B1-c were taken in the Guaranteed Time Observation programs (GTO) 1236 (PI: M. E. Ressler) and 1290 (PI: E. F. van Dishoeck), respectively, as part of the JWST Observations of Young protoStars+ (JOYS+) collaboration\footnote{\url{https://miri.strw.leidenuniv.nl}}. 
MIRI-MRS covers 4.9--27.9~$\mu$m with 
a spectral resolution $R=\lambda/\Delta\lambda$ of 1300-3700 \citep{Rieke2015, Wright2015, Wells2015, Labiano2021, Wright2023, Argyriou2023}.
The dither patterns were optimized for extended sources, with the 2-point pattern used for IRAS~2A and the 4-point pattern for B1-c. Two pointings were observed for B1-c; one centered on the protostar itself, and one covering the blue-shifted outflow. Both programs included dedicated background observations in a 2-point dither pattern that allow for a subtraction of the telescope background and detector artifacts. All observations were taken using the FASTR1 readout mode and cover the full 4.9-27.9~$\mu$m wavelength coverage of MIRI-MRS. The total integration time of 333~s for IRAS~2A was equally divided over the three MIRI-MRS gratings. For B1-c, the total integration time was 8000~s, of which 4000~s were used in grating B (MEDIUM) to get a good SNR in the deep silicate absorption band, and 2000~s each for gratings A (SHORT) and C (LONG).


The data were processed through all three stages of the JWST calibration pipeline version 1.12.5 \citep{Bushouse2023}, using the same procedure as described by \cite{vanGelder2024}. The reference contexts {\tt jwst$\_$1126.pmap} and {\tt jwst$\_$1177.pmap} of the JWST Calibration Reference Data System \citep[CRDS;][]{Greenfield2016} were used for IRAS~2A and B1-c, respectively. The raw data were processed through the {\tt Detector1Pipeline} using the default settings. The dedicated backgrounds were subtracted at the detector level in the {\tt Spec2Pipeline}, as well as applying the fringe flat for extended sources (Crouzet et al. in prep.) and a residual fringe correction (Kavanagh et al. in prep.). An additional bad pixel map was applied to the resulting calibrated detector files using the Vortex Imaging Processing (VIP) package \citep{Christiaens2023}. The final datacubes were constructed using the {\tt Spec3Pipeline} for each band of each channel separately.

\section{Methods}\label{sect:methods}
\subsection{ALMA}
The ALMA spectrum of IRAS~2A was extracted from the peak pixel of \ce{CH3OH} emission at R.A. 03$^\mathrm{h}$28$^\mathrm{m}$55.569$^\mathrm{s}$, Dec. +31$^\mathrm{d}$14$^\mathrm{m}$36.930$^\mathrm{s}$ (J2000). This position was selected based on the integrated intensity maps of a dozen of O-COM lines shown in Fig.~\ref{fig:IRAS2A_ALMA_mom0}, which is slightly offset from the continuum peak at R.A. 03$^\mathrm{h}$28$^\mathrm{m}$55.5735$^\mathrm{s}$, Dec. +31$^\mathrm{d}$14$^\mathrm{m}$36.925$^\mathrm{s}$ (6 pixels offset in R.A. and 2 pixels in Dec; the pixel size is $0.01\arcsec\times0.01\arcsec$). The line emission of IRAS~2A near the continuum peak is attenuated, probably due to the high optical depth of dust in Band 7 \citep[e.g.,][]{DeSimone2020}. B1-c does not show the same attenuation issue and therefore its spectrum was extracted from the peak pixel of the continuum emission at R.A. 03$^\mathrm{h}$33$^\mathrm{m}$17.881$^\mathrm{s}$, Dec. +31$^\mathrm{d}$09$^\mathrm{m}$31.740$^\mathrm{s}$, which is consistent with the emission peak of O-COMs (Fig.~\ref{fig:B1c_ALMA_mom0}).

The ALMA spectra of the continuum window (333.78--335.65 GHz) of IRAS~2A and B1-c are shown in the top panel of Fig.~\ref{fig:spec_overview}. The spectra were converted to brightness temperature (K) scale by averaging over the synthesized beam. Line identification and spectral fitting were performed using the spectral analysis software CASSIS\footnote{\url{http://cassis.irap.omp.eu/}}\citep{Vastel2015CASSIS}. We mainly considered O-COMs in the fitting, including methanol and its isotopologues (\ce{CH3OH}, \ce{^{13}CH3OH}, and \ce{CH3^{18}OH}), acetaldehyde (\ce{CH3CHO}), ethanol (\ce{C2H5OH}), dimethyl ether (\ce{CH3OCH3}), methyl formate (\ce{CH3OCHO}), glycolaldehyde (\ce{CH2OHCHO}), and ethylene glycol (\ce{(CH2OH)2}). The LTE modelled spectra were fit to the observations, and a best-fit column density ($N$), excitation temperature ($T_\mathrm{ex}$), and line width (FWHM) were determined for each species in each source by grid fitting or visual inspection. 

We adopted the grid fitting method when a species has more than five clean lines detected (clean means the line is unblended or partially blended with the line profile still recognizable). For each source and each species, a grid of $N$, $T_\mathrm{ex}$, and FWHM was preset, and each grid point corresponds to an LTE model. The $\chi^2$ between the observations and the LTE model was only calculated around the fully unblended lines. The model with the smallest $\chi^2$ gave the best-fit parameters. The 2$\sigma$ uncertainties on $N$ and $T_\mathrm{ex}$ were estimated from the $N$--$T_\mathrm{ex}$ contour plots. The uncertainties of $N$ are usually 30--40\% of the best-fit values, and the uncertainties of $T_\mathrm{ex}$ varies with species.

The grid-fitting method was mainly applied to B1-c. However, most of the emission lines in the IRAS~2A spectrum show double-peaked features, suggesting two velocity components. This makes the blending issue more severe, and hence makes it more tricky to perform grid fittings. Instead, we fit the spectrum through visual inspection, which is more flexible in this case. For each species and each velocity component, we first determined the $v_\mathrm{lsr}$ and FWHM based on several strong and unblended lines, then manually adjusted $N$ and $T_\mathrm{ex}$ to achieve a better fit between the LTE models and the observed spectrum. All the parameter were fine-tuned once a rough range around the best fit was found. In this way, the uncertainties on $N$ and $T_\mathrm{ex}$ could not be calculated from contour plots, but were estimated at a level of 20--30\%. More details about the fitting strategy for line-rich ALMA spectra can be found in Sect.~3.1 of \cite{Chen2023}.

\subsection{JWST}
The JWST/MIRI-MRS spectrum of IRAS~2A has been presented and analyzed in \cite{Rocha2024}. Here we implemented a similar analysis to B1-c. The B1-c spectrum was manually extracted from the continuum peaks at R.A. 03$^\mathrm{h}$33$^\mathrm{m}$17.8959$^\mathrm{s}$ +31$^\mathrm{d}$09$^\mathrm{m}$31.8578$^\mathrm{s}$, which is only $\sim$0.1$\arcsec$ offset from the ALMA continuum peak at 0.9 mm (see Fig.~\ref{fig:B1c_JWST_spec_extraction}). The diameter was set to four times the size of the point spread function (PSF) of MIRI-MRS \citep[FWHM$_\mathrm{PSF}$ = 0.033$\arcsec\times$ ($\lambda$/$\mu$m) + 0.106$\arcsec$;][]{Law2023}, that is, the extraction aperture is increasing toward longer-wavelength channels. In our interested COM fingerprint range, the aperture size is $\sim$1.4$\arcsec$ in diameter. An additional 1D residual fringe correction was performed especially to remove the high-frequency dichroic noise in channels 3 and 4 (Kavanagh et al. in prep.). Since the photometric calibration between the bands was accurate enough, the 12 sub-bands were stitched together without applying any flux adjustment.

Starting from the original extracted spectrum, we required five steps to reach the final goal of determining the column densities of COM ices: fit a global continuum (Sect.~\ref{sect:method_global_cont}), subtract the silicate features (Sect.~\ref{sect:method_silicates}), fit a local continuum (Sect.~\ref{sect:method_local_cont}), remove the superposed gas-phase lines (which is needed for B1-c but might be skipped for other sources if these lines are weak; Sect.~\ref{sect:method_gas_lines}), and decompose the COM fingerprint ranges between 6.8 and 8.8~$\mu$m using the lab spectra (Sect.~\ref{sect:method_fit_COM_bands}). To calculate the ice column density ratios of COMs with respect to a reference species, we also determined the ice column densities of \ce{CH3OH} and \ce{H2O} with a slightly different routine than that for the 6.8--8.8 $\mu$m range (see Sect.~\ref{sect:method_CH3OH_H2O_band}).

\subsubsection{Global continuum}\label{sect:method_global_cont}
The first step is to fit a global continuum level so that the original spectrum in flux scale ($\lambda F_\lambda$) can be converted into optical depth ($\tau_\lambda$) scale by taking the logarithm between the global continuum and the observed spectrum. 
The goal of this step is not to derive the origin of the mid-IR continuum emission, but to enable the subsequent comparison with the lab spectra that are measured in absorbance ($Abs_\nu$), which is directly linked to optical depth with a scaling factor of 2.303 (i.e., $\ln{10}$):
\begin{equation}\label{eq:tau_abs}
    \tau_\nu = 2.303\times Abs_\nu.
\end{equation}
We fit a third-order polynomial to the observed spectrum in three ranges: 4.9--5.5~$\mu$m at the short-wavelength end, 7.75--7.9~$\mu$m around the `bump' between the \ce{NH4^+} band at 6.8 $\mu$m and the silicate band at 9.8 $\mu$m, and 27.6--27.9~$\mu$m at the long-wavelength end (see Fig.~\ref{fig:fitting_steps}a). The flux values in the last two ranges were manually lifted by a factor of 1.3 to leave some space for the red wings of the \ce{H2O} bending mode at $\sim$6~$\mu$m and the silicate band at $\sim$18~$\mu$m. Both the fit ranges and the lifting factor of 1.3 were tuned carefully to produce a plausible global continuum. The last range of 27.6--27.9~$\mu$m was set to be narrow since using a wider range will create an increasing feature in the polynomial longward of 20~$\mu$m, which will introduce some artifacts in the optical depth spectrum that do not match with the supposed silicate features in the next step.

\subsubsection{Silicate bands}\label{sect:method_silicates}
The second step is to fit and subtract the silicate features at 9.8 and 18~$\mu$m. This step removes the contribution from silicates to the COMs bands in 8.0--8.8~$\mu$m, the \ce{CH3OH} band at 9.74~$\mu$m, and the water libration band at $\sim$13~$\mu$m as much as possible. The silicate features may not be fully removed, but the residuals will be further excluded in the next step of local continuum subtraction.
We tried fitting with two silicate profiles: the observed spectrum toward the galactic center source GCS\,3 \citep{Kamper2004}, and the synthetic silicate spectra of pyroxene (\ce{Mg_xFe_{1-x}SiO3}) and olivine (\ce{MgFeSiO4}) computed by the \texttt{optool} code \citep{OpTool}. The GCS\,3 and the synthetic silicate spectra were scaled manually to match the optical depth spectrum of B1-c. The fitting criterion is to fit the two silicate bands at 9.8 and 18~$\mu$m as well as possible without overfitting any part of the observed spectrum, especially the blue wing of the 9.8~$\mu$m band.

Figure~\ref{fig:fitting_steps}b shows the fitting results of the silicate features. The GCS\,3 profile can be scaled deeper at the 9.8~$\mu$m band than the \texttt{optool} mixture of pyroxene and olivine, but leaves non-negligible residuals in the blue wing of the 9.8~$\mu$m band (roughly between 8 and 9~$\mu$m). This has been seen in previous ice studies \citep[e.g.,][]{Boogert2008} and also our JOYS+ data. The reasons why are still being investigated. 
The scaling factors of the synthetic silicates, pyroxene and olivine, are tuned manually to achieve the best fit. Pyroxene has a more blueshifted 9.8~$\mu$m band than olivine, and its absorbance ratio between the 9.8 and 18~$\mu$m bands is higher. The contribution of pyroxene is constrained by the small bump at $\sim$8.5~$\mu$m in the observed spectrum (see the blow-up in Fig.~\ref{fig:fitting_steps}b), which should not be overfit by the blue wing of the synthetic silicate spectrum. Similarly, olivine has a relatively stronger 18~$\mu$m band and its scaling factor is constrained by not overfitting the 18~$\mu$m band. The synthetic silicate spectrum turned out to fit the B1-c spectrum better than the GCS\,3 profile. This however,  may not be the general case, since the observed silicate bands in independent sources can be quite different from one another. The silicate features in the mid-IR spectra of protostellar sources is worth a thorough investigation which is beyond the scope of this paper. We provide some additional details and relevant discussion in Appendix~\ref{appendix:silicates}.

\subsubsection{Local continuum}\label{sect:method_local_cont}
\begin{figure}[H]
    \centering
    \includegraphics[width=\hsize]{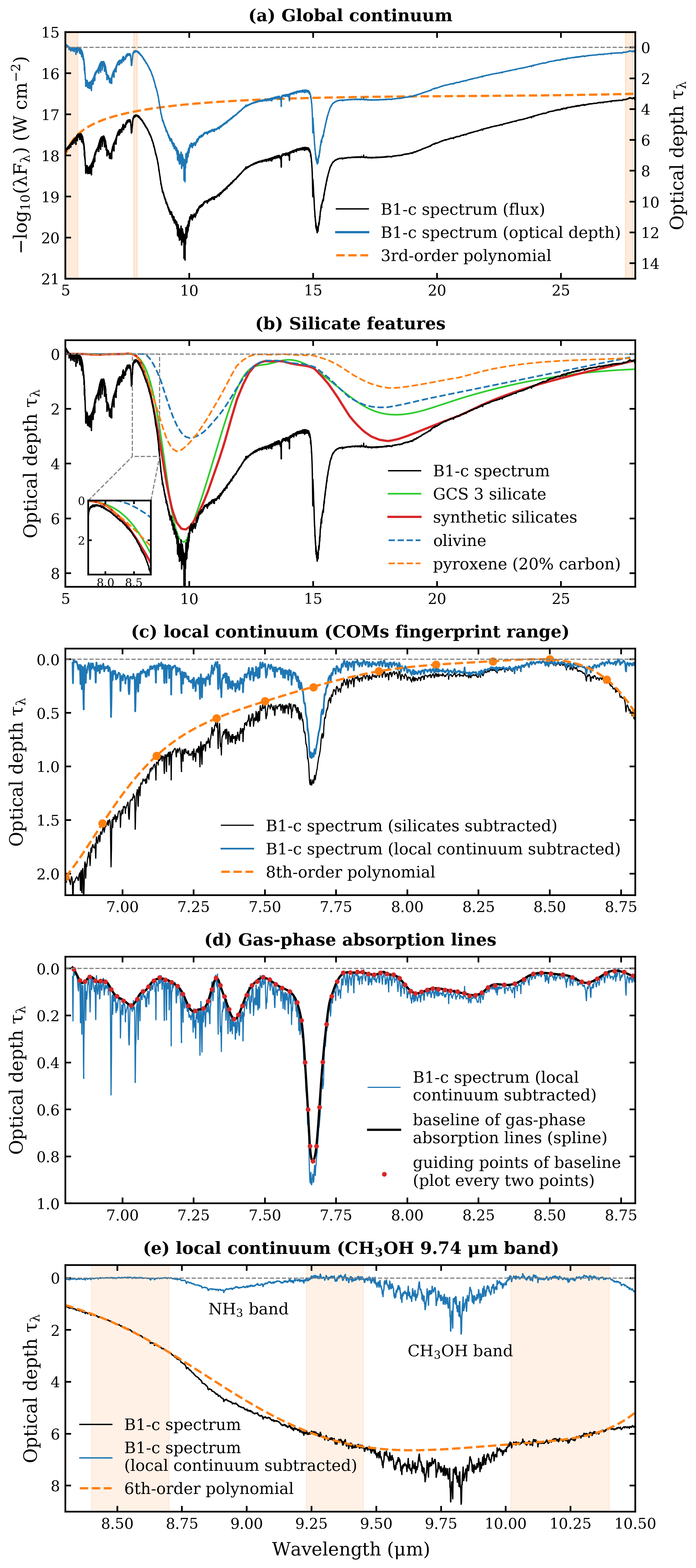}
    \caption{Panels (a)--(d) show the four steps to isolate the fingerprints range of COM ices between 6.8 and 8.8 $\mu$m from the original JWST/MIRI-MRS spectrum of B1-c: (a) fit a global continuum and convert the spectrum to optical depth scale; (b) subtract silicate features at $\sim$9.8 and 18 $\mu$m; (c) trace a local continuum in the 6.8--8.8 $\mu$m range to isolate the weak bands of COM ices from other strong features; (d) trace a baseline of the gas-phase lines in absorption to restore the band profiles of ices. Panel (e) shows the isolation of the \ce{CH3OH} ice band at 9.74 $\mu$m by tracing a local continuum. Orange shaded regions in panels (a) and (e) show the selected wavelength ranges for the polynomial fitting.}
    \label{fig:fitting_steps}
\end{figure}

After subtracting the silicate component from the optical depth spectrum, a local continuum was fit between 6.8 and 8.8~$\mu$m to isolate the absorption bands of COM ices from other strong features \citep[e.g., the \ce{NH4^+} band at 6.8~$\mu$m, the red wing of the \ce{H2O} bending mode at 6~$\mu$m, and the leftover silicate features in the previous step;][]{Schutte1999}. Due to the richness of absorption features in this range, continuous absorption-free regions for fitting hardly exist. Instead, a sequence of `guiding points' are set manually at certain locations \citep[e.g.,][]{Grant2023a}. We fit a seventh-order polynomial to about ten guiding points at positions that were set between two absorption bands 
or in the middle of a broad band to bridge the guiding points on both sides (see Fig.~\ref{fig:fitting_steps}c). Polynomials with lower orders would slightly deviate from some of the guiding points or create artificial features. 

Similar to fitting a local continuum, the choice of local continuum is somewhat subjective. \cite{Rocha2024} show in their Section 4.2.4 and appendix J that the ice column densities of some species can be changed by using a different local continuum. In our case, we traced the local continuum as close to the observed spectrum as possible, so that the ice abundances of the targeted COMs are not likely to be overestimated. 

The noise level in optical depth around the COM fingerprint range was also estimated in this step. A second-order polynomial was fit to a small range between 8.227 and 8.240 $\mu$m that is relatively free of emission or absorption features. The noise level was calculated using:
\begin{equation}\label{eq:obs_rms}
    \sigma = \sqrt{\frac{\sum_{i=1}^N(y_i - \bar{y})^2}{N}}
\end{equation}
where $y_i$ is the polynomial-subtracted optical depth at each channel between 8.227 and 8.240 $\mu$m, and $\bar{y}$ is the mean of $y_i$. The noise level of the B1-c spectrum was estimated as 2.3$\times10^{-3}$ in units of optical depth.

\begin{figure*}[!htbp]
    \centering
    \includegraphics[width=0.95\textwidth]{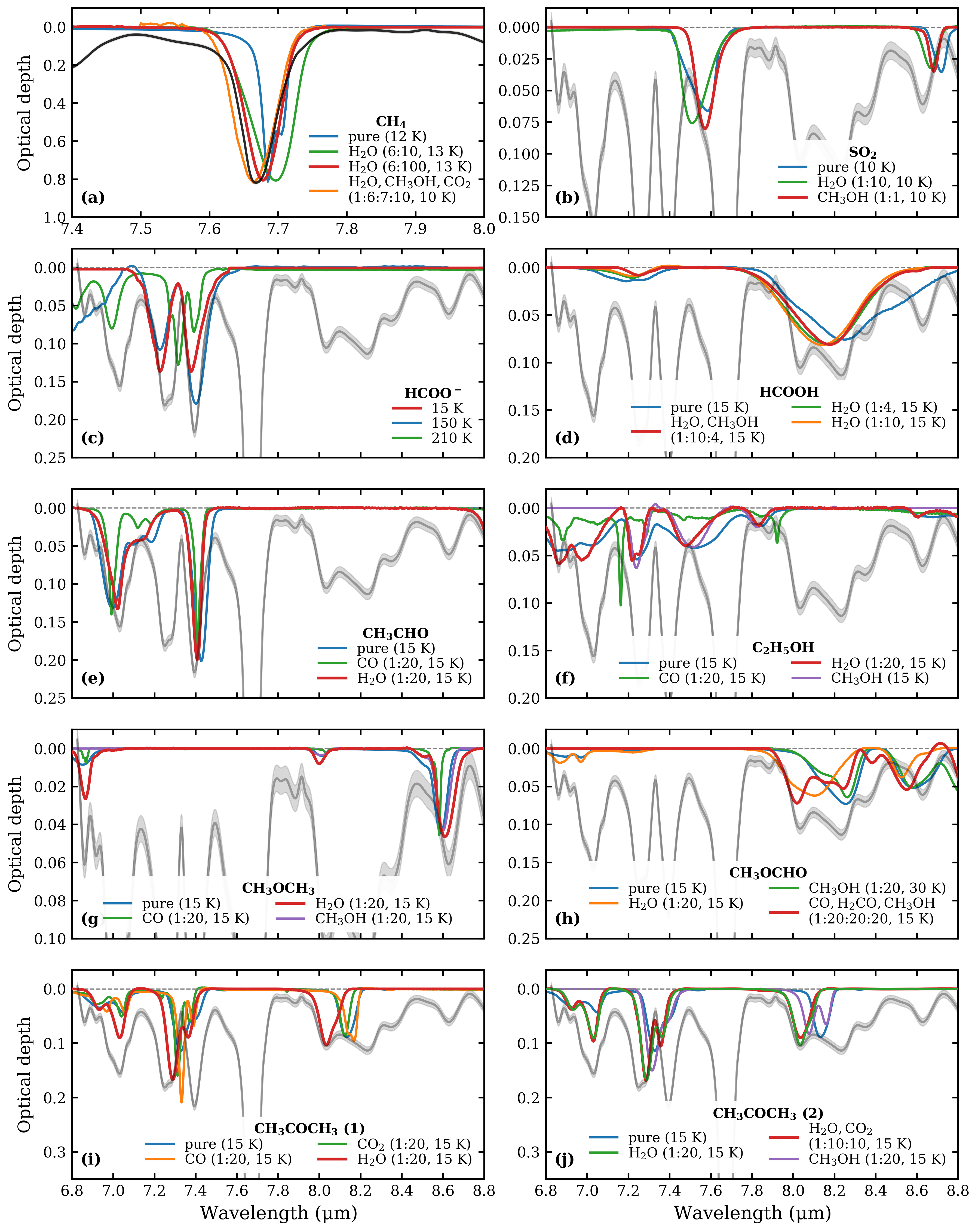}
    \caption{Comparison between the JWST/MIRI spectrum of B1-c (gray) and the lab spectra (colors) in the COMs fingerprints range of 6.8--8.8 $\mu$m. Each panel focuses on one species and shows the comparison between the observed B1-c spectrum and the lab spectra with different mixing constituents; except for panel (c) which shows the lab spectra of \ce{HCOO^-} under different temperatures. The observed spectrum along with the 3$\sigma$ level is shown in light gray; except for panel (a), which blows up the \ce{CH4} band at $\sim$7.7 $\mu$ and the observed spectrum is plotted in black for clarity. In each panel, the spectrum in blue corresponds to the pure ice, and the best-fit spectrum to the observations is highlighted with a thicker red line. Similar comparison but for lab spectra under different temperatures is shown in Figs.~\ref{fig:obs_vs_lab_T_1} and \ref{fig:obs_vs_lab_T_2}.} 
    \label{fig:obs_vs_lab_mix}
\end{figure*}

\subsubsection{Gas-phase lines}\label{sect:method_gas_lines}
There are plenty of gas-phase absorption lines superposed on the broad ice bands that need to be accounted for after subtracting the local continuum. An ideal but non-trivial solution would be simultaneously tracing a baseline and fitting LTE models to retrieve the column density and temperature of the gas-phase components. However, deriving the gas properties is not the focus of this paper, so we only considered tracing a baseline of these gas lines to isolate the ice bands that we are interested in. Similar to the local continuum, the baseline was determined by fitting a spline function to a series of guiding points that were set manually (Fig.~\ref{fig:fitting_steps}d). The smoothing factor of the spline function was also carefully tuned.


\begin{figure*}[!h]
    \centering
    \includegraphics[width=\linewidth]{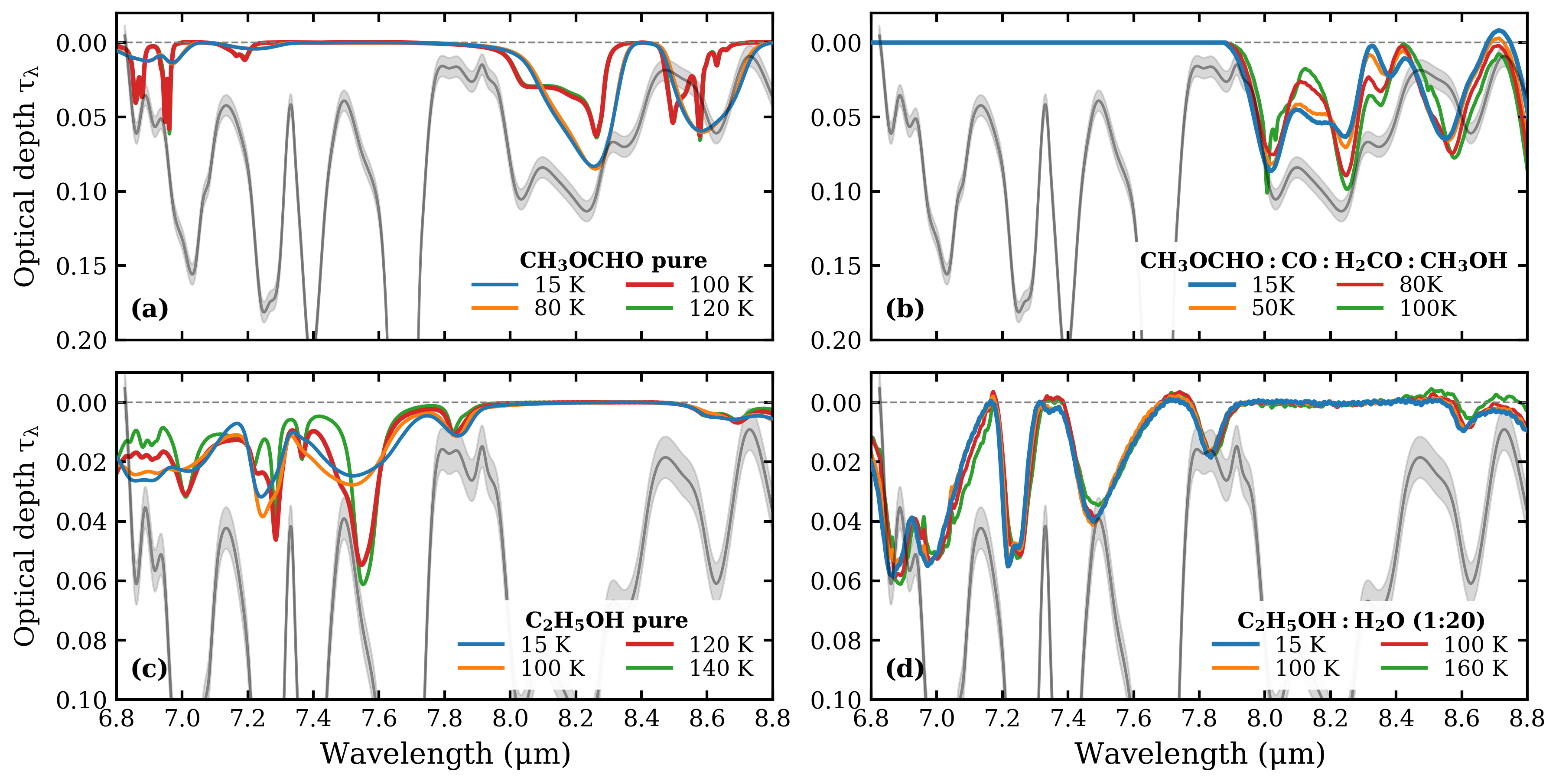}
    \caption{Same as Fig.~\ref{fig:obs_vs_lab_mix} but for comparison among different temperatures. Here we take two species, \ce{CH3OCHO} and \ce{C2H5OH}, as examples; the remaining species (\ce{CH3CHO}, \ce{CH3OCH3}, and \ce{CH3COCH3}) are shown in Fig.~\ref{fig:obs_vs_lab_T_2}. The left and right columns show the lab spectra of pure ices and ice mixtures, respectively. In the pure-ice panels, the spectra with crystalline features are highlighted in thicker red lines. The corresponding temperatures indicate the upper limit of crystallines. In the mixed-ice panels, the spectra with the lowest temperature (15 K) are highlighted in thicker blue lines; they are also the spectra used in the overall fitting.}
    \label{fig:obs_vs_lab_T_1}
\end{figure*}

\subsubsection{Absorption features of COM ices}\label{sect:method_fit_COM_bands}
The final step is to decompose the spectrum between 6.8 and 8.8~$\mu$m by fitting with lab spectra of different ices. However, this is not as straightforward as fitting the ALMA spectra where the LTE models can be analytically computed. When fitting the JWST spectra, the variables are not only the temperature and the ice column density of a species (or equivalently, the scaling factor of the corresponding lab spectrum), but also the mixing conditions of the ices. In the solid phase, a species can be mixed with various constituents in different ratios under different temperatures (e.g., \ce{CH3OH} mixed with \ce{H2O} in an abundance ratio of 1:10 at 15 K).
The relation between the band profile and the mixing condition, that is, mixing constituents, mixing ratio, and temperature, is not linear. Therefore, we need to first select out the mixing condition that matches best with the observations for each species (which is introduced in this subsection), and then do a least-squares fitting with the selected lab spectra of the candidate species to get a best fit on the ice column densities, or the scaling factors (which will be introduced in the next subsection).

These two steps can be executed either simultaneously or separately. The \texttt{ENIIGMA} code implemented in \cite{Rocha2024} is the first case, which calculates the chi-square of all the possible combinations of lab spectra and searches for the global minimum using genetic algorithms.
However, \texttt{ENIIGMA} is limited in 
the stability of fitting results due to the intrinsic randomness of genetic algorithms. In this work, we performed the two steps of selecting the lab spectra and fitting the scaling factors separately, which can also be used to crosscheck the results reported in \cite{Rocha2024}.

\paragraph{Selection of lab spectra.}
In the first step of this process, we selected a list of species that are likely to contribute to the absorption features between 6.8 and 8.8~$\mu$m. These species can be both COMs and simple molecules, and the promising candidates have been explored in \cite{Rocha2024}. In this work, we considered 12 molecules in total, including five simple species (\ce{CH4}, \ce{SO2}, \ce{OCN^-}, \ce{HCOO^-}, and \ce{HCOOH}) and seven COMs (\ce{CH3CHO}, \ce{C2H5OH}, \ce{CH3OCH3}, \ce{CH3OCHO}, \ce{CH3COOH}, \ce{CH3COCH3}, and \ce{CH3CN}).
Each species has a collection of lab spectra measured under different mixing conditions of ices (i.e., mixing constituent, mixing ratio, and temperature).
The next step is to determine under which mixing condition the lab spectrum matches best with the observations. We first did this by superposing all the lab spectra of a certain species on the JWST spectrum to see which one matches the observations best, as indicated by Figs.~\ref{fig:obs_vs_lab_mix}--\ref{fig:obs_vs_lab_T_1} and \ref{fig:obs_vs_lab_T_2}, in which each panel shows the comparison among different mixing constituents and temperatures for one species, respectively. For most candidate species, this is already very informative and efficient. 

\begin{figure*}[!h]
    \centering
    \includegraphics[width=\textwidth]{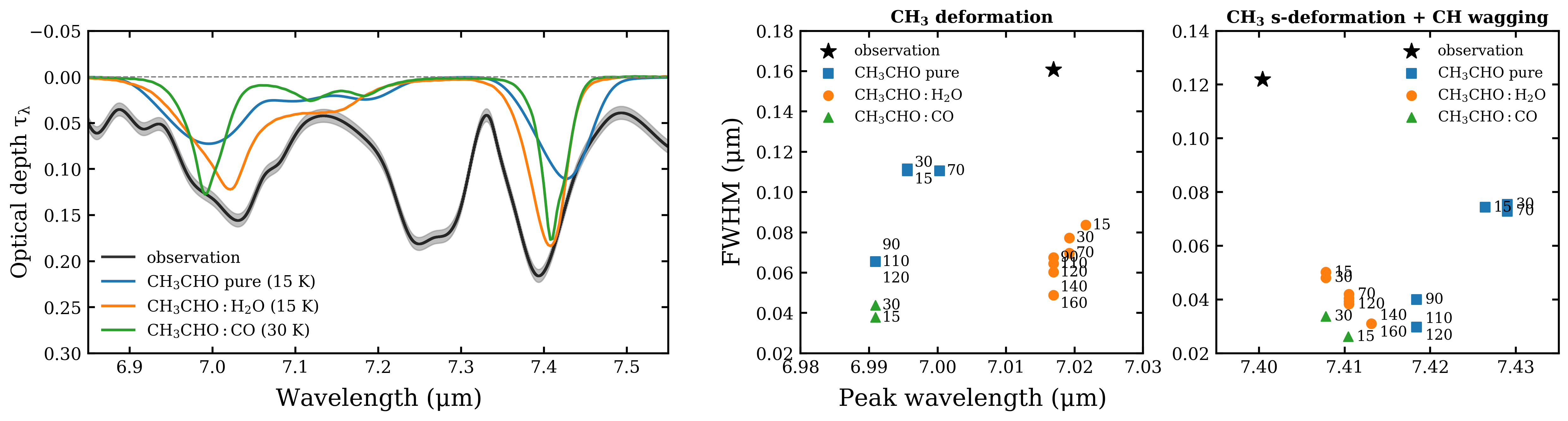}
    \caption{Two methods of selecting the best-fit ice mixture, taking \ce{CH3CHO} as an example. The left panel shows the direct comparison of spectral profiles between observations and experiments. The middle and the right panels show the comparison of peak position and FWHM of absorption bands between lab spectra and observations.}
    \label{fig:CH3CHO_bands}
\end{figure*}

However, if the differences among spectra are small and direct comparison is not straightforward, an alternative way is to compare the peak positions and FWHMs of the characteristic absorption bands of a certain species, when these data are available for the lab spectra \citep[e.g.,][]{Boogert1997, TvS2018, TvS2021, Rachid2020, Rachid2022}. The peak positions and FWHMs of observed bands are derived by fitting Gaussian functions to the observed spectrum.
Taking \ce{CH3CHO} as an example, Fig.~\ref{fig:CH3CHO_bands} shows the comparison between the lab spectra and the observed B1-c spectrum for two \ce{CH3CHO} bands at $\sim$7.0 (\ce{CH3} deformation mode) and $\sim$7.42~$\mu$m (\ce{CH3} s-deformation and CH wagging modes). It is already straightforward to tell from the left panel that the \ce{CH3CHO}:\ce{H2O} mixture fits best with the observations. The two panels on the right show that the observed spectrum has both bands obviously wider than the lab spectra, suggesting that these two bands are not only attributable to \ce{CH3CHO}, but also have contribution from other species. The peak positions of both bands match better with the \ce{H2O} mixture than the pure ice and the CO mixture, which further supports that the \ce{CH3CHO}:\ce{H2O} spectrum is the most suitable one to use in the overall fitting of scaling factors. As for temperature, lower temperatures are favored in the comparison. We chose the spectrum at 15 K for the overall fitting, but 30 K can also be used alternatively, since the difference in band profiles is small under 70 K. The selection of ice mixtures of other species follows the same procedure.

For COMs, there is usually only one mixing ratio that has been measured in laboratories for a specific mixture (e.g., COM:\ce{H2O} = 1:20), so we can only select among different mixing constituents and temperatures. For some simple species such as \ce{CH4}, there are more choices for mixing ratios, but within a limited temperature range. Despite the lack of additional laboratory measurements, in Sect.~\ref{sect:results_JWST} we will show that there is already a lot we can do with the current measurements. 

The lab spectra used in comparison had been baseline corrected to isolate the absorption bands of the targeted species from the strong features of the mixing constituents (e.g., \ce{H2O} and \ce{CH3OH}). This process is important but sometimes not trivial, especially for the ice mixtures with \ce{CH3OH}. The details are elaborated in Appendix~\ref{appendix:baseline_correction}.

\paragraph{Overall fitting of the COM fingerprint range 6.8--8.8 $\mu$m.}
In the second step, we performed an overall fitting on the scaling factors of all the selected lab spectra (each spectrum corresponding to the best-fit ice mixture of one species), as described in:
\begin{equation}\label{eq:linear_fitting}
    \tau_{\mathrm{obs}} = \sum_i^N a_i\,\tau_{i, \mathrm{lab}},
\end{equation}
where $\tau_\mathrm{obs}$ is the observed spectrum between 6.8 and 8.8~$\mu$m,  $a_i$ is the scaling factor of the $i$-th component, that is, the selected best-fit lab spectrum of the $i$-th species.  $N$ is the total number of the spectra (or species) that are considered in the fitting. 
We first performed a normal least-squares fitting to the observations using the \textit{Omnifit} code\citep{Omnifit} with all the candidate species taken into account, in order to get a general idea of the best-fit ranges. Then we selected a subgroup of significant candidates for a following Markov chain Monte Carlo \citep[MCMC;][]{emcee} fitting to better constrain the uncertainties on their scaling factors. To reduce the computation time, we set boundaries on the scaling factors to make sure that the fitting results fall in a reasonable range. The lower limits and the initial values are set as zero, and the upper limits are determined by manually comparing the lab spectra with observations (as done in Figs.~\ref{fig:obs_vs_lab_mix} and \ref{fig:obs_vs_lab_uplim}). 

\subsubsection{Ice bands of \ce{CH3OH} and \ce{H2O}}\label{sect:method_CH3OH_H2O_band}
As the simplest and the most abundant COM, \ce{CH3OH} is usually taken as the reference species to calculate relative abundances (i.e., ratios) of other COMs. These abundance ratios are generally compared among different sources to see if there are similarities or differences. The abundance of \ce{CH3OH} ice was derived separately since its strongest mid-IR band is located at 9.74~$\mu$m, outside of the 6.8--8.8~$\mu$m range. The \ce{CH3OH} band at 9.74~$\mu$m can be isolated by fitting a local continuum on either the original optical depth spectrum \citep{Bottinelli2010}, or the silicate subtracted spectrum. However, the SNR of this band is usually much lower than other parts of the spectrum due to the strong extinction by the silicate 9.8~$\mu$m band.
Here we fit a sixth-order polynomial to the original optical depth spectrum in three wavelength ranges: 8.40--8.80, 9.23--9.45, and 10.02--10.40~$\mu$m (see Fig.~\ref{fig:fitting_steps}d). 
The local continuum subtracted spectrum between 9.5 and 10.0~$\mu$m was then fit by a Gaussian function. 
We chose to use a Gaussian function instead of the lab spectra for simplicity, since this band is isolated and the fitting results would be similar. 

\ce{H2O} is also a reference species commonly used to calculate relative abundances, especially for simple molecules. We fit the \ce{H2O} band at 13~$\mu$m in the silicate subtracted spectrum using the lab spectra recently measured by \cite{Slavicinska2024_HDO}. The weaker 6~$\mu$m band was taken as a secondary reference. Since this band is very broad and strong, and its spectral profile changes prominently with temperature, it is easy to select the best fitting \ce{H2O} spectrum to the observations and adjust the scaling factor accurately through visual inspection. 

\subsubsection{Ice column density}
Once the best-fit scaling factors are found for the selected lab spectra (for COMs in Sect.~\ref{sect:method_fit_COM_bands}), we can calculate the ice column densities of each species by picking a vibrational band, usually the strongest one in the interested wavelength range, and using:
\begin{equation}\label{eq:N_ice}
    N_\mathrm{ice} = \frac{1}{A} \int_{\nu_1}^{\nu_2} a\,\tau_{\nu,\mathrm{lab}}\,\mathrm{d}\nu,
\end{equation}
where $A$ is the strength of this band that extends from wavenumber $\nu_1$ to $\nu_2$, $\tau_{\nu,\mathrm{lab}}$ is the lab spectrum at wavenumber $\nu$ in optical depth scale, and $a$ is the scaling factor of this spectrum compared to the observations. In reality, the IR spectra are measured in absorbance ($Abs$), and the conversion to optical depth is given by Eq.~\ref{eq:tau_abs}.
In cases that we use a Gaussian function to represent the absorption band (e.g., for the \ce{CH3OH} band in Sect.~\ref{sect:method_CH3OH_H2O_band}), $\tau_{\nu,\mathrm{lab}}$ is replaced by the best-fit Gaussian function and the scaling factor $a$ is not needed.

\section{Results}\label{sect:results}
\subsection{ALMA}
\subsubsection{Emission maps}\label{sect:result_ALAM_images}
The continuum emission of IRAS~2A and B1-c appears to be roughly symmetric and round, even at the high angular resolution of $\sim$0.1$\arcsec$ (i.e., 32 au at distance of 320 pc; see contours in Figs.~\ref{fig:IRAS2A_ALMA_mom0}--\ref{fig:B1c_ALMA_mom0}), which is consistent with previous observations \citep{Taquet2015, Segura-Cox2018, vanGelder2020, Yang2021_PEACHES}. 
The COM emission shows consistent morphologies among different species and transitions. In B1-c, the COM emission is symmetric and follows the morphology of the continuum, suggesting this protostellar system is fairly quiescent. Conversely, the morphologies in IRAS~2A are asymmetric, with the emission attenuated at the continuum peak and stronger at an offset position to the southwest. This asymmetry was not revealed in previous observations with lower angular resolution, and may be related to the dynamics of the protostar and the circumstellar disk \citep[e.g., a recent outburst;][]{Hsieh2019}.

The emission maps of multiple COM lines shown in Figs.~\ref{fig:IRAS2A_ALMA_mom0}--\ref{fig:B1c_ALMA_mom0} also reflect how large the hot cores are. The size of a hot core is usually considered as the radius at which $T$ = 100 K, at which temperature most of the volatile species have sublimated from ice mantles into the gas phase. This radius can be estimated analytically by the following equation \citep{Bisschop2007, vHoff2022}:
\begin{equation}\label{eq:R_T100K}
    R_\mathrm{T=100\ K} \approx 15.4 \sqrt{L/L_\odot}\ \mathrm{au},
\end{equation}
Taking the luminosity of 91 and 5.9 $L_\odot$ for IRAS~2A and B1-c \citep{vanGelder2022a}, the $R_\mathrm{T=100\ K}$ are estimated as 147 au and 37 au (i.e., $\sim$4 and 1 beams), respectively. 
The spectra of both sources were extracted within $R_\mathrm{T=100\ K}$, hence the derived column densities are also expected to be representative of the gas-phase COM abundances inside the hot cores.

\begin{table*}[!h]
    \setlength{\tabcolsep}{0.08cm}
    \caption{Column densities and excitation temperatures of gas-phase molecules derived from the ALMA Band 7 spectra.}
    \centering
    \begin{tabular}{lccccccccccc}
    \toprule
    \multicolumn{2}{c}{Sources} & \multicolumn{5}{c}{NGC 1333 IRAS~2A} & \multicolumn{5}{c}{B1-c}\\
    \cmidrule(lr{0.3em}){3-7}\cmidrule(lr{0.3em}){8-12}
     & & $N_\mathrm{gas}$ & $T_\mathrm{ex}$ & FWHM & $v_\mathrm{lsr}$ & X/\ce{CH3OH} & $N_\mathrm{gas}$ & $T_\mathrm{ex}$ & FWHM & $v_\mathrm{lsr}$ & X/\ce{CH3OH}\\
    Species & Catalog & (cm$^{-2}$) & (K) & \multicolumn{2}{c}{(km s$^{-1}$)} & (\%) & (cm$^{-2}$) & (K) & \multicolumn{2}{c}{(km s$^{-1}$)} & (\%) \\
    \midrule
    \ce{CH3OH} & CDMS & (1.1$\pm$0.3)$\times10^{19}$ & 165$\pm$10 & 3.5 & 5.0 & $\equiv$100 & (9.0$\pm$3.0)$\times10^{18}$ & 175$\pm$20 & 3.5 & 6.0 & $\equiv$100 \\
     & & (7.0$\pm$2.0)$\times10^{18}$ & 210$\pm$10 & 3.0 & 8.5 & & & & & & \\
    \ce{^{13}CH3OH} & CDMS & >5.0$\times10^{16}$ & 170$\pm$20 & 3.0 & 5.3 & >0.44 & >2.5$\times10^{16}$ & [175] & 3.3 & 5.8 & >0.28\\
    & & >3.0$\times10^{16}$ & 210$\pm$20 & 3.0 & 8.3  & & & & & & \\
    \ce{CH3^{18}OH} & CDMS & $\leq$3.0$\times10^{16}$ & [170] & [3.0] & [5.0] & $\leq$0.39 & $\leq$3.0$\times10^{16}$ & 100--200 & [3.0] & [5.8] & $\leq$0.33 \\
     & & $\leq$4.0$\times10^{16}$ & [210] & [3.0] & [8.5] & & & & & & \\
    \ce{CH2DOH} & JPL & (1.3$\pm$1.0)$\times10^{17}$& 180$\pm$30 & 3.0 & 4.8 & 1.7 & (5.0$\pm$1.0)$\times10^{17}$ & 200$\pm$40 & 3.0 & 6.0 & 5.6\\
    & & (1.7$\pm$1.0)$\times10^{17}$& 220$\pm$20 & 3.0 & 7.9 & & & & & & \\
    \ce{CH3CHO} & JPL & (1.0$\pm$0.3)$\times10^{16}$ & 170$\pm$50 & 2.8 & 5.3 & 0.14 & (2.0$\pm$0.4)$\times10^{16}$ & 210$\pm$50 & 3.3 & 6.0 & 0.22 \\
    & & (1.5$\pm$0.5)$\times10^{16}$ & 220$\pm$50 & 2.8 & 8.0 & & & & & & \\
    \ce{C2H5OH} & CDMS & (4.5$\pm$1.5)$\times10^{16}$ & 220$\pm$50 & 3.0 & 5.0 & 0.58 & (7.0$\pm$2.0)$\times10^{16}$ & 250$\pm$50 & 3.0 & 6.0 & 0.78 \\
     & & (6.0$\pm$1.5)$\times10^{16}$ & 220$\pm$50 & 3.0 & 8.0 & & & & &  \\
    \ce{CH3OCH3} & CDMS & $\leq$2.0$\times10^{17}$ & [200] & [3.0] & -- & $\leq$1.1 & $\leq$6.0$\times10^{16}$ & [200] & [3.0] & -- & $\leq$0.67 \\
    \ce{CH3OCHO} & JPL & (7.0$\pm$2.0)$\times10^{16}$ & 170$\pm$30 & 3.0 & 5.0 & 0.83 & (1.5$\pm$0.4)$\times10^{17}$ & 150$\pm$30 & 3.3 & 6.0 & 1.7\\
     & & (8.0$\pm$2.0)$\times10^{16}$ & 170$\pm$30 & 3.0 & 8.0 &  &  &  &  & \\
     \ce{CH2OHCHO} & CDMS & $\leq$1.0$\times10^{16}$ & [200] & [3.0] & 6.5 & $\leq$0.06 & $\leq$2.0$\times10^{16}$ & [200] & [3.0] & 6.0 & $\leq$0.22 \\
    a-\ce{(CH2OH)2} & CDMS & (5.0$\pm$1.5)$\times10^{16}$ & 220$\pm$50 & 3.5 & 5.8 & 0.47 & (2.5$\pm$0.5)$\times10^{16}$ & 200$\pm$40 & 3.0 & 6.0 & 0.28 \\
     & & (3.5$\pm$1.0)$\times10^{16}$ & 220$\pm$50 & 2.0 & 8.0 &  &  &  &  &  & \\
    g-\ce{(CH2OH)2} & CDMS & (2.0$\pm$0.5)$\times10^{16}$ & 220$\pm$30 & 2.5 & 5.5 & 0.33 & (3.0$\pm$0.5)$\times$10$^{16}$ & 200$\pm$40 & 2.8 & 6.0 & 0.33 \\
     & & (4.0$\pm$1.0)$\times10^{16}$ & 220$\pm$30 & 2.5 & 8.0 &  &  &  &  &  & \\
    t-\ce{HCOOH} & CDMS & (9--12)$\times10^{16}$ & 100--300 & 4.5--5.5 & 7.0 & 0.5--0.67 & (1.8--3.2)$\times10^{16}$ & 100--300 & 4.0 & 6.5 & 0.2--0.36\\
    \ce{H2CCO} & CDMS & (3.8--6.0)$\times10^{15}$ & 100--300 & 2.5 & 4.8 & 0.11--0.17 & (5.0--7.5)$\times10^{15}$ & 100--300 & 3.0 & 6.0 & 0.06--0.08\\
     & & (6.5--9.5)$\times10^{15}$ & 100--300 & 2.8 & 7.7 & & & & & & \\
    \ce{CH2DCN} & CDMS & (2.0$\pm$0.5)$\times10^{15}$ & 200$\pm$50 & 4.0 & 7.2 & 0.01 & (1.3$\pm$0.3)$\times10^{15}$ & 200$\pm$60 & 3.0 & 6.0 & 0.014 \\
    \ce{NH2CHO} & JPL & (1.3--1.9)$\times10^{16}$ & 100--300 & 4.2--5.0 & 6.9 & 0.07--0.1 & (1.5--2.0)$\times10^{15}$ & 100--300 & 3.0 & 6.0 & 0.01--0.02 \\
    \ce{C2H5CN} & JPL & (2.0--3.5)$\times10^{15}$ & 100--300 & 3.0 & 5.0 & 0.04--0.08 & (3.5--6.0)$\times10^{15}$ & 100--300 & 3.0 & 6.0 & 0.04--0.07\\
     & & (2.0--3.5)$\times10^{15}$ & 100--300 & 3.0 & 8.0 & & & & & & \\
    \bottomrule
    \end{tabular}
    \label{tab:ALMA_results}
\end{table*}

\subsubsection{Spectra}
In the nine spectral windows between 333.8 and 347.6 GHz in ALMA Band 7, rich molecular lines are observed in both sources. We determined column densities ($N$) and excitation temperatures ($T_\mathrm{ex}$) for six O-bearing COMs and one N-bearing COM that have enough clean lines detected; they are \ce{CH3OH}, \ce{CH2DOH}, \ce{CH3CHO}, \ce{C2H5OH}, \ce{CH3OCHO}, \ce{(CH2OH)2}, and \ce{CH2DCN}. In addition, we detect several strong lines of \ce{CHD2OH} and \ce{CD3OH} in both IRAS~2A and B1-c, implying a high deuteration rate of \ce{CH3OH} in low-mass protostars \citep[e.g., ][]{Drozdovskaya2022}, but this will not be studied in this paper.

For other species that are detected but do not have enough clean lines to constrain $N$ and $T_\mathrm{ex}$ independently, we provide either upper limits or a range of $N$ assuming $T_\mathrm{ex}$ = 100--300 K, which is typical for hot cores. These species include the \ce{^{13}C} and \ce{^{18}O} isotopologues of \ce{CH3OH}, \ce{CH3OCH3}, and \ce{CH2OHCHO}. \ce{^{13}CH3OH} only has two strong lines covered in our spectral setup, and they are likely to be optically thick given their high Einstein A coefficients ($A_\mathrm{ij}>4\times10^{-4}$ s$^{-1}$) and the low column density ratio with respect to the main isotopologue. The derived \ce{CH3OH}/\ce{^{13}CH3OH} ratios are about 200, larger than the $^{12}$C/$^{13}$C ratio in the vicinity of the Solar System (60--70), which implies that the column densities of \ce{^{13}CH3OH} were underestimated due to optically thick lines. It is also difficult to constrain the $T_\mathrm{ex}$ with both of their upper energy levels $E_\mathrm{up}$ > 190 K. Therefore, a lower limit of $N$ is provided for \ce{^{13}CH3OH} at the same $T_\mathrm{ex}$ as the main isotopologue. \ce{CH3^{18}OH} and \ce{CH3OCH3} only have weak transitions ($A_\mathrm{ij}$<10$^{-6}$ s$^{-1}$) covered in our data, therefore upper limits were estimated assuming a fixed $T_\mathrm{ex}$ and line width (FWHM) at 3~km~s$^{-1}$. The $T_\mathrm{ex}$ of \ce{CH3^{18}OH} is fixed at the same value of \ce{CH3OH}, and the $T_\mathrm{ex}$ of \ce{CH3OCH3} is fixed at 200 K given that the covered transitions have high $E_\mathrm{up}$ of > 500~K. \ce{CH2OHCHO} have $\sim$10 lines detected, but most of them are blended with other stronger lines, thus its column densities are reported as upper limits as well.

Besides \ce{CH2DCN}, we also detected two N-COMs, \ce{NH2CHO} and \ce{C2H5CN}, and each of them have 2--3 strong lines covered in our data. Although these lines are unblended, they share similar $E_\mathrm{up}$ and therefore $N$ and $T_\mathrm{ex}$ are degenerate with each other. In this case, we report $N$ under $T_\mathrm{ex}$ = 100--300 K.
Two 5-atom molecules, ketene (\ce{H2CCO}) and trans-formic acid (t-HCOOH), each have one strong line detected. They are often studied along with O-COMs given that they may serve as precursors of O-COMs in their formation routes. The column densities of \ce{H2CCO} and t-HCOOH are also estimated under $T_\mathrm{ex}$ = 100--300 K because of the degeneracy between $N$ and $T_\mathrm{ex}$.
Other species such us the isotoplogues of abundant simple molecules (HDO, HDCO, \ce{H^{13}CN}), sulfur-bearing molecules (SO, \ce{SO2}, and their isotopologues), and carbon-chain molecules (\ce{HC3N}, $c$- and $l$-\ce{C3H2}), also have one or two strong lines detected in the spectra. 

The best-fit column densities and excitation temperatures of the aforementioned species along with several other COMs and simple molecules are listed in Table~\ref{tab:ALMA_results}. The LTE-modelled spectra along with the observed ALMA spectra of IRAS~2A and B1-c are displayed in Fig.~\ref{fig:ALMA_fit_IRAS2A}--\ref{fig:ALMA_fit_B1-c}. The transitions that were considered in the LTE fitting of ALMA spectra are listed in Table~\ref{table:transition}.

There are two special cases worth mentioning. The first is the determination of the \ce{CH3OH} column densities. Usually, \ce{CH3^{18}OH} is used to infer the column density of \ce{CH3OH} assuming a $^{16}$O/$^{18}$O ratio. This is because \ce{CH3OH} and \ce{^{13}CH3OH} tend to be optically thick, and directly fitting their lines may lead to underestimation of their column densities. In our data, \ce{CH3^{18}OH} was not robustly detected, but several \ce{CH3OH} lines with very high $E_\mathrm{up}$ (> 1000 K) or very low Einstein A coefficients ($A_\mathrm{ij}\sim$10$^{-7}$ s$^{-1}$) were detected. These lines are intrinsically much weaker and therefore more likely to be optically thin. The column densities of \ce{CH3OH} were determined based on those high-$E_\mathrm{up}$ or low-$A_\mathrm{ij}$ lines. Our fitting results show column density ratios of \ce{CH3^{18}OH}/\ce{CH3OH} < 0.4\% for IRAS~2A and B1-c (Table~\ref{tab:ALMA_results}), which are consistent with the expected ratio of 0.18\% assuming $^{16}$O/$^{18}$O$\sim$560 in the local ISM \citep{Wilson1994}.

The second case is the double-peaked features observed in the emission lines of IRAS~2A, which are not rarely seen in hot cores \citep[e.g.,][]{Chen2023}, and probably due to the dynamics of a circumstellar disk \citep{Nazari2024_diskwinds}. The spectrum can be well fit by two velocity components with different $N$ (and sometimes different $T_\mathrm{ex}$ and FWHM). By contrast, B1-c only shows one velocity component, even at the continuum peak.

The column densities ratios with respect to \ce{CH3OH} were calculated to facilitate the comparison with other sources or results in other studies. As famous hot corino sources, IRAS~2A and B1-c have been targeted in previous observations \citep{Taquet2015, Yang2021_PEACHES, vanGelder2020}. In particular, the COM ratios in B1-c reported by \cite{vanGelder2020} are consistent with our fitting results within 50\% for O-COMs, and even within 10\% for \ce{CH3CHO} and \ce{C2H5OH}. For IRAS~2A, the COM ratios reported by \cite{Taquet2015} and \cite{Yang2021_PEACHES} are factors of 2--3 higher than our results, which is likely because their angular resolution or sensitivity was not as good as ours (insufficient angular resolution may lead to beam dilution, and insufficient sensitivity would hamper robust detection of some species). 

\subsection{JWST}\label{sect:results_JWST}
The continuum emission of IRAS~2A and B1-c is spatially unresolved with JWST/MIRI-MRS (as shown in Fig.~\ref{fig:B1c_JWST_spec_extraction} for B1-c), therefore we only focus on spectral analysis. Studies on COM ices in IRAS~2A have been carried out by \cite{Rocha2024}, and here we perform a similar analysis for B1-c.

Based on the selection described in \ref{sect:method_fit_COM_bands}, we first select the lab spectra with the mixing conditions that fit best with the observed B1-c spectrum (one spectrum for each candidate species), and then perform overall fittings to constrain the scaling factors of these lab spectra. A certain mixing condition refers to a certain combination of mixing constituent, mixing ratio, and temperature, among which mixing constituent has the most important influence on band profiles. The availability of mixing ratios in lab databases is limited for some species (especially COMs), therefore, we only focus on mixing constituents (Sect.~\ref{sect:results_JWST_mixing_consituent}) and temperatures (Sect.~\ref{sect:results_JWST_temperature}).

\subsubsection{Constituents of ice mixtures}\label{sect:results_JWST_mixing_consituent}
Figure~\ref{fig:obs_vs_lab_mix} compares the lab spectra of different mixing constituents with the observations (panels a--j correspond to \ce{CH4}, \ce{SO2}, \ce{HCOO^-}, \ce{HCOOH}, \ce{CH3CHO}, \ce{C2H5OH}, \ce{CH3OCH3}, \ce{CH3OCHO}, and \ce{CH3COCH3}). \ce{OCN^-} and \ce{CH3COOH} are not shown in this figure since they only have lab data for one mixing constituent. \ce{CH3CN} is also excluded because its three bands between 6.8 and 7.4 $\mu$m have very similar profiles among different mixing constituents. The selection of \ce{CH3CN} mixture was based on the results of \cite{Nazari2024_CH3CN}, who reported tentative detections of \ce{CH3CN} ice in the Near Infrared Spectrograph (NIRSpec) data, and the \ce{CH3CN}:\ce{H2O}:\ce{CO2} mixture is the main contributor of the observed band at 4.43 $\mu$m. The comparison between the B1-c spectrum and the lab spectra of \ce{OCN^-}, \ce{CH3COOH}, and \ce{CH3CN} ices are shown in Fig.~\ref{fig:obs_vs_lab_uplim}.
In the following paragraphs, we introduce the comparison results for each species following the order of display in Fig.~\ref{fig:obs_vs_lab_mix}.

\paragraph{\ce{CH4}.} In B1-c, the \ce{CH4} band at 7.67 $\mu$m is superposed by gas-phase \ce{CH4} lines in absorption, which have been removed before fitting with lab spectra (Sect.~\ref{sect:method_gas_lines}). However, none of the existing lab spectra fits perfectly with the observations in terms of peak position. The best two candidates are the \ce{H2O} mixture with a mixing ratio of 6:100, and a more complex mixture with \ce{H2O}, \ce{CH3OH}, and \ce{CO2}. The 7.67~$\mu$m band of these two mixtures is slightly redshifted and blueshifted from the observations, respectively. The same band of the pure \ce{CH4} ice and the \ce{CH4}:\ce{H2O} (6:10) mixture is more redshifted and too narrow as well. This suggests that in reality the surrounding of \ce{CH4} ice is dominated by \ce{H2O}, and there are also other species present. We finally chose the \ce{CH4}:\ce{H2O} (6:100) mixture to fit the observations considering that the mixing ratio of the \ce{CH4}:\ce{H2O}:\ce{CH3OH}:\ce{CO2} (1:6:7:10) mixture is not very reasonable (too little \ce{H2O} and too much \ce{CH3OH}), but the fitting results are expected to be similar using either of the spectra.

\paragraph{\ce{SO2}.} The 7.6 $\mu$m band of the \ce{CH3OH} mixture fits best with the blue wing of the observed \ce{CH4} band. Half of this band is overlapping with the \ce{OCN^-} band at 7.62 $\mu$m, and may lead to degeneracy in the overall fittings (see Sect.~\ref{sect:results_fit_stats}).

\paragraph{\ce{HCOOH}.} the \ce{H2O} mixtures fit the observations better than the pure HCOOH. The difference between the \ce{H2O} and \ce{H2O}:\ce{CH3OH} mixtures at $\sim$8.2 $\mu$m is tiny, only that the \ce{H2O}:\ce{CH3OH} mixture is slightly redshifted and fits the observed band better. The absorption features shortward of 7.0~$\mu$m in the HCOOH:\ce{H2O}:\ce{CH3OH} spectrum belong to \ce{CH3OH}, and was excluded during baseline correction (see Appendix~\ref{appendix:baseline_correction}).

\paragraph{\ce{CH3CHO}.} It has been discussed as an example in Sect.~\ref{sect:method_fit_COM_bands} that the \ce{H2O} mixture fits best with the observations. The pure ice and CO mixture have their 7.0 and 7.42 $\mu$m bands deviated from observations.

\paragraph{\ce{C2H5OH}.} It is a bit hard to select among the pure ice, the \ce{H2O} mixture, and the \ce{CH3OH} mixture. The pure \ce{C2H5OH} matches the observations better at the 7.2~$\mu$m band, but its other bands are too wide. The \ce{H2O} mixture fits better in 6.8--7.2~$\mu$m, but its 7.24 $\mu$m band is slightly blueshifted from the observations. The \ce{CH3OH} mixture has similar band profiles to the \ce{H2O} mixture, but the \ce{C2H5OH} bands between 6.8 and 7.2 $\mu$m are blended with the \ce{CH3OH} bands, and it is difficult to accurately separated in the \ce{CH3OH} mixture. We finally chose the \ce{C2H5OH}:\ce{H2O} mixture considering that \ce{H2O} is the dominant species in ice mantles, and the cases for some other O-COMs also show that \ce{H2O}-rich mixtures suit better than pure ices. The \ce{CH3OH} is a promising candidate as well, but not selected for technical reasons.

\paragraph{\ce{CH3OCH3}.} Except for the CO mixture, the pure ice and other two mixtures (with \ce{H2O} and \ce{CH3OH}) have similar band profiles at 8.59 $\mu$m. The \ce{H2O} has the best match with observations in terms of peak position and band width, while the \ce{CH3OH} mixture can not be fully excluded.

\paragraph{\ce{CH3OCHO}.} The mixture with CO, \ce{H2CO}, and \ce{CH3OH} fits the observations at 8.1--8.4 $\mu$m better than the \ce{H2O} mixture (panel h of Fig.~\ref{fig:obs_vs_lab_mix}). In the \ce{H2O} mixture, the C--O stretching band of \ce{CH3OCHO} at 8.25 $\mu$m is significantly blueshifted and smoothed, which does not reproduce the observed profile. This suggests that \ce{CH3OCHO} is more likely to be formed in a CO-rich environment other than a \ce{H2O}-rich one. However, a caveat exists that the 8.02~$\mu$m band in the lab spectrum of the \ce{CH3OCHO}:CO:\ce{H2CO}:\ce{CH3OH} mixture is mainly contributed by \ce{H2CO}, not by \ce{CH3OCHO}. This raises the question how realistic is this mixing ratio between \ce{CH3OCHO} and \ce{H2CO} (1:20), considering that the observed 8.03 $\mu$m band also likely has a contribution from the \ce{CH3COCH3}:\ce{H2O} mixture (panel i of Fig.~\ref{fig:obs_vs_lab_mix}). The relative strength of the \ce{H2CO} band will affect our estimation on the ice abundance of \ce{CH3COCH3}. A more detailed discussion on how to deal with the \ce{H2CO} band blended in the lab spectrum of the \ce{CH3OCHO}:CO:\ce{H2CO}:\ce{CH3OH} mixture is provided in Appendix~\ref{appendix:H2CO}.

\paragraph{\ce{CH3COCH3}.} There are more mixtures measured in laboratories than other O-COMs \citep{Rachid2020}, and the comparison is separated into two panels (i and j) in Fig.~\ref{fig:obs_vs_lab_mix}. Panel i shows that the \ce{H2O} mixture has all the bands blueshifted from the pure ice and the CO or \ce{CO2} mixtures. In particular, the CCC asymmetric stretching band is significantly blueshifted from 8.14 to 8.03 $\mu$m. The double-peaked \ce{CH3} symmetric deformation band at $\sim$7.33 $\mu$m is also blueshifted. These blueshifts make the \ce{H2O} mixture match the observations much better. Panel j shows that the \ce{CH3OH} mixture has the 8.14 $\mu$m band split into two peaks, not matching the observations. The mixture with \ce{H2O} and \ce{CO2} has almost the same spectrum as the \ce{H2O} only mixture, but the 8.03 $\mu$m band is slightly weaker compared to the 7.3 $\mu$m band. The relatively weaker 8.03 $\mu$m band is favored in the overall fittings since the observed B1-c spectrum tends to be overfit at this position by a combination of \ce{H2CO} and \ce{CH3COCH3} (see Sect.~\ref{sect:results_JWST_decomposition}). We finally chose the \ce{CH3COCH3}:\ce{H2O}:\ce{CO2} (1:10:10) mixture for the overall fittings, but the difference would be small if using \ce{CH3COCH3}:\ce{H2O} (1:20).\\

In summary, our comparison between the lab spectra and the observations reveals that most COM ices (except for \ce{CH3OCHO}) are surrounded by a \ce{H2O}-rich environment. For some species such as \ce{C2H5OH} and \ce{CH3OCH3}, mixing with \ce{CH3OH} cannot be ruled out. The possibility of CO-dominated mixtures is low, probably due to its desorption above 20 K.
\ce{CH3OCHO} is an outlier that its surrounding is not dominated by \ce{H2O}, but rich in CO, \ce{H2CO}, and \ce{CH3OH}, implying a formation route of CO hydrogenation. However, this set of mixing constituents (CO+\ce{H2CO}+\ce{CH3OH}) is only measured for \ce{CH3OCHO}, hence it is too early to conclude that \ce{CH3OCHO} has a different formation route than other COMs.\\


\subsubsection{Temperature of ice mixtures}\label{sect:results_JWST_temperature}
Unlike mixing constituents, varying temperature only induces very small differences in band profiles as long as the temperature is below the crystallization point. 
Figure~\ref{fig:obs_vs_lab_T_1} shows the comparison between the observations and the lab spectra of COM ices under different temperatures. Two species, \ce{CH3OCHO} and \ce{C2H5OH} are shown as examples of two scenarios, and the remaining three COMs (\ce{CH3CHO}, \ce{CH3OCH3}, and \ce{CH3COCH3}) are shown in Fig.~\ref{fig:obs_vs_lab_T_2}. For all the considered COMs, the pure ices show significant differences in the band profiles when transiting from amorphous to crystalline state. As temperature increases, the bands become narrower and sharper, some even split into two bands. For some species such as \ce{CH3OCHO}, the changes in band profiles during this transition are also distinct in the ice mixtures. Because of the dilution of other constituents (e.g., \ce{H2O}), the band width may remain similar, but the relative intensity or peak position of each band will change significantly after crystallization. On the other hand, the band profiles of \ce{C2H5OH}:\ce{H2O} mixture only show small changes after crystallization, which are very difficult to distinguish when compared with observations. The details of other three COMs are described in Appendix~\ref{appendix:obs_vs_lab}. In general, \ce{CH3OCH3} and \ce{CH3COCH3} are of the same type as \ce{CH3OCHO}; and \ce{CH3CHO} is more like \ce{C2H5OH}.

By comparing the lab spectra under different temperatures, we can infer the range of crystalline temperature under laboratory conditions ($T_\mathrm{crystal,\,lab}$) for pure and mixed COM ices . We also compared the lab spectra with the JWST spectrum of B1-c and constrained the temperature ranges of the detected COM ices (summarized in Table~\ref{tab:T_crystal}). For pure COM ices,  $T_\mathrm{crystal,\,lab}$ is $\lesssim$100 K. Under astrophysical conditions, $T_\mathrm{crystal}$ is usually 20\%--40\% lower. For mixed COM ices, $T_\mathrm{crystal,\,lab}$ is slightly higher, $\gtrsim$100 K. Despite noticeable changes in band profiles of crystalline ices, the observations can only constrain the laboratory temperature $T_\mathrm{lab}$ of the detected COM ices up to $\sim$100 K (equivalent to 60--80 K in space).

Similar degeneracy is also found for $T_\mathrm{lab}<70$ K in IRAS~2A using the \texttt{ENIIGMA} fitting code \citep{Rocha2024}. Although IRAS~2A shows evidence of more thermal processing than B1-c by its double-peaked \ce{CO2} band at 15.2 $\mu$m \citep[e.g.,][]{Brunken2024}, the difference in thermal processing is hardly manifested in the band profiles of COM ices. This means that we cannot tell how many COM ices are as cold as 10 K and how many are as warm as 60 K; even if we know, we are still not able to distinguish whether these ices are formed in the cold collapse stage ($\sim$10 K in space) or the subsequent warm-up stage, since the ices could form in cold environment but then be heated. The only known information is that the observed bands of COM ices are relatively broad and smooth, which is not in favor of the sharp spectral features of crystalline ices. This does not rule out of the presence of crystalline ices, but suggests the observed ices to be mainly amorphous ($T<T_\mathrm{crystal}$), which is reasonable considering that only a small part of the envelope is heated by the protostar to a temperature as high as $T_\mathrm{crystal}$.

\begin{figure*}[!h]
    \centering
    \begin{subfigure}[b]{\textwidth}
        \centering
        \includegraphics[width=\textwidth]{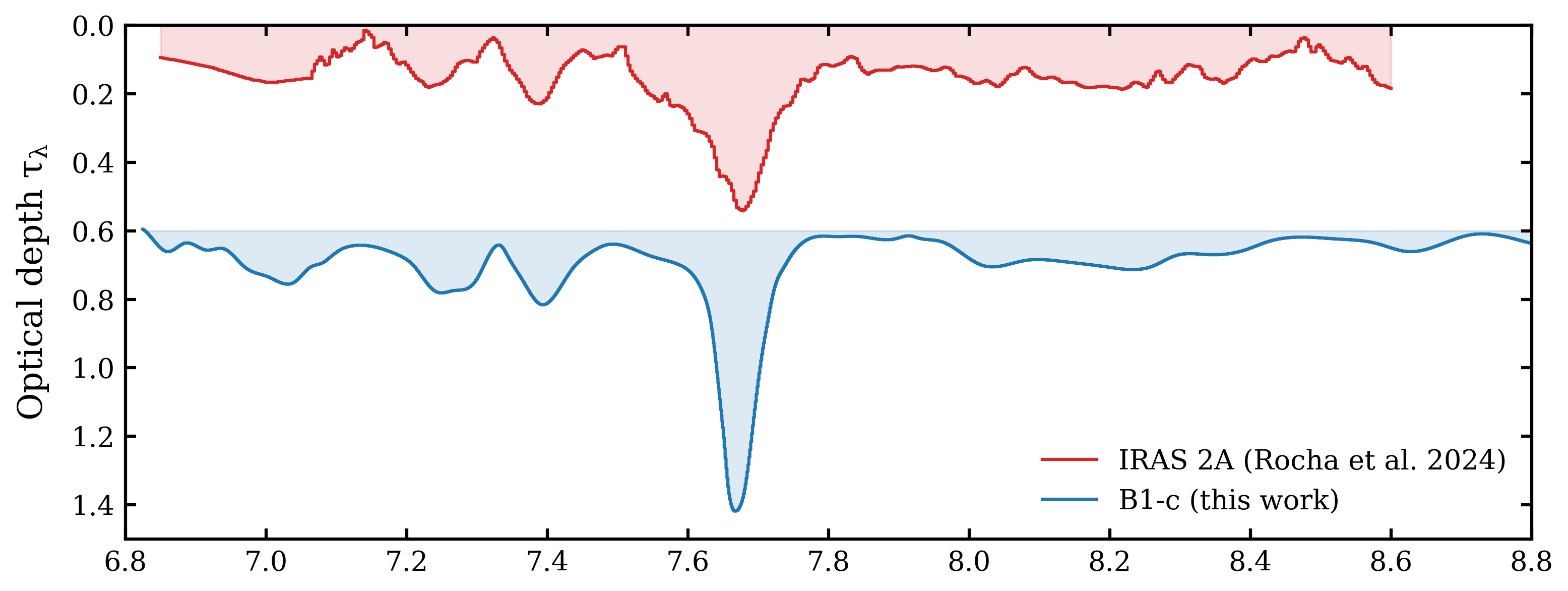}
    \end{subfigure}
    \hfill
    \begin{subfigure}[b]{\textwidth}
        \centering
        \includegraphics[width=\textwidth]{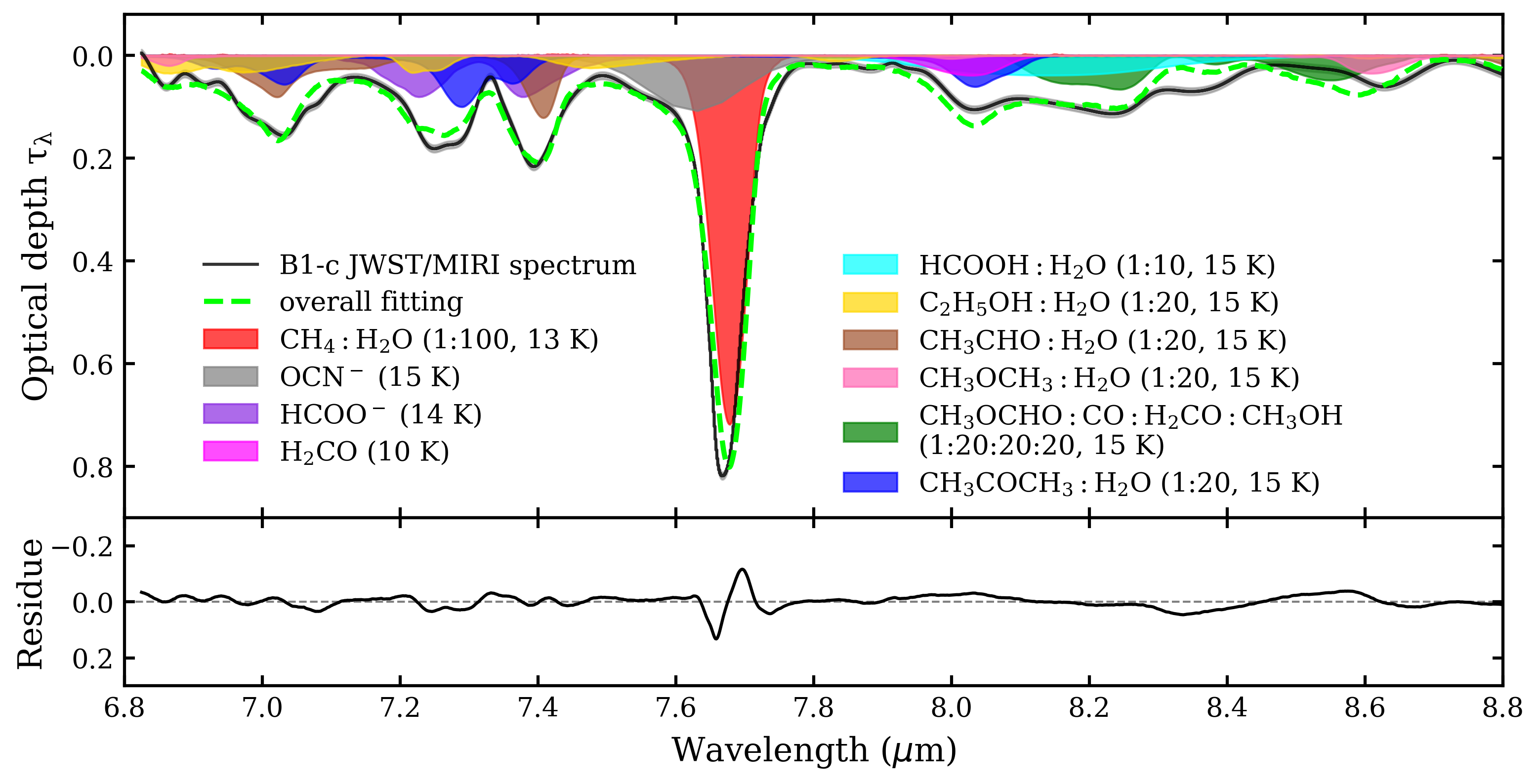}
    \end{subfigure}
    \caption{Top: JWST/MIRI-MRS spectrum between 6.8 and 8.8~$\mu$m of IRAS~2A \citep{Rocha2024} and B1-c (this work). The IRAS~2A spectrum was binned by a factor of 4, therefore shows lower spectral resolution. The B1-c spectrum is the version after removing the gas-phase lines (i.e., the black line in Fig.~\ref{fig:fitting_steps}d). Bottom: Best-fit combination of lab spectra to the B1-c JWST/MIRI-MRS spectrum between 6.8 and 8.8~$\mu$m. Nine out of 12 candidate species that have statistically significant contribution to the 6.8--8.8 $\mu$m range are displayed, with the other three species (\ce{SO2}, \ce{CH3COOH}, and \ce{CH3CN}) excluded. The best-fit mixing conditions (constituents, ratio, and temperature) selected in Sect.~\ref{sect:method_fit_COM_bands} are labelled in the legend. The residue is shown below.}
    \label{fig:JWST_fit_B1c_COMs}
\end{figure*}

\subsubsection{Decomposition of the B1-c spectrum in 6.8--8.8~$\mu$m}\label{sect:results_JWST_decomposition}
As introduced in Sect.~\ref{sect:method_fit_COM_bands}, after selecting the most suitable ice mixture of each candidate species, we performed least-squares and MCMC fittings to search for the best-fit scaling factors of lab spectra to the observed B1-c spectrum between 6.8 and 8.8~$\mu$m. We focus on the spectral decomposition in this subsection, and leave the statistics to the next one. The spectral fitting of the 6.8--8.8 $\mu$m range for B1-c is similar to that for IRAS~2A, therefore our analysis is generally based on the results of \cite{Rocha2024}. 
The top panel of Fig.~\ref{fig:JWST_fit_B1c_COMs} displays the isolated JWST/MIRI spectrum in the COM fingerprint range (6.8--8.8 $\mu$m) of IRAS~2A and B1-c. Despite some difference introduced by the different methods of removing the superposed gas-phase lines, the absorption bands show very similar profiles in these two sources (see elaboration below). The bottom panel of Fig.~\ref{fig:JWST_fit_B1c_COMs} shows the best-fit results and residues of the B1-c spectrum, and we reserve the statistical details to Sect.~\ref{sect:results_fit_stats}.

\paragraph{7.5--7.8~$\mu$m.} This range mainly contains absorption bands of simple molecules, and therefore is introduced first. The strongest band at 7.67~$\mu$m is attributed to the \ce{CH4} deformation mode. The residues around this band is due to the deviation of peak position between the observations and the selected lab spectrum, which is the best among the candidates but still not optimal.
The blue wing of the \ce{CH4} band $\sim$7.6~$\mu$m is much weaker in B1-c than in IRAS~2A. It corresponds to one \ce{OCN^-} band and one \ce{SO2} band, which overlap with each other by $\sim$50\%. The \ce{SO2} component is not shown in Fig.~\ref{fig:JWST_fit_B1c_COMs} since \ce{OCN^-} was favored in the overall fittings by least-squares and MCMC, instead an upper limit was estimated for \ce{SO2}.
There is also a red wing of the \ce{CH4} band in IRAS~2A, and is mainly made up of a broad \ce{CH3COOH} band at $\sim$7.75~$\mu$m. This red wing is not present in B1-c after we traced a local continuum that is close to the observed spectrum. \cite{Rocha2024} also discussed in their Sect. 4.2.4 and appendix J that using a different local continuum could remove the \ce{CH3COOH} component from the overall fit, and \ce{CH3COOH} is only considered as tentatively detected in IRAS~2A.

\paragraph{6.8--7.5~$\mu$m.} This range contains four absorption bands at 6.85, 7.02, 7.26, and 7.40~$\mu$m. \cite{Rocha2024} attribute them to two O-COMs (\ce{CH3CHO} and \ce{C2H5OH}) and \ce{HCOO^-}, which also applies to B1-c. The \ce{CH3CHO}:\ce{H2O} mixture is one of the contributors to the 7.02 and 7.40~$\mu$m bands. \ce{HCOO^-} has two bands at 7.23 and 7.38 $\mu$m, making up the red part of the observed bands at 7.26 and 7.40~$\mu$m. \ce{C2H5OH} has three bands at 6.86, 6.97, and 7.24~$\mu$m, of which the contribution is less significant but not negligible.

The peak position of the third band is slightly different between IRAS~2A and B1-c. In IRAS~2A, it is observed to peak at 7.24 $\mu$m; more blueshifted than the same band in B1-c which peaks at 7.26 $\mu$m. The offset between the \ce{HCOO^-} 7.23~$\mu$m band and the observed 7.26~$\mu$m band in B1-c leaves an underfit area at 7.3~$\mu$m. 
There are a few candidate species that have absorption bands at around 7.3~$\mu$m: \ce{CH3CN}, \ce{CH3COOH}, and \ce{CH3COCH3}. \ce{CH3CN} has recently been reported to be tentatively detected in NIRSpec spectra of several protostellar sources \citep{Nazari2024_CH3CN}, but its abundance is mainly constrained by the other two stronger bands at $\sim$6.9 and 7.1~$\mu$m, where the absorption is weak in observations. Similarly, the abundance of \ce{CH3COOH} is constrained by the stronger band at 7.75~$\mu$m, which is degenerate with the local continuum subtraction. Even if \ce{CH3COOH} is present, its 7.3~$\mu$m band also tends to be too weak and broad to fill the gap at 7.3 $\mu$m in the observed B1-c spectrum. The best candidate turns out to be the \ce{H2O}-rich mixtures of \ce{CH3COCH3}, of which the \ce{CH3} symmetric deformation band at 7.33 $\mu$m can help solve the problem, and the other two bands at 7.03 and 8.03~$\mu$m also fit well with the observations (see panel i of Fig.~\ref{fig:obs_vs_lab_mix}). In \cite{Rocha2024}, \ce{CH3COCH3} is not considered as firmly detected in IRAS~2A based on their recurrence analysis (see their Sect.~4.2.2). A possible explanation is that they trace a local continuum that is not close enough to the observed spectrum when isolating the COM fingerprint range, and therefore leave some space for \ce{CH3COOH} bands at 7.7 and 7.3 $\mu$m. However, if they trace a local continuum close to the observed spectrum as we did, there will also be an unfit area at $\sim$7.3 $\mu$m (shown in their Fig.~J.2), which can be attributed to \ce{CH3COCH3}. 

\paragraph{7.8--8.8~$\mu$m.} This range is composed of several blended bands between 7.9 and 8.45~$\mu$m and a small band at 8.63~$\mu$m. The band at 8.03~$\mu$m is contributed and slightly overproduced by a combination of \ce{H2CO} band at 8.02 $\mu$m and \ce{CH3COCH3}:\ce{H2O} band at 8.03 $\mu$m. The contribution of \ce{H2CO} was fixed by fitting its another band at 6.67 $\mu$m (see Appendix~\ref{appendix:H2CO}). To alleviate the overestimation, we chose the \ce{CH3COCH3} mixture with \ce{H2O} and \ce{CO} instead of the \ce{H2O}-only mixture, since the \ce{H2O}:\ce{CO2} mixture has a relatively weaker 8.03 $\mu$m band (Fig.~\ref{fig:obs_vs_lab_mix} j), although the difference is small.
The broad band peaking at 8.24 $\mu$m is composed of the broad HCOOH band at 8.17 $\mu$m and the double-peaked \ce{CH3OCHO} band at $\sim$8.2~$\mu$m. The observed band at 8.63~$\mu$m matches best with the \ce{CH3OCH3}:\ce{H2O} band at 8.6~$\mu$m, with potential contribution from the \ce{SO2}:\ce{H2O} band at 8.67~$\mu$m. Similar to \ce{CH3COCH3}, \ce{CH3OCH3} is also considered as tentative detection for IRAS~2A based on the recurrence analysis; however, this is likely because \cite{Rocha2024} only perform the fittings over 6.8--8.6 $\mu$m, missing a large portion of the observed 8.63 $\mu$m band. After extending the considered wavelength range to 8.8 $\mu$m, \ce{CH3OCH3} is likely to be considered as a firm detection (as shown in Fig.~\ref{fig:obs_vs_lab_mix}g).
Besides the contribution of the aforementioned species, there is still an unfit band between 8.3 and 8.4 $\mu$m, also seen in other JOYS+ sources. It is recently found to be attributable to \ce{CH2OH} (priv. comm. with W. Rocha), and will be studied in a future paper. On the other hand, the region at $\sim$8.5~$\mu$m is overfit by the \ce{CH3} rocking band of \ce{CH3OCHO}. The reason is unclear, and could be related to the local continuum subtraction.\\

The decomposition results of the COM fingerprint range between 6.8 and 8.8 $\mu$m are generally the same for B1-c (this work) and IRAS~2A \cite{Rocha2024}, in spite of different fitting strategies. The only difference is that we tend to consider \ce{CH3OCH3} and \ce{CH3COCH3} as firmly detected and provide constraints on their ice column densities instead of upper limits (see Sect.~\ref{sect:results_fit_stats}).


\begin{sidewaystable*}[!ph]
    \setlength{\tabcolsep}{0.03cm}  
    \renewcommand{\arraystretch}{1.6}  
    \caption{Ice column densities of candidate species derived from the least-squares and MCMC fittings to the JWST/MIRI-MRS spectra of NGC 1333 IRAS~2A and B1-c.} 
    \centering
    \begin{tabular}{lcccccccccccccc}
    \toprule
    \multicolumn{5}{c}{Source} & \multicolumn{5}{c}{B1-c} & \multicolumn{5}{c}{NGC 1333 IRAS~2A $^a$}\\
    \cmidrule(lr{0.3em}){6-10}\cmidrule(lr{0.3em}){11-15}
    Species & Band $\lambda$ & Mode $^b$ & $A$ & Ref. $^c$ & $N_\mathrm{ice}$ & X/\ce{CH3OH} & X/\ce{H2O} & $T$ & mixture & $N_\mathrm{ice}$ & X/\ce{CH3OH} & X/\ce{H2O} & $T$ & mixture \\
     & ($\mu$m) & & (10$^{-18}$ cm$^{-2}$) & & (10$^{17}$cm$^{-2}$) & (\%) & (\%) & (K) & & (10$^{17}$ cm$^{-2}$) & (\%) & (\%) & (K) & \\
    \midrule
    \ce{H2O} & 13.2 & libration & 32 & [1][2] & 250$\pm$17 & -- & $\equiv$100 & 15 & pure & 300$\pm$12 & -- & $\equiv$100 & 15, 160 & pure\\
    \ce{CH3OH} & 9.74 & CO str. & 15.6$^d$ & [3] & 30$\pm$10 & $\equiv$100 & 11.9$\pm$4.1 & -- & (Gaussian) & 19$\pm$4 & $\equiv$100 & 6.3$\pm$1.4 & -- & (Gaussian) \\
    \ce{CH4} & 7.67 & \ce{CH4} def. & 8.4 & [1][2][4] & 9.3$\pm$5.1 & 30.9$\pm$19.8 & 3.7$\pm$2.0 & 13 & w/\ce{H2O} & 4.9$_{-3.2}^{+7.5}$ & 26$_{-18}^{+40}$ & 1.6$_{-1.1}^{+2.5}$ & 15 & w/\ce{H2O}\\
    \ce{SO2} & 7.60 & \ce{SO2} str. & 34 & [4] & <0.47 & <1.6 & <0.19 & 10 & w/\ce{CH3OH} & 0.6$_{-0.4}^{+1.9}$ & 3.2$_{-2.2}^{+10}$ & 0.2$_{-0.1}^{+0.6}$& 10 & w/\ce{CH3OH}\\
    \ce{OCN^-} & 7.62 & 2$\nu_2$ & 7.45 & [5] & 4.1$\pm$2.2 & 13.6$\pm$8.7 & 1.6$\pm$0.9 & 15 & \ce{NH3}+\ce{HNCO} $^e$ & 3.7$_{-3.3}^{+6.6}$ & 19$_{-17}^{+35}$ & 1.2$_{-1.1}^{+2.2}$ & 15 & \ce{NH3}+\ce{HNCO} $^e$ \\
    \ce{H2CO} & 8.02 & \ce{CH2} rock. & 1.5 & [1] & 4.5$\pm$1.3 & 14.9$\pm$6.7 & 1.8$\pm$0.6 & 15 & pure & 12.4$_{-6.6}^{+19.7}$ & $65_{-37}^{+105}$ & 4.1$_{-2.2}^{+6.6}$ & 15 & \makecell{w/CO, \ce{CH3OH},\\ \ce{CH3OCHO}} \\
    \ce{HCOO^-} & 7.38 & CO str. & 17 & [7] & 0.93$\pm$0.51 & 3.1$\pm$2.0 & 0.37$\pm$0.21 & 14 & \makecell{\ce{H2O}+\ce{NH3}\\+\ce{HCOOH} $^e$} & 1.4$_{-0.4}^{+2.4}$ & 7.4$_{-2.6}^{+12.7}$ & 0.5$_{-0.1}^{+0.80}$ & 14 & \makecell{\ce{H2O}+\ce{NH3}\\+\ce{HCOOH} $^e$} \\
    \ce{HCOOH} & 8.23 & CO str. & 29 & [1] & 1.1$\pm$0.6 & 3.7$\pm$2.4 & 0.44$\pm$0.25 & 15 & w/\ce{H2O} & 3.0$_{-1.7}^{+5.3}$ & 16$_{-10}^{+28}$ & 1.0$_{-0.6}^{+1.8}$& 15 & w/\ce{H2O}, \ce{CH3OH}\\
    \ce{CH3CHO} & 7.42 & \makecell{\ce{CH3} sym. def. \\+ CH wag.} & 4.1 & [8][9] & 1.9$\pm$1.0 & 6.2$\pm$4.0 & 0.74$\pm$0.41 & 15 & w/\ce{H2O} & 2.2$^{+2.8}_{-1.4}$ & 12$_{-8}^{+15}$ & 0.7$_{-0.5}^{+0.9}$ & 16 & w/\ce{H2O} \\
    \ce{C2H5OH} & 7.24 & \ce{CH3} sym. def. & 2.4 & [10] & 2.1$\pm$1.2 & 7.1$\pm$4.6 & 0.85$\pm$0.47 & 15 & w/\ce{H2O} & 3.7$_{-0.5}^{+4.5}$ & 19$_{-5}^{+24}$ & 1.2$_{-0.2}^{+1.5}$ & 15 & w/\ce{H2O}\\
    \ce{CH3OCH3} & 8.59 & COC str. + \ce{CH3} rock. & 5.55$^d$ & [9] & 0.51$\pm$0.28 & 1.7$\pm$1.0 & 0.20$\pm$0.11 & 15 & w/\ce{H2O} & <5.8 & <13 & <0.8 & -- \\ 
    \ce{CH3OCHO} & 8.25 & CO str. & 24.9 & [11] & 0.48$\pm$0.26 & 1.6$\pm$1.0 & 0.19$\pm$0.11 & 15 & \makecell{w/CO, \ce{H2CO},\\ \ce{CH3OH}} & 0.2$_{-0.1}^{+0.4}$ & 1.0$_{-0.6}^{+2.1}$ & 0.07$_{-0.03}^{+0.13}$ & 15 & \makecell{w/CO, \ce{H2CO},\\ \ce{CH3OH}} \\ 
    \ce{CH3COOH} & 7.82 & OH bend. & 45.7 & [5] & <0.21 & <0.7 & <0.08 & 10 & w/\ce{H2O} & 0.9$_{-0.6}^{+1.3}$ & 4.7$_{-3.3}^{+6.9}$ & 0.3$_{-0.2}^{+0.4}$ & 10 & w/\ce{H2O} \\
    \ce{CH3COCH3} & 7.33 & \ce{CH3} sym. str. & 10.2$^d$ & [12][13] & 1.7$\pm$1.0 & 5.8$\pm$3.7 & 0.69$\pm$0.38 & 15 & w/\ce{H2O}, \ce{CO2} & <1.1 & <6 & <0.4 & 15 & w/\ce{H2O}\\
    \ce{CH3CN} & 7.27 & \ce{CH3} sym. def. & 1.2 & [14] & <3.1 & <10 & <1.2 & 15 & w/\ce{H2O}, \ce{CO2} & <5.0 & <26 & <1.7 & 15 & w/\ce{H2O}\\
    \bottomrule
    \end{tabular}
    \begin{minipage}{0.99\textwidth}
    $^a$ Results obtained from \cite{Rocha2024}.\\
    $^b$ Abbreviation of vibrational modes: def. = deformation; str. = stretching; rock. = rocking; sym. = symmetric; wag. = wagging; bend. = bending.\\
    $^c$ References of band strengths $A$: [1] \cite{Bouilloud2015}; [2] \cite{Hudgins1993}; [3] \cite{Luna2018}; [4] \cite{Boogert1997}; [5] \cite{Rocha2024}; [7] \cite{Schutte1999}; [8] \cite{Hudson2020}; [9] \cite{TvS2018}; [10] \cite{Boudin1998}; [11] \cite{TvS2021}; [12] \cite{Hudson2018}; [13] \cite{Rachid2020}; [14] \cite{Rachid2022}.\\
    $^d$ See Appendix~\ref{appendix:band_strength} for additional details. \\
    $^e$ These constituents are not added intentionally, but are ingredients to form the target ions (i.e., \ce{OCN^-} and \ce{HCOO^-}. \\
    \end{minipage}
    \label{tab:JWST_results}
\end{sidewaystable*}

\subsubsection{Fitting statistics}\label{sect:results_fit_stats}
We adopted least-squares and MCMC fittings to find the best-fit values and uncertainties of the scaling factors of lab spectra (see Table~\ref{tab:JWST_B1c_fit_statistics}). In the least-squares fitting, we considered all the 12 candidate species: \ce{CH4}, \ce{SO2}, \ce{OCN^-}, \ce{HCOO^-}, \ce{HCOOH}, \ce{CH3CHO}, \ce{C2H5OH}, \ce{CH3OCH3}, \ce{CH3OCHO}, \ce{CH3COOH}, \ce{CH3COCH3}, and \ce{CH3CN} (i.e., $N=12$ in Eq.~\ref{eq:linear_fitting}). The scaling factor of \ce{H2CO} was determined from the band at 6.67 $\mu$m, and was fixed when fitting the 6.8--8.8 $\mu$m range (see details in Appendix~\ref{appendix:H2CO}).

The least-squares fitting results show that the lab spectrum scaling factors of most of the candidate species were well constrained with relative errors smaller than 10\% (see Table~\ref{tab:JWST_B1c_fit_statistics}). There are four exceptions: \ce{SO2}, \ce{CH3OCH3}, \ce{CH3COOH}, and \ce{CH3CN}, and particularly, \ce{SO2} and \ce{CH3COOH} have very little contribution. The absence of \ce{CH3COOH} is because we traced a local continuum close to the observed spectrum, which eliminated the red wing of the \ce{CH4} band at 7.67 $\mu$m. However, the lack of \ce{SO2} is more likely because its band at 7.6 $\mu$m is too close to the \ce{OCN^-} band at 7.62 $\mu$m, and hence highly degenerate with each other. The blue wing of the \ce{CH4} band in B1-c is less prominent than in IRAS~2A, making it more difficult to distinguish between the contribution from \ce{SO2} and \ce{OCN^-}. The degeneracy origin prevents us to draw any conclusion that \ce{SO2} is not present or more depleted in B1-c than in IRAS~2A. Instead, an upper limit was estimated for \ce{SO2} by only scaling the lab spectrum of \ce{SO2} to the observations (Fig.~\ref{fig:obs_vs_lab_mix}b). 
For \ce{CH3OCH3} and \ce{CH3CN}, the best-fit scaling factors are not negligible, but the relative errors are slightly larger than other species, mainly because their bands correspond to weak and blended absorption features in the observations. 

In the least-squares fitting, eight out of 12 candidate species have fairly small relative uncertainties ($<$10\%). In the next step, we performed an additional MCMC fitting on these eight species plus \ce{CH3OCH3} (i.e., nine species in total) to get a better understanding of the uncertainty level. \ce{CH3OCH3} was taken into consideration because it is the main contributor to the observed band at 8.63 $\mu$m, and it is one of the most abundant O-COMs observed in the gas phase. We did not include \ce{CH3CN} in the MCMC fitting since it does not have a characteristic band in the 6.8--8.8 $\mu$m range; all the three bands are weak and blended with others. Including \ce{CH3CN} would also make the MCMC fitting less convergent, suggesting that its contribution is less significant. Instead, we report upper limits for \ce{CH3CN} like for \ce{SO2} and \ce{CH3COOH}.

The best-fit scaling factors derived from the MCMC fitting are consistent with the least-squares within 5\%. Fig.~\ref{fig:B1c_mcmc_corner_plot} displays the posterior distributions of each component. Most components are rather independent from each other, except for pairs that have similar band locations (e.g., \ce{CH4} and \ce{OCN^-}, HCOOH and \ce{CH3OCHO}). The relative uncertainties generally increase, but still at a low level of $\sim$10\%, except for \ce{CH3OCH3}, whose relative uncertainty is $\sim$20\%. For the left three species, \ce{SO2}, \ce{CH3COOH}, and \ce{CH3CN}, we report only upper limits. We manually scaled the lab spectrum of each species to the observed B1-c spectrum (e.g., Fig.~\ref{fig:obs_vs_lab_uplim}), and the maximum scaling factors allowed by the observations were converted to upper limits of ice column densities. 

Based on the fitting results of scaling factors, the ice column densities of each species can be calculated using Eq.~\ref{eq:N_ice}. The absolute ice column densities are on the same order of magnitude in B1-c (this work) and IRAS~2A \citep{Rocha2024}.
The uncertainties of the ice column densities are propagated from scaling factors ($\sim$10\%), band strengths ($\sim$20\%), and the steps taken to isolate the COM bands between 6.8 and 8.8 $\mu$m (Sects.~\ref{sect:method_global_cont}--\ref{sect:method_gas_lines}). The uncertainties of the isolation steps should be dominant but also difficult to quantify, since that they are more or less subjective. Here we estimated a conservative uncertainty level of 50\% for isolating the COM fingerprint range, which resulted in a $\sim$55\% total uncertainties of the ice column densities. This uncertainty level is consistent with that of IRAS~2A reported in \cite{Rocha2024}.
The derived ice column densities and uncertainties, along with the temperatures and mixing constituents of the best-fit lab spectra are summarized in Table~\ref{tab:JWST_results} for both B1-c (this work) and IRAS~2A \citep{Rocha2024}.

\begin{figure}[!h]
    \centering
    \includegraphics[width=0.5\textwidth]{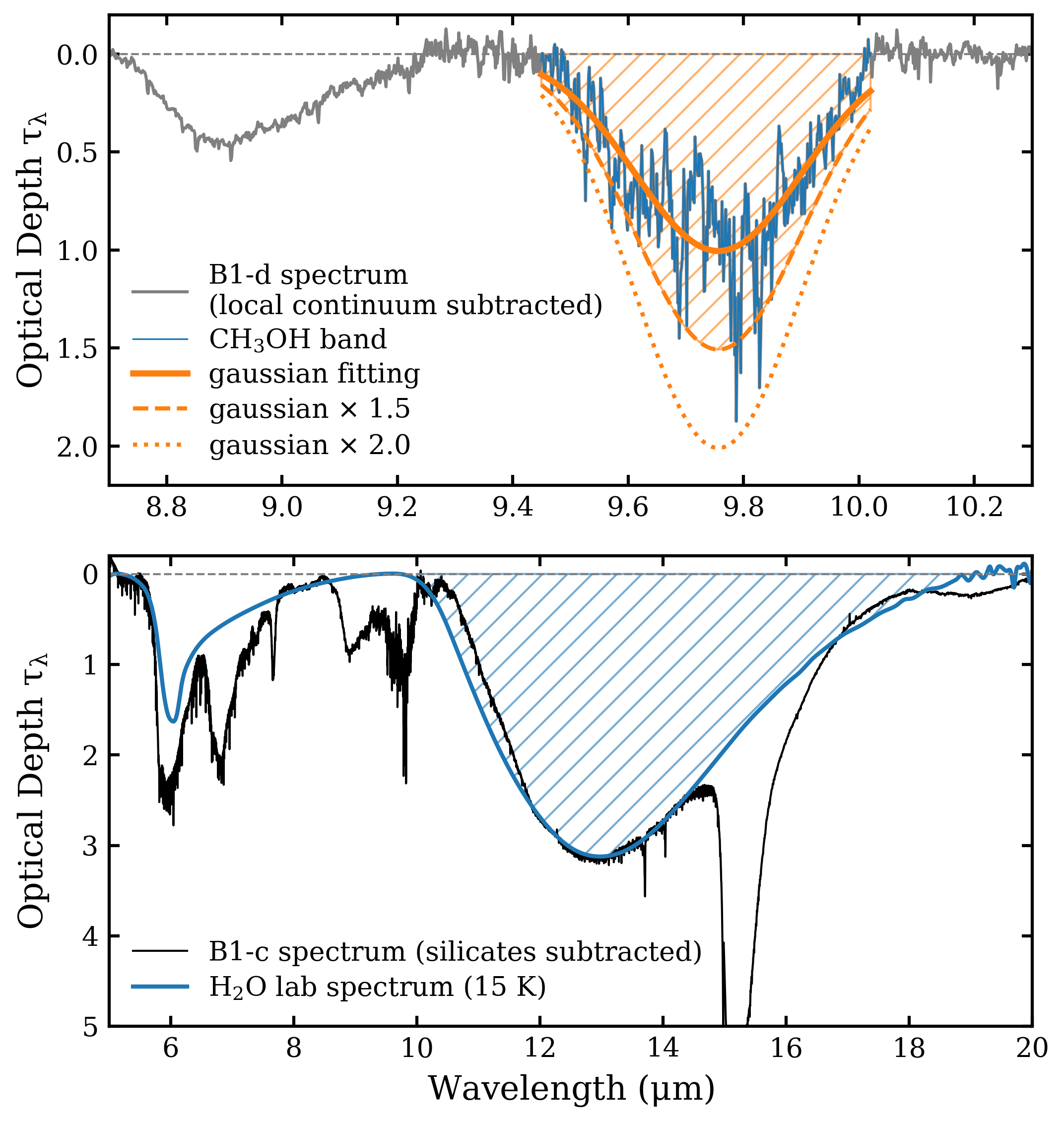}
    \caption{Fitting results of the \ce{CH3OH} band at 9.74 $\mu$m (top panel) and the \ce{H2O} band at 13 $\mu$m (bottom panel). The hatched regions indicate the integrated areas for calculating the ice column densities of \ce{CH3OH} and \ce{H2O}.}
    \label{fig:CH3OH_H2O_fit}
\end{figure}

\subsubsection{\ce{CH3OH} and \ce{H2O} bands}\label{sect:result_CH3OH_H2O}
The strategies for fitting the \ce{CH3OH} band at 9.74 $\mu$m and the \ce{H2O} and at 13 $\mu$m have been introduced in Sect.~\ref{sect:method_CH3OH_H2O_band}. Here we report the fitting results and the derived ice column density of \ce{CH3OH} and \ce{H2O}.
The \ce{CH3OH} band at 9.74 $\mu$m is isolated and was fit by a Gaussian function (see top panel of Fig.~\ref{fig:CH3OH_H2O_fit}). However, it is likely that the best-fit Gaussian function underestimates the real intensity of this band because of the silicate extinction. The extinction correction factor is not trivial to estimate, and a better strategy would be combining the two \ce{CH3OH} bands in the NIRSpec range (at 3.4 and 3.9 $\mu$m) and fit all three bands simultaneously. Unfortunately, the \ce{CH3OH} band at 3.4~$\mu$m is below the detection limit of our observations of B1-c, probably due to the strong extinction by \ce{H2O} at $\sim$3.05~$\mu$m. 

The strength of the 3.9 $\mu$m band is also not well constrained, since there is strong indication that a grain shape correction is needed to properly trace the continuum of this band \citep{Dartois2022, Dartois2024}. Considering that NIRSpec data are not the focus of this paper and will be reserved for following studies, we performed a preliminary analysis of the 3.9 $\mu$m band and derived an upper limit of 4.0$\times10^{18}$ cm$^{-2}$ for the \ce{CH3OH} ice column density (see details in Appendix~\ref{appendix:NIRSpec_CH3OH}). For comparison, the best-fit Gaussian to the 9.74~$\mu$m band resulted in a column density of 2.0$\times10^{18}$ cm$^{-2}$. We finally adopted a value of 3.0$\times10^{18}$ cm$^{-2}$, which corresponds to 1.5 times the best-fit Gaussian. As shown in the top panel of Fig.~\ref{fig:CH3OH_H2O_fit}, this correction factor of 1.5 is reasonable when considering the extinction of the silicate band at 9.8 $\mu$m. The uncertainty was estimated as $\pm$1.0 $\times10^{18}$ cm$^{-2}$. 

The fitting result of the \ce{H2O} band at 13 $\mu$m is shown in the bottom panel of Fig.~\ref{fig:CH3OH_H2O_fit}. A scaling factor of 14.5 was estimated through visual inspection for the lab spectrum of \ce{H2O} ice at 15 K (Sect.~\ref{sect:method_CH3OH_H2O_band}), giving a column density of 2.5$\times10^{19}$ cm$^{-2}$. The slight excess between 10 and 12 $\mu$m is likely due to the over-subtraction of silicate features, as the synthetic silicates have a wider red wing of the 9.8 $\mu$m band than the GCS~3 profile (Fig.~\ref{fig:fitting_steps}b).

With the ice column densities of \ce{H2O} and \ce{CH3OH} derived, we are able to calculate the relative abundances of other species, which are presented in Table~\ref{tab:JWST_results}. The total ice abundance of the detected six COMs, namely \ce{CH3OH}, \ce{CH3CHO}, \ce{C2H5OH}, \ce{CH3OCH3}, \ce{CH3OCHO}, and \ce{CH3COCH3}, is 14.6\% with respect to \ce{H2O} ice in B1-c. As the most abundant COM, the relative abundance of \ce{CH3OH} is 11.9\%, dominates more than 80\% of all COMs. The upper limit of the most abundant N-COM, \ce{CH3CN}, is 1.2\% with respect to \ce{H2O}. According to \cite{Rocha2024}, the ice abundances of all the O-COMs and \ce{CH3OH} with respect to \ce{H2O} are 9.8\% and 6.3\% in IRAS~2A, respectively; the upper limit of \ce{CH3CN} is 1.7\%. The ice ratios between \ce{CH3OH} and other COMs will be discussed in Sect.~\ref{sect:discussion}. We provide the statistics but leave out the discussion on the simple species as they are beyond the scope of this study.

\begin{figure*}[!h]
    \centering
    \includegraphics[width=\textwidth]{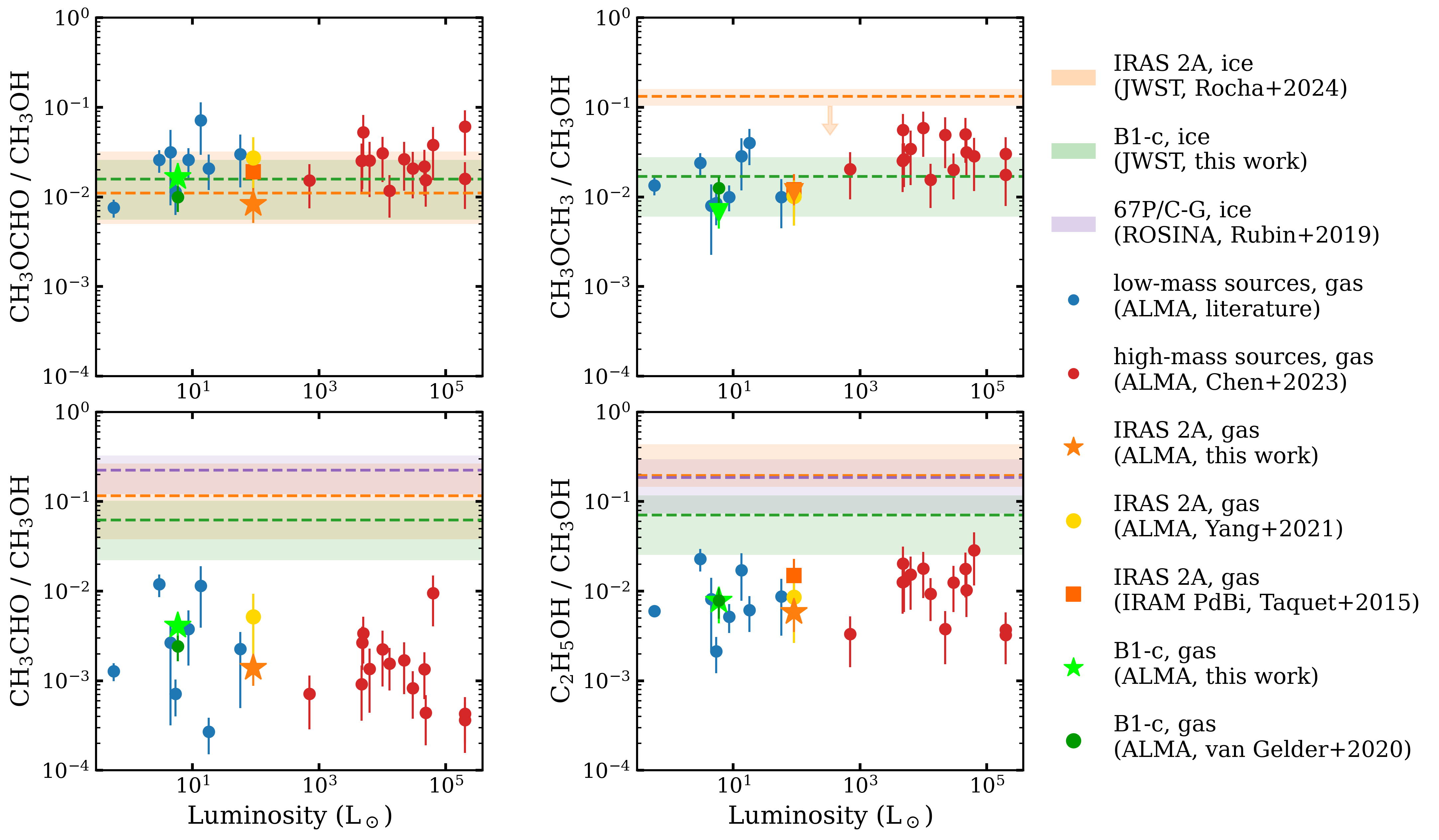}
    \caption{Ratios of four O-COMs with respect to \ce{CH3OH} in gas (data points) and ice (shaded regions). Gas ratios are collected from a total sample of nine low-mass sources from this work and literature \citep[blue circles;][]{Jorgensen2018, Manigand2020, vanGelder2020, Yang2021_PEACHES}, and 14 high-mass sources from the CoCCoA survey \citep{Chen2023}. Ice ratios are available for only two protostars \citep[IRAS~2A and B1-c;][this work]{Rocha2024} and one comet \citep[67P/C-G;][]{Rubin2019}. All data of the two focused low-mass sources, NGC 1333 IRAS~2A and B1-c, are colored in orange and green, respectively. The gas ratios of IRAS~2A and B1-c derived in this work and previous observations are highlighted in stars and squares \citep{Taquet2015, vanGelder2020}, respectively. Upper limits are denoted by downward triangles or arrows.}
    \label{fig:COM_ratios_gas_vs_ice}
\end{figure*}

\section{Discussion}\label{sect:discussion}
\subsection{Column density ratios: ice vs gas}\label{sect:discuss_COM_ratios}
The detection of COM ices by JWST \citep[this work and][]{Rocha2024} strongly suggests that the formation of COMs starts in the solid phase, as illustrated in Fig.~\ref{fig:COM_schematic}. However, it remains unclear to what extent the gas-phase chemistry will alter the composition of COMs after their sublimation from dust grains. Now with COMs observable in both gas and ice, we are able to gain insights into this phase transition by comparing the COM ratios in both phases. 

Our comparison is focused on four O-COMs: \ce{CH3CHO}, \ce{C2H5OH}, \ce{CH3OCH3}, and \ce{CH3OCHO}, which are detectable in both gas and ice. Their ratios with respect to \ce{CH3OH} in the two phases are summarized in Fig.~\ref{fig:COM_ratios_gas_vs_ice}. The number of sources that have COM ice ratios derived is limited. Therefore, in addition to the two case-studied sources IRAS~2A and B1-c, we also include the comet 67P/C-G for comparison, of which the ice abundances of detected molecules are reported in \cite{Rubin2019}. For the gas ratios, in addition to previous observations of IRAS~2A and B1-c \citep{Taquet2015, Yang2021_PEACHES, vanGelder2020}, we also collected literature data of other sources with different masses and luminosities from previous line surveys \citep[i.e., PILS, PEACHES, and CoCCoA;][]{Jorgensen2018, Manigand2020, vanGelder2020, Yang2021_PEACHES, Chen2023}, in order to better represent the gas-phase trend using a larger sample. For the PEACHES sources, we adopted the \ce{CH3OH} column density derived from optically thin minor isotopologues (i.e., \ce{^{13}CH3OH} or \ce{CH3^{18}OH}) by \cite{vanGelder2022a}.

Figure~\ref{fig:COM_ratios_gas_vs_ice} shows two different cases: \ce{CH3OCHO} and \ce{CH3OCH3} show constant gas-phase ratios among the ALMA sample, which also matches well with the ice ratios. In contrast, \ce{CH3CHO} and \ce{C2H5OH} show larger scatter in the gas-phase ratios, and the ice ratios are higher than the gas ones by 1--2 orders of magnitude. 
The possible reasons are discussed in the following subsections.


\subsection{Origin of similarity in COM ratios between ice and gas}\label{sect:discuss_similarities}
The top panels in Fig.~\ref{fig:COM_ratios_gas_vs_ice} show constant gas-phase ratios of \ce{CH3OCHO} and \ce{CH3OCH3} with respect to \ce{CH3OH} in both low- and high-mass sources, and the good consistency in the ratios between gas and ice. These can be explained by direct inheritance or strong chemical links among \ce{CH3OCHO}, \ce{CH3OCH3}, and \ce{CH3OH}.
Direct inheritance means that the bulk of \ce{CH3OCHO}, \ce{CH3OCH3}, and \ce{CH3OH} remain intact during the transition from ice to gas, and their relative abundances also stay on the same level in the gas phase. In the hot core phase, either their absolute abundances remain stable, or their abundances are altered similarly due to a strong chemical link.
For instance, they are involved in a similar chemical network and share similar routes of formation and destruction. The numerical simulations on grain-surface chemistry by \cite{Simons2020} suggest that \ce{CH3OH} and \ce{CH3OCHO} could share the same precursor (\ce{CH3O} radical) in their formation pathways. Although \ce{CH3OCH3} is not discussed in \cite{Simons2020}, studies by \cite{Garrod2022} who present simulation results of a three-phase chemical network (bulk ice, grain surface, and gas phase) show that the \ce{CH3O} radical plays an important role in the formation of \ce{CH3OCH3} ice. The possibility of having chemical links among \ce{CH3OH}, \ce{CH3OCHO}, and \ce{CH3OCH3} is also supported by the remarkably constant ratio between \ce{CH3OCHO} and \ce{CH3OCH3} observed in a large sample of protostellar sources with different masses and luminosities \citep{Coletta2020}. However, it is difficult to clearly distinguish between the cases of direct inheritance and strong chemical links only by observations; more studies by simulations and experiments are needed to break this degeneracy.

There is also observational evidence that COMs other than \ce{CH3OH} can be inherited to later stages \citep[e.g., the recent detection in the protoplanetary disk around Oph IRS 48; ][]{Brunken2022, Yamato2024, Booth2024b}. The trend of O-COM abundances in the protoplanetary disks seem to be consistent with those observed in protostars (e.g., there are more \ce{CH3OCH3} and \ce{CH3OCHO} than \ce{CH3CHO} and \ce{C2H5OH} in the gas phase), though the exact ratios with respect to methanol are different. Theoretically, the gas-phase COMs observed in the Class 0 stage can be accreted to the circumstellar disk, and some of them will freeze out in the disk midplane at the Class II stage.  However, it is not clear to what extent the high temperature environment generated by the accretion shock at the disk-envelope interface would destroy these molecules.

\subsection{Origin of difference in COM ratios between ice and gas}\label{sect:discuss_differences}
\subsubsection{Gas-phase reprocessing}
In contrast to \ce{CH3OCHO} and \ce{CH3OCH3}, \ce{CH3CHO} and \ce{C2H5OH} show not only large scatters in their gas-phase ratios with respect to \ce{CH3OH}, but also inconsistent ratios between ice and gas (bottom panels in Fig.~\ref{fig:COM_ratios_gas_vs_ice}). The large scatter in gas-phase ratios among different sources may be caused by the physical structures \citep[e.g., the presence of a circumstellar disk can lower the temperature in the hot core;][]{Nazari2022_lm, Nazari2023_hm}. It is also possible that gas-phase chemistry after ice sublimation plays an important role in the chemical evolution of \ce{CH3HO} and \ce{C2H5OH}, in which case the variance in physical properties (density, temperature, and UV intensity) are reflected in the chemical abundances.
\cite{Chen2023} suggest that \ce{C2H5OH} may go through gas-phase destruction in the hot core phase. For example, \ce{C2H5OH} can be H-abstracted by OH radicals or halogen atoms and finally converted into HCOOH, \ce{H2CO}, and two other O-COMs, \ce{CH2OHCHO} and \ce{CH3COOH} \citep{Skouteris2018}. 
The situation of \ce{CH3CHO} is more complicated. Simulation results reported by \cite{Garrod2022} show that a significant proportion of \ce{CH3CHO} is formed in the gas phase, leading to an increasing gas-phase abundance of \ce{CH3CHO} during the warm-up stage, compared to the ice abundance in the cold collapse stage. However, our observations reveal the opposite case. The \ce{CH3CHO} ratios with respect to \ce{CH3OH} are actually lower in gas than in ice by more than one order of magnitude, which implies more \ce{CH3CHO} is consumed (either destructed into smaller species or converted into larger COMs) than reproduced in the gas phase.

It is not clear to what extent the gas-phase chemistry in hot cores can change the abundances of COMs after they sublimate from dust grains \citep[e.g.,][]{Balucani2015}. The influence of gas-phase chemistry in the COM abundances can be a few factors or orders of magnitude, and it is likely to vary among different species. Astrochemical models and experiments are required to further investigate the gas-phase chemical network in hot cores.

\begin{figure}[!h]
    \centering
     \begin{subfigure}[b]{0.5\textwidth}
         \centering
         \includegraphics[width=\textwidth]{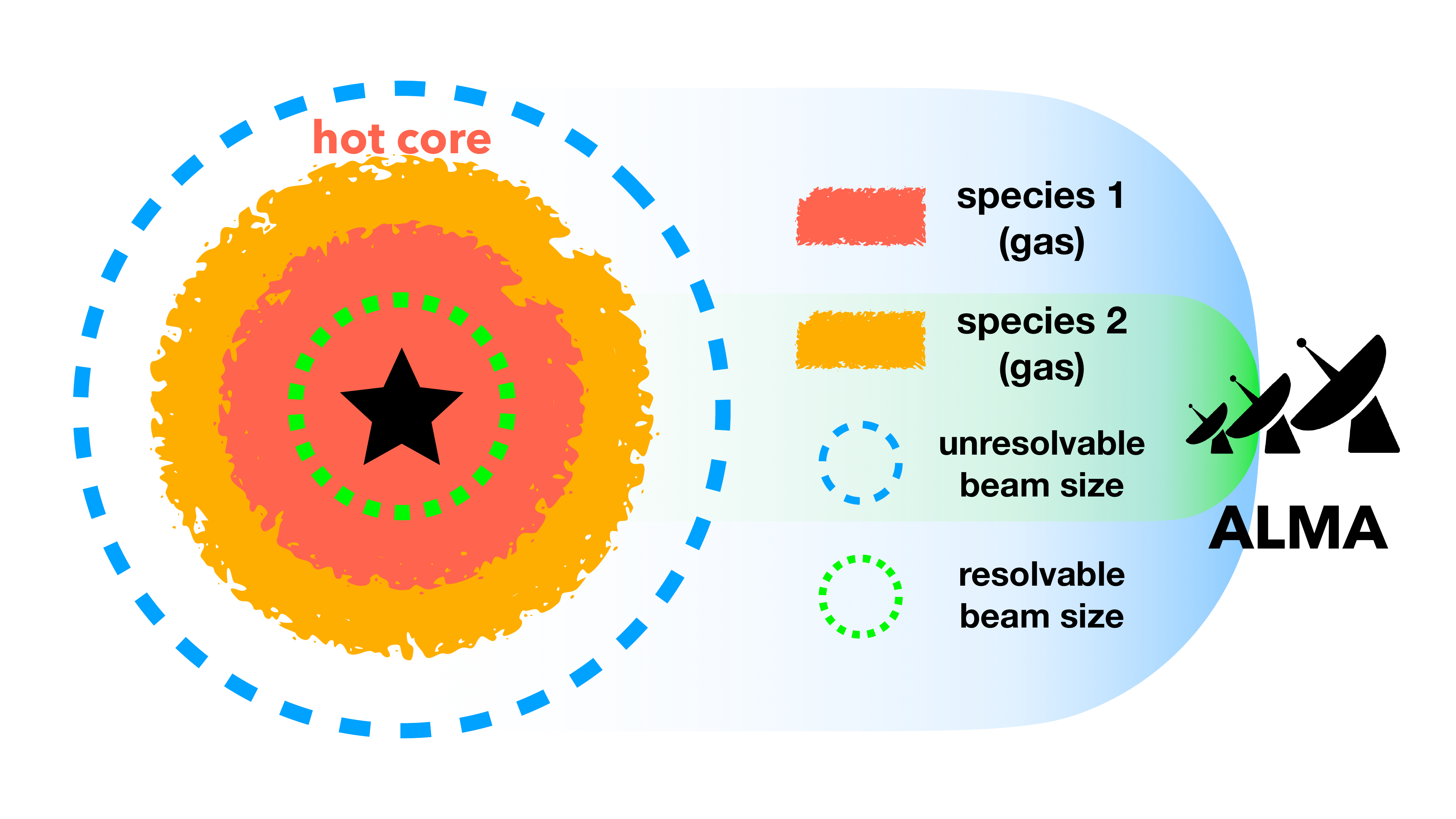}
     \end{subfigure}
     \hfill
     \begin{subfigure}[b]{0.5\textwidth}
         \centering
         \includegraphics[width=\textwidth]{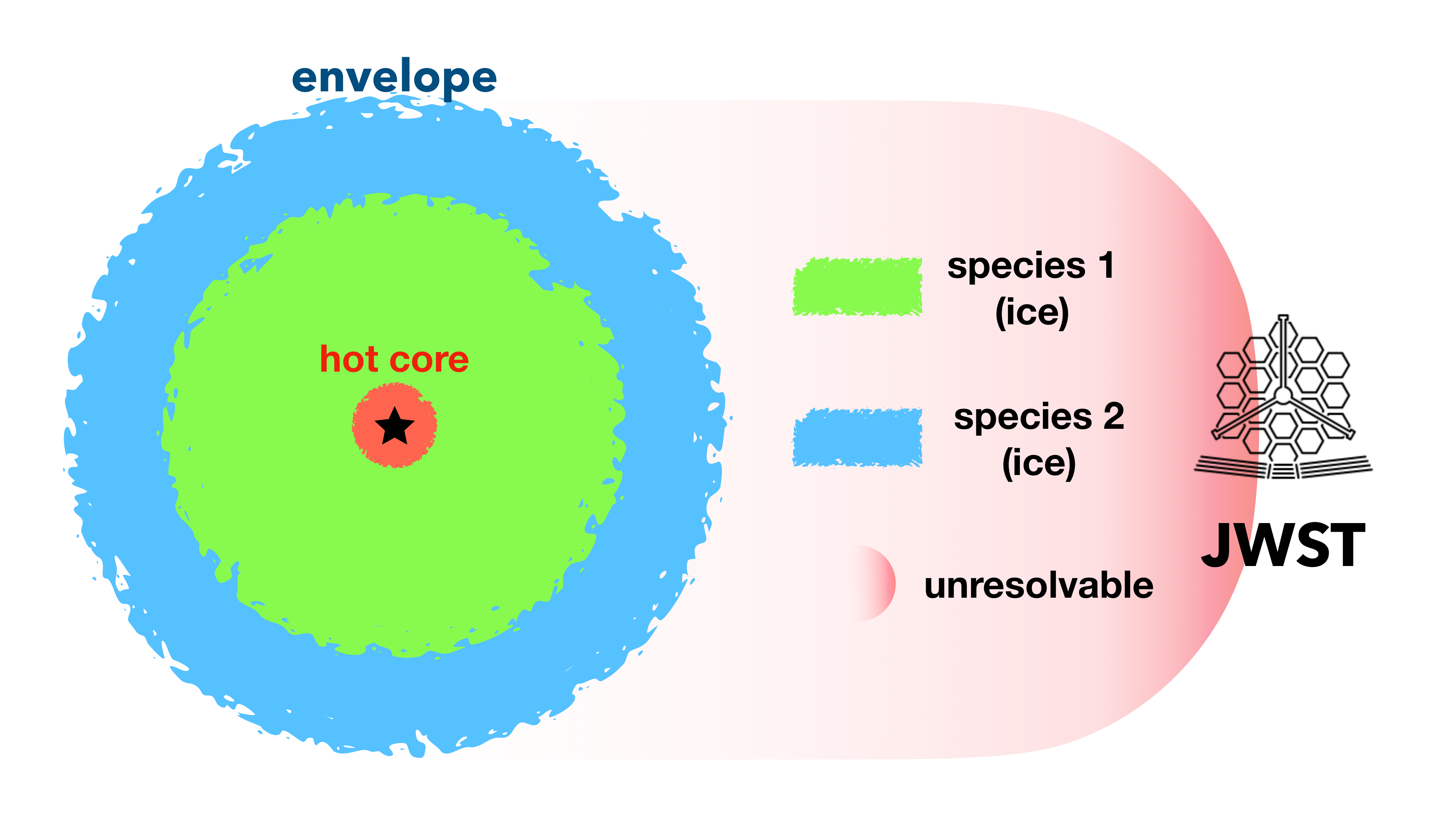}
     \end{subfigure}
    \caption{Schematics of how different spatial distributions of two species can affect the column density ratios estimated from ALMA (top) and JWST (bottom) observations. In each panel, two colors are used to indicate two species with different spatial distribution. In reality, the distance between telescopes and sources is very large, and the beam should be like a thin cylinder. Here for visualization, the beam sizes are exaggerated.}
    \label{fig:spatial_distribution_cartoon}
\end{figure}

\begin{figure*}[!h]
    \centering
    \includegraphics[width=\textwidth]{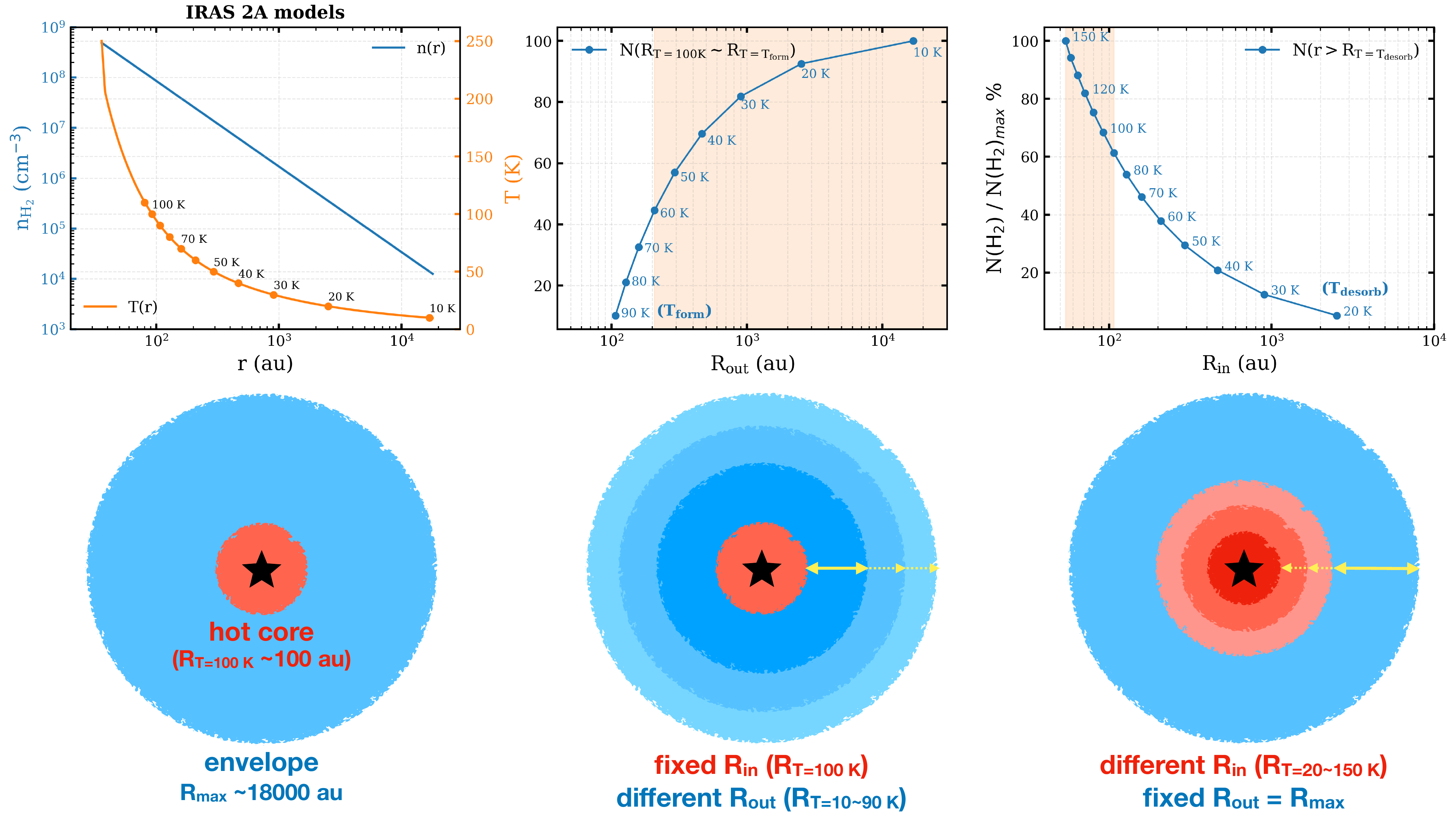}
    \caption{Top left: \ce{H2} number density (blue) and temperature (orange) profiles of the physical model of NGC 1333 IRAS~2A \citep{Kristensen2012}. Top middle: normalized \ce{H2} column densities in envelopes with a fixed inner radius (where $T$ = 100 K) and different outer radii (where $T$ = 10--90 K).  Top right: normalized \ce{H2} column density in an envelope with a fixed outer radius (the max radius in the model, $\sim$18000 au) and different inner radii (where $T$ = 20--150 K). In the top middle and right panels, each point corresponds to an \ce{H2} column density, that is, integrated \ce{H2} number density (top left panel) from an inner radius to an outer one. The orange shaded regions indicate the likely inner or outer radii in real cases. Bottom panels show corresponding cartoons of the toy model. Hot cores and envelopes are colored in red and blue, respectively. In the bottom middle and right panels, yellow arrows indicate the span of envelopes with different outer or inner radii.} 
    \label{fig:IRAS2A_profiles}
\end{figure*}

\subsubsection{Spatial distribution}\label{sect:discuss_observation}
Besides the influence of gas-phase chemistry, it is also important to consider observational effects (e.g., beam dilution). ALMA and JWST are tracing different parts of protostellar systems: ALMA only observes the emission (or absorption) lines of gas-phase molecules, mostly from the central hot core region; in comparison, JWST traces materials along the line of sight in a pencil-like beam, and the observed materials can be in both phases, as long as their spectral features fall between 0.6 and 27.9 $\mu$m (i.e., the wavelength coverage of NIRSpec plus MIRI). The molecules detected by JWST, in particular those in the solid phase, are mostly coming from the extended envelope instead of the central hot core.

Figure~\ref{fig:spatial_distribution_cartoon} shows how spatial distribution can affect the observed column density ratios between two species. In ALMA observations (top panel of Fig.~\ref{fig:spatial_distribution_cartoon}), it is possible to spatially resolve a hot core, that is, the spatial scale corresponding to the beam size is smaller than the region where most of the ice mantles has sublimated into the gas phase. Assume species 1 has higher sublimation temperature than species 2, then its gas material will be present and emit in a more compact region (i.e., within a smaller radius). If the beam size is small enough to resolve the emitting regions of both species 1 and 2, the observed ratio between species 1 and 2 will be representative of the actual ratio. On the other hand, if the beam is not able to resolve the emitting region of any of the species, then its (or their) emission will be diluted within the beam, and the ratio between species 1 and 2 will be underestimated, since the beam dilution effect is more severe for the more compactly distributed species.
In the case of our ALMA observations, the angular resolution is $\sim$0.1$\arcsec$, equivalent to $\sim$32 au at the distance of 320 pc for IRAS~2A and B1-c. As already estimated in Sect.~\ref{sect:result_ALAM_images} using Eq.~\ref{eq:R_T100K}, the hot core sizes of IRAS~2A and B1-c are 147 au and 37 au, respectively, both spatially resolvable. This means that the gas-phase COM ratios derived from our ALMA data are likely to be free from beam dilution effects.

In JWST observations (bottom panel of Fig.~\ref{fig:spatial_distribution_cartoon}) the observed ice column densities of COMs are integrated along the line of sight in the protostellar envelope, and therefore dependent on the spatial distribution of these species. 
Assume that the ice of species 1 is present in a less extended region in the envelope than species 2. This situation is possible when species 2 starts to form in the solid phase at lower density than species 1 due to different formation mechanisms. 
Since JWST observes an integrated abundance (i.e., column density) along the line of sight, the integrate is done over a more extended region for species 2, and as a result, the ice column density ratio between species 1 and 2 estimated from JWST observations is lower than the value if we only pick a small region inside the envelope. If species 1 appears to be \ce{CH3OH} and species 2 corresponds to \ce{CH3CHO} or \ce{C2H5OH}, and the resolution of ALMA observations is high enough to resolve the emitting regions of both species, than the ice ratios between \ce{CH3CHO} (or \ce{C2H5OH}) and \ce{CH3OH} estimated from JWST observations will be higher than the gas ratios derived from ALMA observations, as we find in this work.


More quantitatively, we test the influence of spatial distribution on column density ratios between two species with a toy model. We took the 1D spherical model created by \cite{Kristensen2012} for IRAS~2A. The temperature and \ce{H2} number density profiles of this model are shown in the top left panel of Fig.~\ref{fig:IRAS2A_profiles}, with corresponding cartoons shown in the bottom panels. We calculated the \ce{H2} column density (i.e., number density integrated over radius) of an envelope where ices are present between an inner and an outer radius ($R_\mathrm{in}$ and $R_\mathrm{out}$). For each species, $R_\mathrm{in}$ and $R_\mathrm{out}$ can be different; $R_\mathrm{in}$ is considered as the sublimation boundary of ices, and $R_\mathrm{out}$ corresponds to the formation boundary of ices, that is, ices are only present within the envelope; inside $R_\mathrm{in}$, ices have sublimated into the gas phase, and outside $R_\mathrm{out}$, ices have not formed yet. We assumed that the density distributions of all the species (in both gas and ice) are proportional to that of \ce{H2}, and we compared the \ce{H2} column densities of envelopes with different sizes.

Specifically, we considered two cases. In the first case, envelopes have a fixed $R_\mathrm{in}$ but different $R_\mathrm{out}$ (as shown in the bottom middle panel of Fig.~\ref{fig:IRAS2A_profiles}). $R_\mathrm{in}$ is fixed as the radius where $T$ = 100 K, and $R_\mathrm{out}$ are set as radius where $T$ = 10, 20, ..., and 90 K. These temperatures are also noted as $T_\mathrm{form}$ since they by design correspond to the beginning of ice formation. The top middle panel of Fig.~\ref{fig:IRAS2A_profiles} shows how the \ce{H2} column density changes ($N(\mathrm{H_2})$) with $R_\mathrm{out}$. As introduced in Sect.~\ref{sect:intro} and also supported by studies on gas-phase observations \citep[e.g.,][]{Coletta2020, Nazari2022_NCOM, Chen2023}, O-COMs are suggested to start forming in the cold dense pre-stellar phase through CO hydrogenation. According to the constraints on temperature set by the observations (last column in Table~\ref{tab:T_crystal}), the COM ices must have been formed under 100 K in laboratories, which is equivalent to $\sim$60 K in space. The difference in $N(\mathrm{H_2})$ introduced by different $T_\mathrm{form}$ is 50--60\% (i.e., about a factor of two).

In the second case, we compared $N(\mathrm{H_2})$ in envelopes with a fixed $R_\mathrm{out}$ but different $R_\mathrm{in}$. This is to study the influence of desorption temperatures ($T_\mathrm{desorb}$) on ice abundances of different species, as shown in the right panels in Fig.~\ref{fig:IRAS2A_profiles}. For the O-COMs that we focused on in the gas-to-ice comparison (i.e., \ce{CH3OH}, \ce{CH3CHO}, \ce{C2H5OH}, \ce{CH3OCH3}, and \ce{CH3OCHO}), $T_\mathrm{desorb}$ are similar to each other, about 130--160 K under lab conditions \citep{Fedoseev2015, Fedoseev2022}, and will be $\gtrsim$ 90 K under astrophysical conditions. For some larger COMs such as \ce{(CH2OH)2}, $T_\mathrm{desorb}$ can be higher up to 200 K under laboratory conditions (equivalent to $\sim$150 K in space). Although the density is higher toward the protostar, the column densities of envelopes with different $R_\mathrm{in}$ can only vary by $\sim$40\% assuming $T_\mathrm{desorb}\sim$90--150 K for O-COMs.

In reality, each species have different $T_\mathrm{form}$ and $T_\mathrm{desorb}$ ($R_\mathrm{in}$ and $R_\mathrm{out}$, equivalently). In the extreme case, species 1 in Fig.~\ref{fig:spatial_distribution_cartoon} has $T_\mathrm{form}$ = 60 K and $T_\mathrm{desorb}$ = 90 K, and species 2 has $T_\mathrm{form}$ = 10 K and $T_\mathrm{desorb}$ = 150 K, then the column density of species 2 traced by JWST will be a factor of 4.3 as that of species 1. 
For \ce{CH3CHO} and \ce{C2H5OH}, the difference should be smaller given their similar $T_\mathrm{form}$ and $T_\mathrm{desorb}$, in which case different spatial distributions is not able to fully explain the observed order-of-magnitude difference in the column density ratios with respect to \ce{CH3OH} between gas and ice. However, all the estimations were based on the assumption that the density distributions of COMs are proportional to that of \ce{H2}. This may not be true if the formation or evolution mechanisms of COMs are reacting differently to the \ce{H2} density as well as the temperature. The details behind the discrepancy cannot be pinned down only by observations, but require more studies on the spatial distribution of COM ices in protostellar envelopes from the side of simulations and experiments.

\begin{figure}
    \centering
    \includegraphics[width=1.0\hsize]{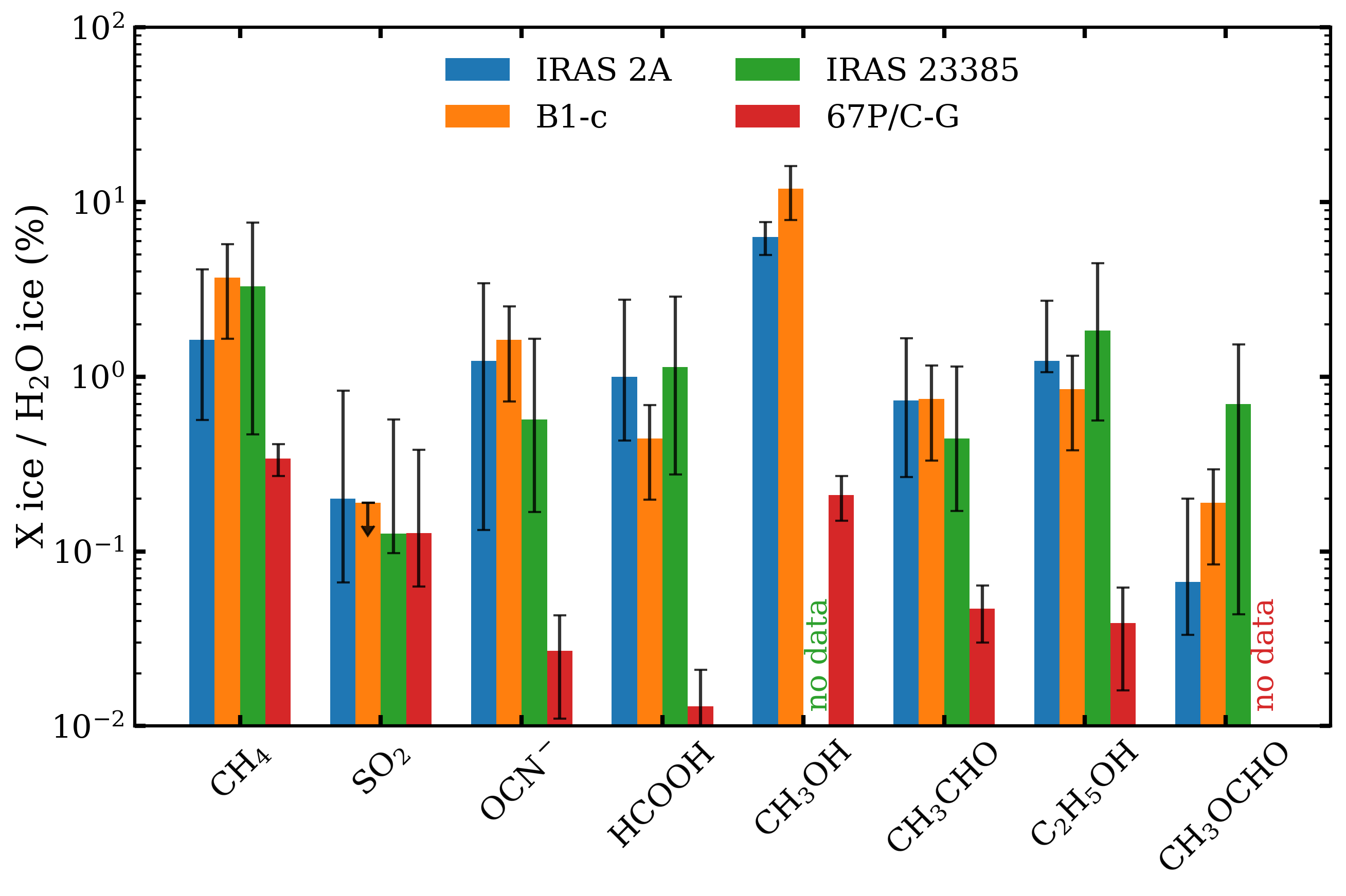}
    \caption{Ice column density ratios of seven species with respect to \ce{H2O} in three protostars \citep[IRAS 23385, IRAS~2A, and B1-c;][this work]{Rocha2024} and one comet \citep[67P/C-G;][]{Rubin2019}. The ice column densities of \ce{CH3OH} for IRAS 23385 and \ce{CH3OCHO} for 67P/C-G are not reported in the references.}
    \label{fig:H2O_ratios}
\end{figure}

\subsection{Ice ratios in protostars and comets}
Additional to the COM ratios with respect to \ce{CH3OH}, we also compared the ice column density ratios of simple molecules and COMs with respect to \ce{H2O} among the three protostellar sources \citep[IRAS~2A, B1-c, and IRAS 23385;][]{Rocha2024} and the comet 67P/C-G \citep{Rubin2019}. We find similar results to those in \cite{Rocha2024} that the ice ratios of most species with respect to \ce{H2O} in protostars are similar and generally higher than the comet, except for \ce{SO2} (Fig.~\ref{fig:H2O_ratios}). The COM ratios match better between protostars and the comet when they are calculated with respect to \ce{CH3OH} instead of \ce{H2O} (Fig.~\ref{fig:COM_ratios_gas_vs_ice}). This may suggest different evolution pathways between simple molecules and COMs, but more sources and studies are needed to draw robust conclusions.

\section{Conclusions}\label{sect:conclusions}
We studied COMs and several simple molecules in both the gas and solid phases for two low-mass protostars, NGC 1333 IRAS~2A and B1-c. We derived the column densities in gas and ice by fitting the gas emission lines in the ALMA Band 7 spectra and the ice absorption bands in the JWST/MIRI-MRS spectrum, respectively. We made the first direct comparison in the column density ratios of four O-COMs with respect to \ce{CH3OH} between the gas and ice, which helps us gain a better understanding of the chemical evolution of these molecules in protostellar systems. Our main conclusions are:

\begin{enumerate}
  \item We derived the gas-phase COM column densities for B1-c and IRAS~2A using a set of high-resolution ALMA Band 7 data, and the COM ratios with respect to \ce{CH3OH} are consistent with previous studies within a factor of two. The small difference may be introduced by the lower spatial resolution and sensitivity in previous observations.
  \item Ices of COMs and other simple molecules are detected in the JWST/MIRI-MRS spectrum of B1-c, following the detection in NGC 1333 IRAS~2A and IRAS 23385+6053 \citep{Rocha2024}. We constrained the total abundance of the detected COM ices with respect to \ce{H2O} ice, $N_\mathrm{ice}$(COM)/$N_\mathrm{ice}$(\ce{H2O}), as $\sim$15\%. In particular, \ce{CH3OH} is the dominate species, with $N_\mathrm{ice}$(\ce{CH3OH})/$N_\mathrm{ice}$(\ce{H2O}) $\sim$ 12\%. The upper limit of the most abundant N-COM, \ce{CH3CN}, is estimated as $N_\mathrm{ice}$(\ce{CH3CN})/$N_\mathrm{ice}$(\ce{H2O}) < 1.2\%.
  \item By directly comparing the lab spectra of different COM ice mixtures and the observed JWST spectrum of B1-c, we find for most COM ices are likely present in a \ce{H2O}-rich environment.
  For some COMs such as \ce{CH3OCH3}, \ce{CH3OH}-rich mixtures cannot be ruled out. \ce{CH3OCHO} is special case that its \ce{H2O}-rich mixture is significantly less matching than the CO:\ce{H2CO}:\ce{CH3OH} mixture with the observations, which implies a formation route of CO hydrogenation.
  \item The temperature of COM ices cannot be well constrained by the observations, since the band profiles of COM ices are highly degenerate before crystallization, which usually occurs at around 100 K under laboratory conditions ($\sim$60--80 K in space). For mixed ices of some COMs such as \ce{C2H5OH}, the degeneracy remains even after crystallization.
  \item The fitting results of the COM fingerprint range (6.8--8.8 $\mu$m) using least-squares and MCMC for B1-c are similar to those reported in \cite{Rocha2024} for IRAS~2A and IRAS 23385 using genetic algorithms. The derived column densities and uncertainties are consistent bewteen B1-c and IRAS~2A.
  The only difference is that we consider \ce{CH3OCH3} and \ce{CH3COCH3} as firmly detected in B1-c, which are considered as tentatively detected in IRAS~2A. We expect to further confirm the detection of these COMs in a larger sample.
  \item The comparison of COM ratios with respect to \ce{CH3OH} between gas and ice show two cases: \ce{CH3OCHO} and \ce{CH3OCH3} have consistent ratios in both phases, while the ice ratios of \ce{CH3CHO} and \ce{C2H5OH} are higher than the gas ones by about one order of magnitude.
  \item The consistency in COM ratios between gas and ice suggests direct inheritance from ice to gas, and probably a common formation history for \ce{CH3OCH3}, \ce{CH3OCHO}, and \ce{CH3OH} ices.
  On the other hand, the difference in COM ratios between gas and ice implies the participation of gas-phase chemistry after ices sublimate in hot cores. Another possible but less dominant cause is the different spatial distributions of COM ices in protostellar envelopes, which are related to the formation time and desorption temperatures of each species.
\end{enumerate}

Thanks to the successful operation of JWST and the high-quality mid-infrared spectra it provides, it is now becoming feasible to have robust detections and quantitative analyses of COMs in the solid phase, allowing direct comparison to the gas-phase counterparts that have been intensively studied by ALMA and other telescopes. As JWST opens a new door to the ice world, it is both intriguing and important to connect it with the already well-explored gas world. The pilot study of two famous hot corinos have shed some light on how COMs evolve through the transition from ice to gas, from the cold envelope to the hot core region. We look forward to further verify our hypotheses in a larger sample, and additional interpretations from simulation or laboratory studies are highly welcome.

\begin{acknowledgements}
This paper makes use of the following ALMA data: ADS/JAO.ALMA\#2021.1.01578.S. ALMA is a partnership of ESO (representing its member states), NSF (USA) and NINS (Japan), together with NRC (Canada), MOST and ASIAA (Taiwan), and KASI (Republic of Korea), in cooperation 74 with the Republic of Chile. The Joint ALMA Observatory is operated by ESO, AUI/NRAO, and NAOJ. 
This work is also based on observations made with the NASA/ESA/CSA James Webb Space Telescope. The data were obtained from the Mikulski Archive for Space Telescopes at the Space Telescope Science Institute, which is operated by the Association of Universities for Research in Astronomy, Inc., under NASA contract NAS 5-03127 for JWST. These observations are associated with program \#1290.
Astrochemistry in Leiden is supported by the Netherlands Research School for Astronomy (NOVA), by funding from the European Research Council (ERC) under the European Union’s Horizon 2020 research and innovation programme (grant agreement no. 101019751 MOLDISK), by the Dutch Research Council (NWO) grants TOP-1 614.001.751 and 618.000.001, and by the Danish National Research Foundation through the Center of Excellence 81 “InterCat” (grant agreement no.: DNRF150).
The work of M.E.R. was carried out at the Jet Propulsion Laboratory, California Institute of Technology, under a contract with the National Aeronautics and Space Administration (80NM0018D0004).
H.B. acknowledges support from the Deutsche Forschungsgemeinschaft in the Collaborative Research Center (SFB 881) “The Milky Way System” (subproject B1).
P.J.K. acknowledges support from the Science Foundation Ireland/Irish Research Council Pathway programme under Grant Number 21/PATH-S/9360.
L.M. acknowledges the financial support from DAE and DST-SERB research grants (SRG/2021/002116 and MTR/2021/000864) of the Government of India.
T.H. acknowledges support from the ERC grant no. 832428 Origins.

\end{acknowledgements}




\bibliographystyle{aa} 
\bibliography{references} 

\newpage
\begin{appendix}
\onecolumn
\section{Images of ALMA and JWST observations}\label{appendix:maps}
\begin{figure*}[!h]
    \centering
    \includegraphics[width=0.85\textwidth]{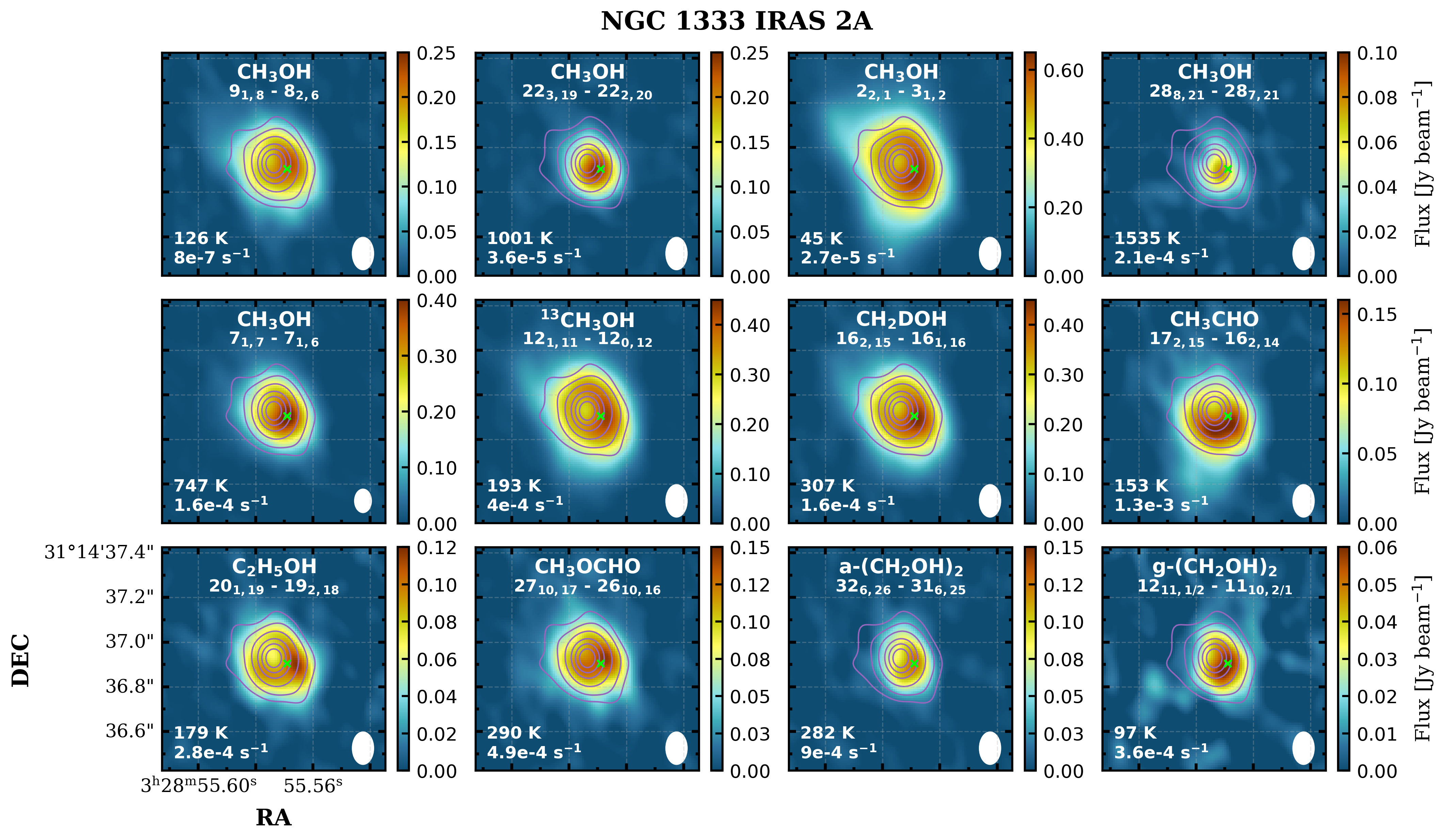}
    \caption{Integrated intensity maps (i.e., moment 0 maps) of 12 emission lines of eight O-COMs in NGC 1333 IRAS~2A. The quantum numbers, $E_\mathrm{up}$, $A_\mathrm{ij}$ of each transition are labelled in white text. The contours in light purple indicate the continuum emission at the 30, 50, 100, 200, 300, 400, and 500$\sigma$ levels ($\sigma$ = 0.2 mJy beam$^{-1}$). The beam size ($0.1\arcsec\times0.15\arcsec$) is denoted by white ellipses in the lower right, and the pixel where the spectrum was extracted is marked by a green cross in each panel. The beam size in the fifth panel is slightly smaller ($0.08\arcsec\times0.11\arcsec$), since this transition lies in a different spectral window than the others.}
    \label{fig:IRAS2A_ALMA_mom0}
\end{figure*}

\begin{figure*}[!h]
    \centering
    \includegraphics[width=0.85\textwidth]{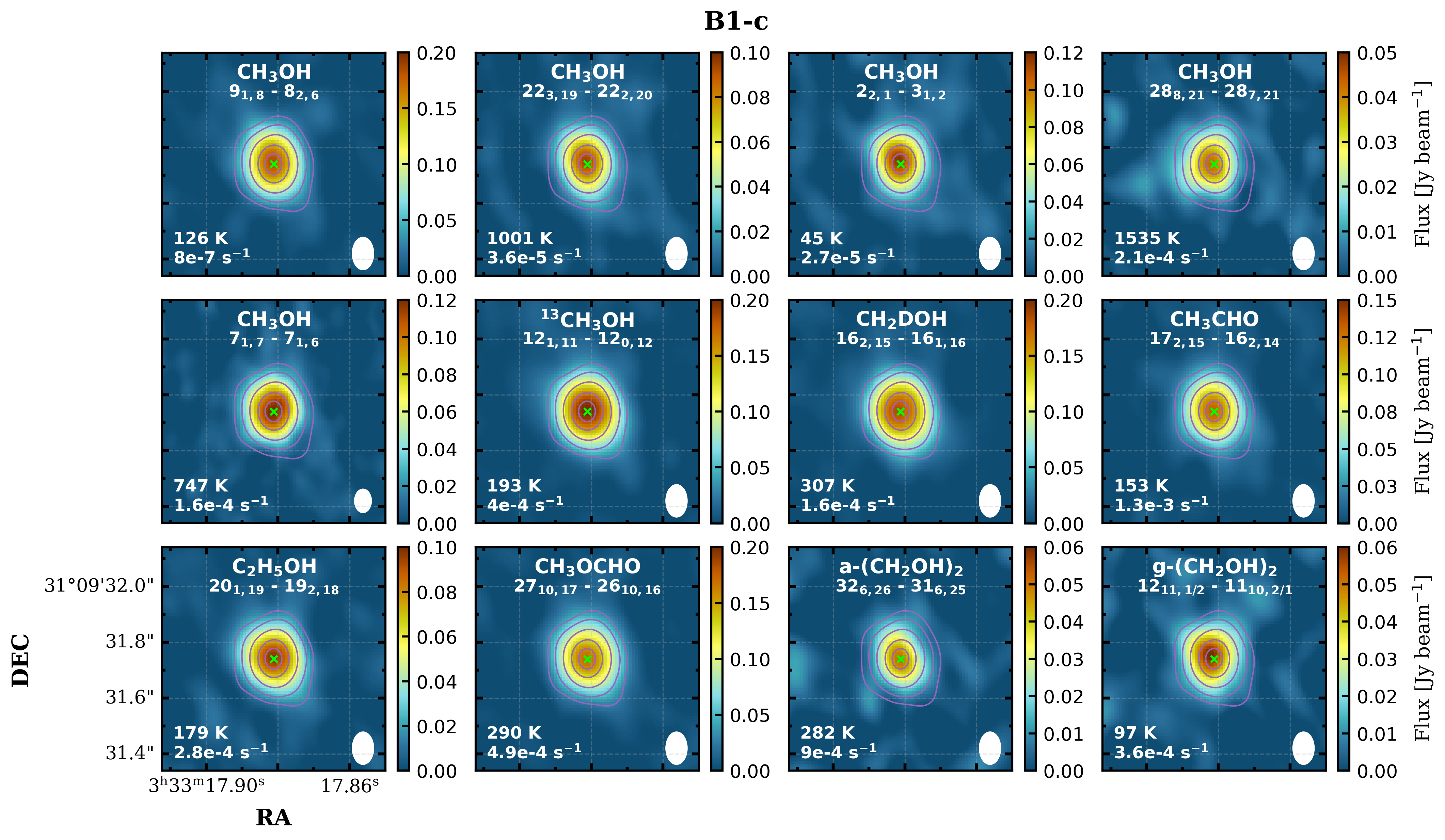}
    \caption{Same as Fig.~\ref{fig:IRAS2A_ALMA_mom0} but for B1-c. The continuum contours are set at 3, 5, 10, 20, 30, and 40$\sigma$ levels ($\sigma$ = 2 mJy beam$^{-1}$).}
    \label{fig:B1c_ALMA_mom0}
\end{figure*}

\begin{figure*}[!h]
    \centering
    \includegraphics[width=\textwidth]{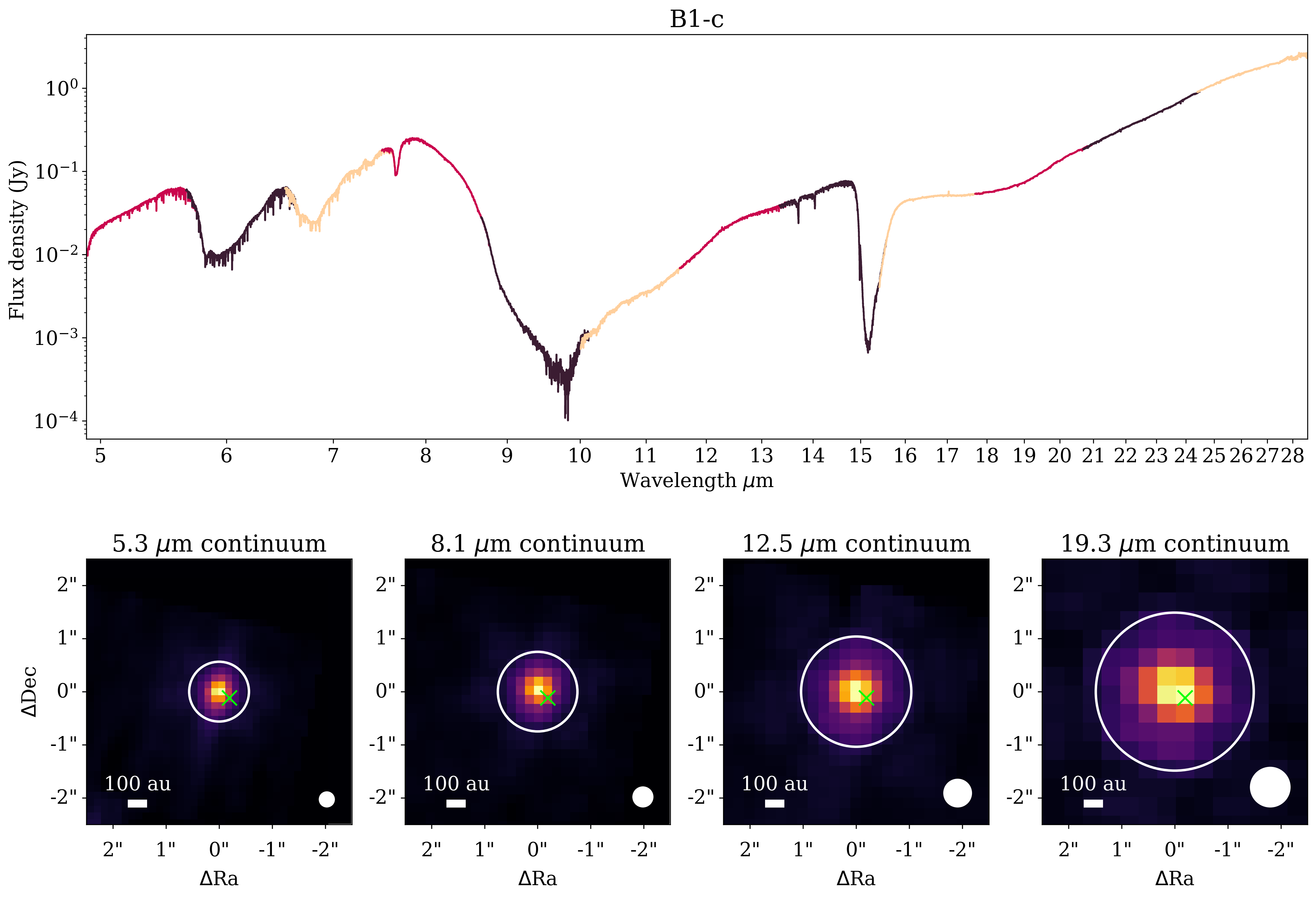}
    \caption{Top panel: JWST/MIRI-MRS spectrum of B1-c extracted from the reduced datacubes. The spectrum slices of three sub-bands in each of the four MIRI-MRS Channels are distinguished in three different colors. Bottom panels: continuum images in four Channels (1: 4.9--7.65 $\mu$m; 2: 7.51--11.7 $\mu$m; 3: 11.55--17.98 $\mu$m; 4: 17.7--27.9 $\mu$m). The sizes of the PSF and the aperture for spectrum extraction increase as a function of wavelength, and are denoted by white circles, respectively. The extraction position of the ALMA spectrum (also the continuum peak) in Fig.~\ref{fig:B1c_ALMA_mom0} is marked here using the same green cross. The ALMA extraction position is well within the MIRI apertures, and the small offset to the aperture center is likely due to the lower accuracy of the JWST pointing.}
    \label{fig:B1c_JWST_spec_extraction}
\end{figure*}

Figures~\ref{fig:IRAS2A_ALMA_mom0}--\ref{fig:B1c_ALMA_mom0} display the integrated intensity maps (i.e., moment 0 maps) of 12 COM lines in IRAS~2A and B1-c. Five lines are from \ce{CH3OH} with different upper energy levels ($E_\mathrm{up}$) and Einstein A coefficients ($A_\mathrm{ij}$). For the other seven O-COMs, the strongest unblended line of each species was selected to display.
Figure~\ref{fig:B1c_JWST_spec_extraction} shows the continuum images in 12 MRS channels along with the extracted spectrum.

\twocolumn
\clearpage
\section{The silicate features}\label{appendix:silicates}
\begin{figure}[!h]
    \centering
    \includegraphics[width=0.5\textwidth]{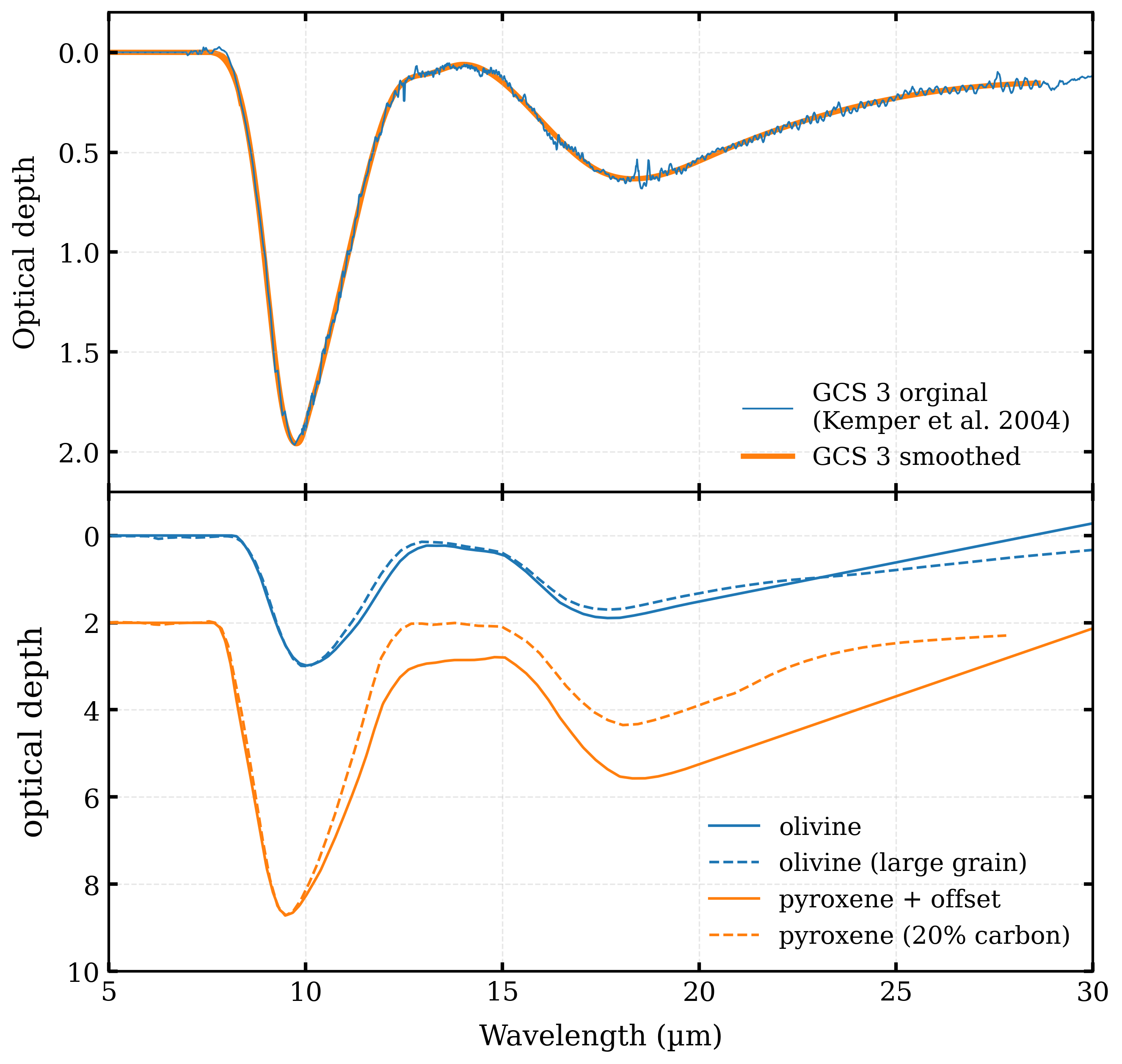}
    \caption{Top: original and smoothed silicate spectra of GCS~3. Bottom: computed spectra of olivine (blue) and pyroxene (orange) with increased grain size or mass fraction of carbon (dashed).}
    \label{fig:appendix_silicates}
\end{figure}

All the silicate spectra used to fit the observed spectrum of B1-c in Sect.~\ref{sect:method_silicates} have no features shortward of 7.6~$\mu$m.
The GCS\,3 silicate spectrum shown in Fig.~\ref{fig:fitting_steps}b is a smoothed version of the original spectrum, in which the emission and absorption lines that are not attributed to silicates were removed (top panel of Fig.~\ref{fig:appendix_silicates}).

When fitting the silicate features, we primarily focused on whether the profiles of the 9.8 and 18~$\mu$m bands and their optical depth ratios match between the observations and the silicate spectra used for fitting. Figure~\ref{fig:appendix_silicates} shows that the optical depth ratio between the two bands in the GCS\,3 profile is $\sim$3.3, larger than those of the computed spectra of olivine and pyroxene ($\lesssim2$). However, the relative intensity of the 18~$\mu$m band in the pyroxene spectrum can be reduced by increasing the mass fraction of carbon. The computed spectrum of pyroxene is by default made up of \ce{Mg_{0.7}Fe_{0.3}SiO3} and mixed with carbon in a mass fraction of 87\% and 13\%, respectively. We finally used the spectrum of 80\% pyroxene mixed with 20\% carbon to get a better fit for the 18~$\mu$m band. The 18~$\mu$m band in the olivine spectrum can also be widened by increasing the grain size (bottom panel of Fig.~\ref{fig:appendix_silicates}). In Sect.~\ref{sect:method_silicates} we only came up with a plausible fitting of the silicate features, but potentially there is a huge parameter space to explore.

Another caveat is that even small difference in the global continuum fitting (Sect.~\ref{sect:method_global_cont}) can change the spectral profile at long wavelengths, which makes it degenerate to fit the global continuum and the 18~$\mu$m silicate band. Fortunately, our analysis on the COMs fingerprints between 6.8 and 8.8~$\mu$m is little affected by the fitting of the 18~$\mu$m band, and we are safe to proceed with a plausible fitting shown in Fig.~\ref{fig:fitting_steps}b.


\section{Selection of ice mixtures}\label{appendix:obs_vs_lab}
\begin{figure*}[!h]
    \centering
    \includegraphics[width=0.95\textwidth]{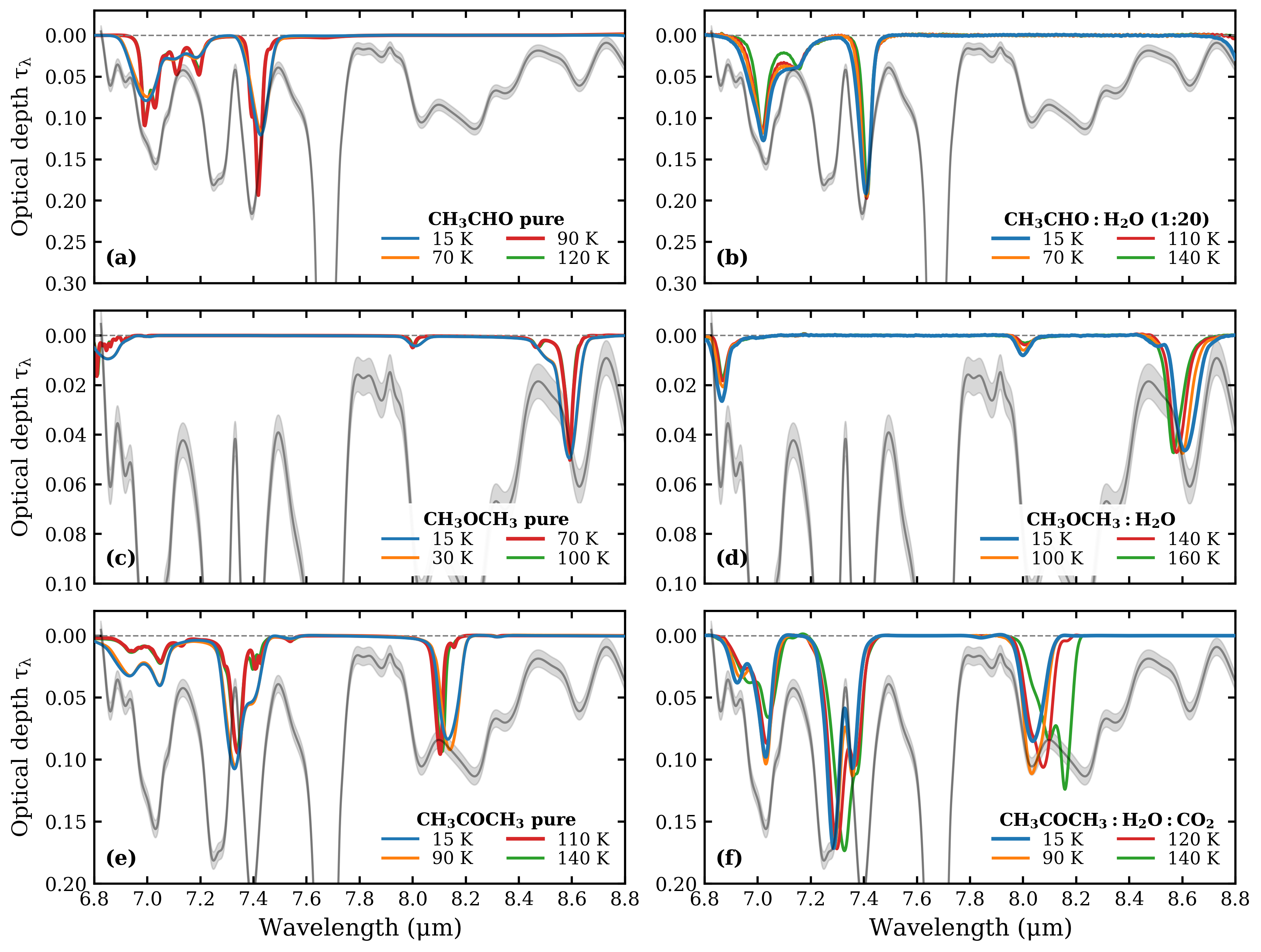}
    \caption{Same as Fig.~\ref{fig:obs_vs_lab_T_1} but for \ce{CH3CHO}, \ce{CH3OCH3}, and \ce{CH3COCH3}.}
    \label{fig:obs_vs_lab_T_2}
\end{figure*}

Figure~\ref{fig:obs_vs_lab_T_2} is the same as Fig.~\ref{fig:obs_vs_lab_T_1} but shows the temperature comparison for the lab spectra of \ce{CH3CHO}, \ce{CH3OCH3}, and \ce{CH3COCH3}. Same as \ce{C2H5OH} and \ce{CH3OCHO}, the band of pure ices will become significantly sharper after crystallization. For the mixed ices, the bands of \ce{CH3CHO}:\ce{H2O} mixture become narrower as temperature increases, but the peak positions remain similar and cannot be ruled out by the observations. The major change of \ce{CH3OCH3}:\ce{H2O} mixture is the peak position of the 8.63 $\mu$m band. The band is blue-shifted as temperature increases, and anti-correlates with the observations when $T>120$ K. For \ce{CH3COCH3}:\ce{H2O} mixture, as temperature increases, the two band at 7.3 and 8.03 $\mu$m are redshifted, and the latter splits into two bands. The anti-correlation occurs at 7.3 $\mu$m when $T>$70--90 K. 

\begin{figure*}[!h]
    \centering
    \includegraphics[width=0.95\textwidth]{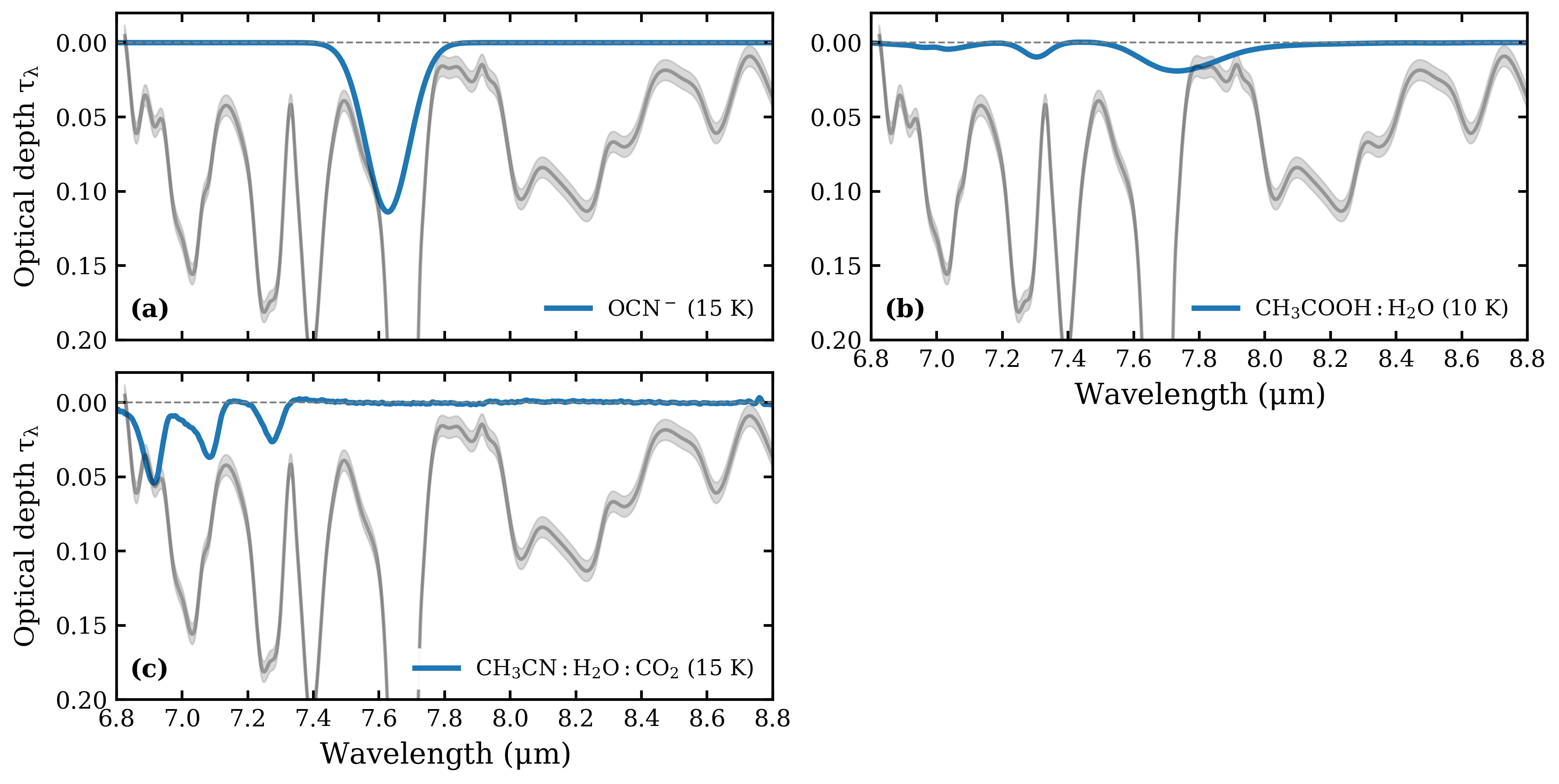}
    \caption{Same as Fig.~\ref{fig:obs_vs_lab_mix} but for \ce{OCN^-}, \ce{CH3COOH}, and \ce{CH3CN}.}
    \label{fig:obs_vs_lab_uplim}
\end{figure*}

\section{Baseline correction of lab spectra}\label{appendix:baseline_correction}
The lab spectra of COMs mixed with \ce{H2O} or \ce{CH3OH} require baseline correction to isolate the weak COM bands from the strong \ce{H2O} band at 6~$\mu$m or the \ce{CH3OH} band at 6.75~$\mu$m. A polynomial was fit to wavelength regions that are free of absorption bands. The baseline correction was only applied between $\sim$6.0 and 10.0~$\mu$m, considering that we only fit the observation between 6.8 and 8.8~$\mu$m.

\begin{figure}[!h]
    \centering
    \includegraphics[width=0.49\textwidth]{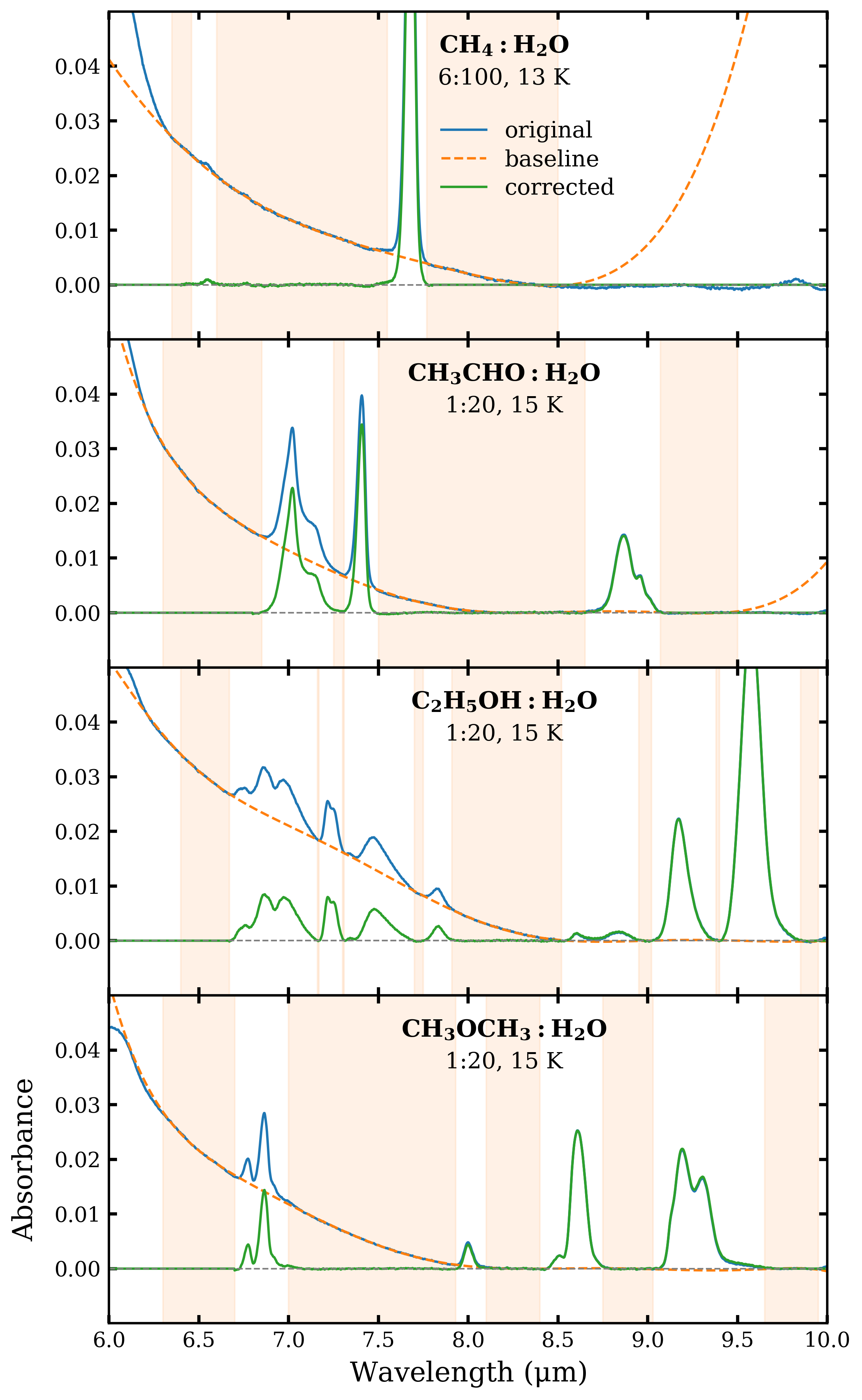}
    \caption{Baseline correction to lab spectra of ice mixtures. We take the \ce{H2O}-rich mixtures of \ce{CH4} and three O-COMs (\ce{CH3CHO}, \ce{C2H5OH}, \ce{CH3OCH3}) as examples. The original and the corrected spectra are plotted in solid blue and green, respectively. The polynomial is plotted in dashed orange, and the wavelength ranges that were selected to fit the polynomial are highlighted in orange panels.}
    \label{fig:blcorr_H2Omix}
\end{figure}

For \ce{H2O}-rich mixtures, the baseline is just the red wing of the \ce{H2O} bending mode at 6.0 $\mu$m, and it is easy to select and fit a polynomial to those ranges that do not have absorption features of the interested species, as shown in Fig.~\ref{fig:blcorr_H2Omix}.

\begin{figure}[!h]
    \centering
    \includegraphics[width=0.49\textwidth]{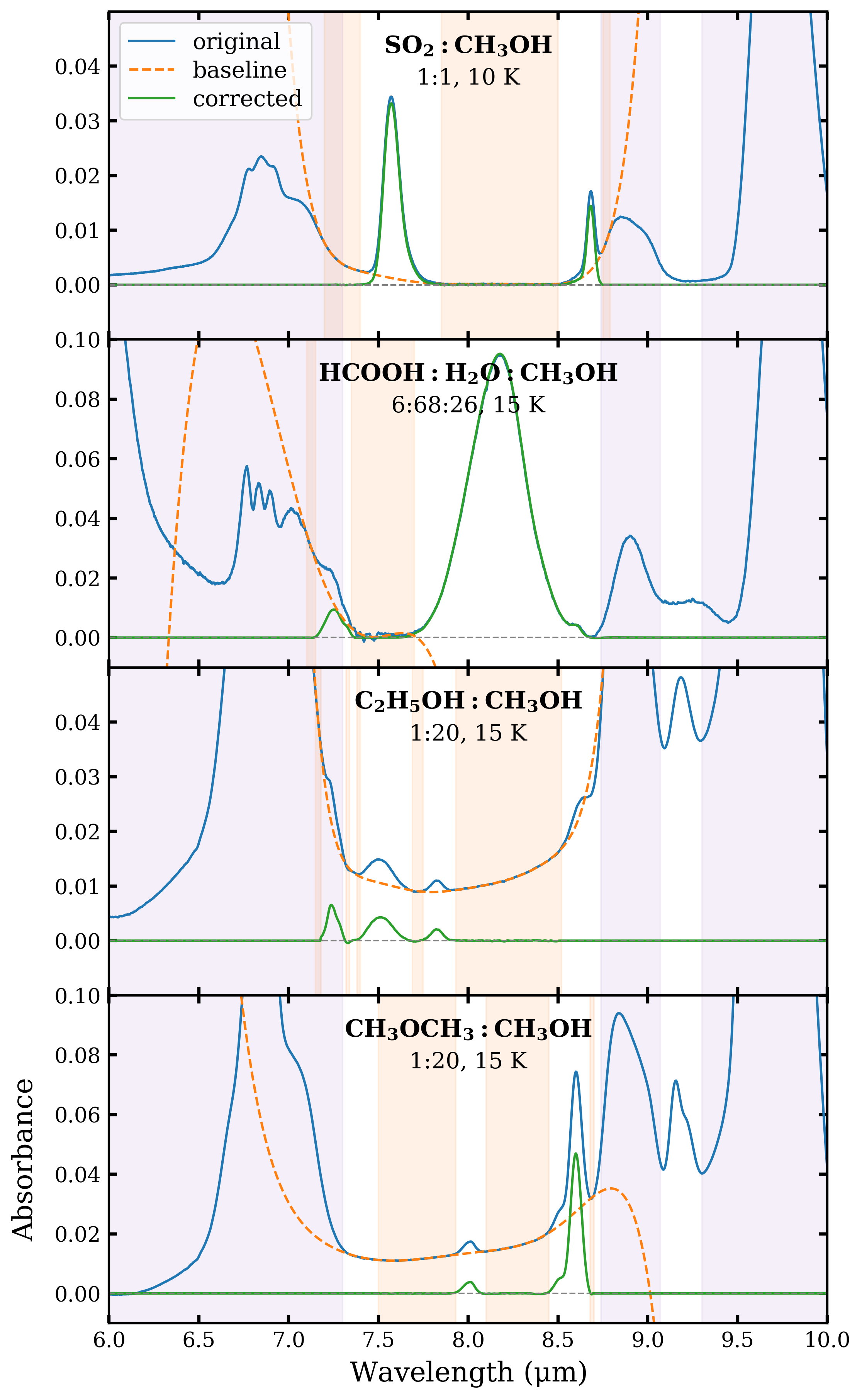}
    \caption{Similar to Fig.~\ref{fig:blcorr_H2Omix} but for \ce{CH3OH} mixtures. \ce{SO2}, \ce{HCOOH}, \ce{C2H5OH}, and \ce{CH3OCH3} are taken as examples. The shaded regions in purple indicate the frequency ranges of \ce{CH3OH} bands.}
    \label{fig:blcorr_CH3OHmix}
\end{figure}

However, it is usually not straightforward to do baseline correction for \ce{CH3OH}-rich mixtures, especially in 6.5--7.3 $\mu$m where there is a \ce{CH3OH} band with wavy features (Fig.~\ref{fig:blcorr_CH3OHmix}; the feature shown in the top panel are \ce{CH3OH}-only). If the interested species has no overlapping band with \ce{CH3OH} (e.g., \ce{SO2}), the routine is the same as for \ce{H2O}-rich mixtures, except that a higher order polynomial may be needed. Unfortunately, most O-COMs have weak bands around 7.0 $\mu$m, and it is difficult to isolate those features from the uneven \ce{CH3OH} features. We tried to reserve the weak bands between 7.1 and 7.3 $\mu$m where the \ce{CH3OH} spectrum is still smooth and it is feasible to trace a local continuum; this was done for HCOOH and \ce{C2H5OH} which have a weak band at $\sim$7.2 $\mu$m. For those bands shortward of 7.1 $\mu$m, we discard them in the corrected spectra (e.g., for \ce{C2H5OH}, \ce{CH3OCH3}, and \ce{CH3OCHO}).

\section{Fitting results of the ALMA spectra of IRAS~2A and B1-c}\label{appendix:ALMA_fitting}
\begin{figure*}[!h]
    \centering
    \includegraphics[width=\textwidth]{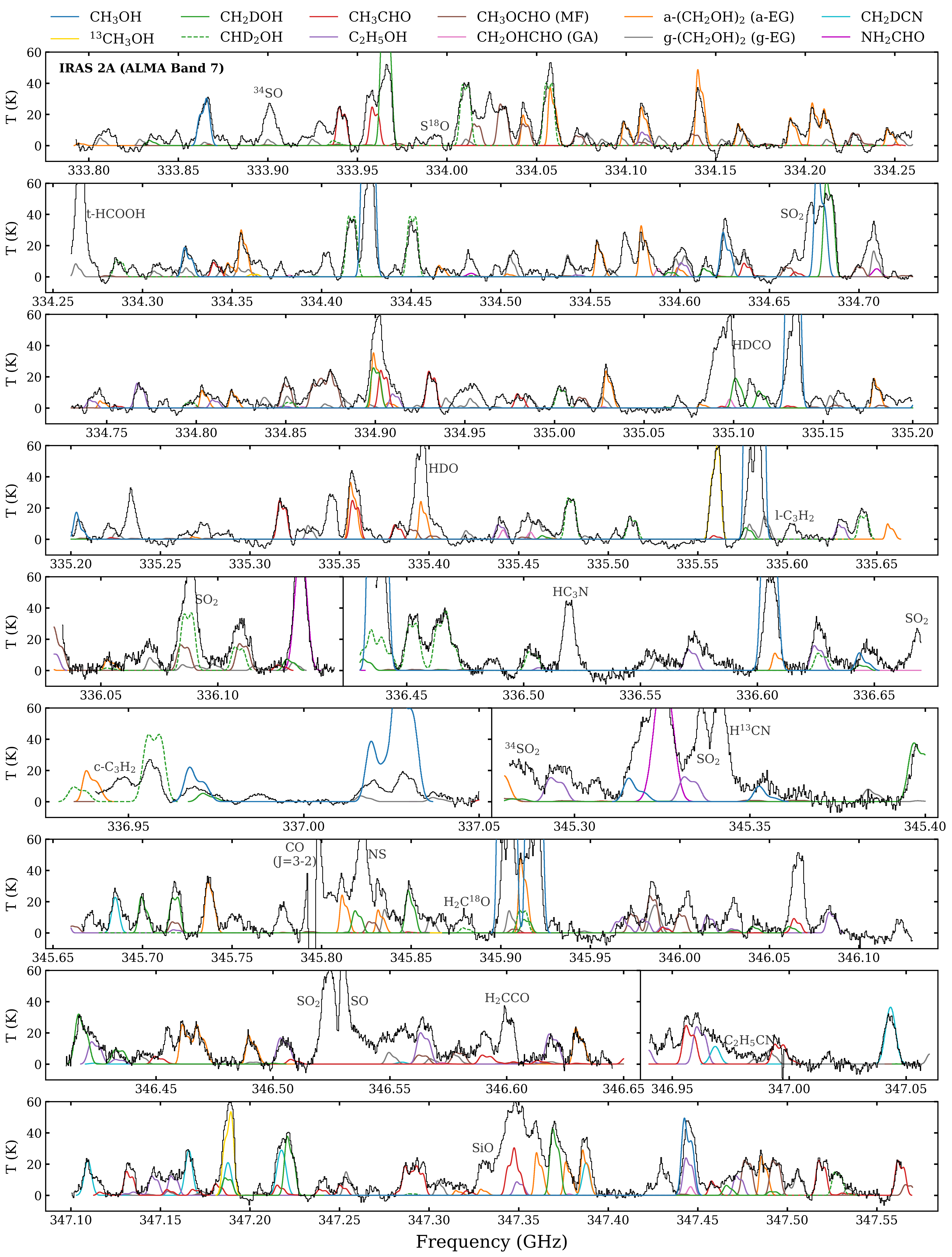}
    \caption{Best-fit LTE models overlaid on the observed ALMA spectrum of IRAS~2A.}
    \label{fig:ALMA_fit_IRAS2A}
\end{figure*}

\begin{figure*}[!h]
    \centering
    \includegraphics[width=1\textwidth]{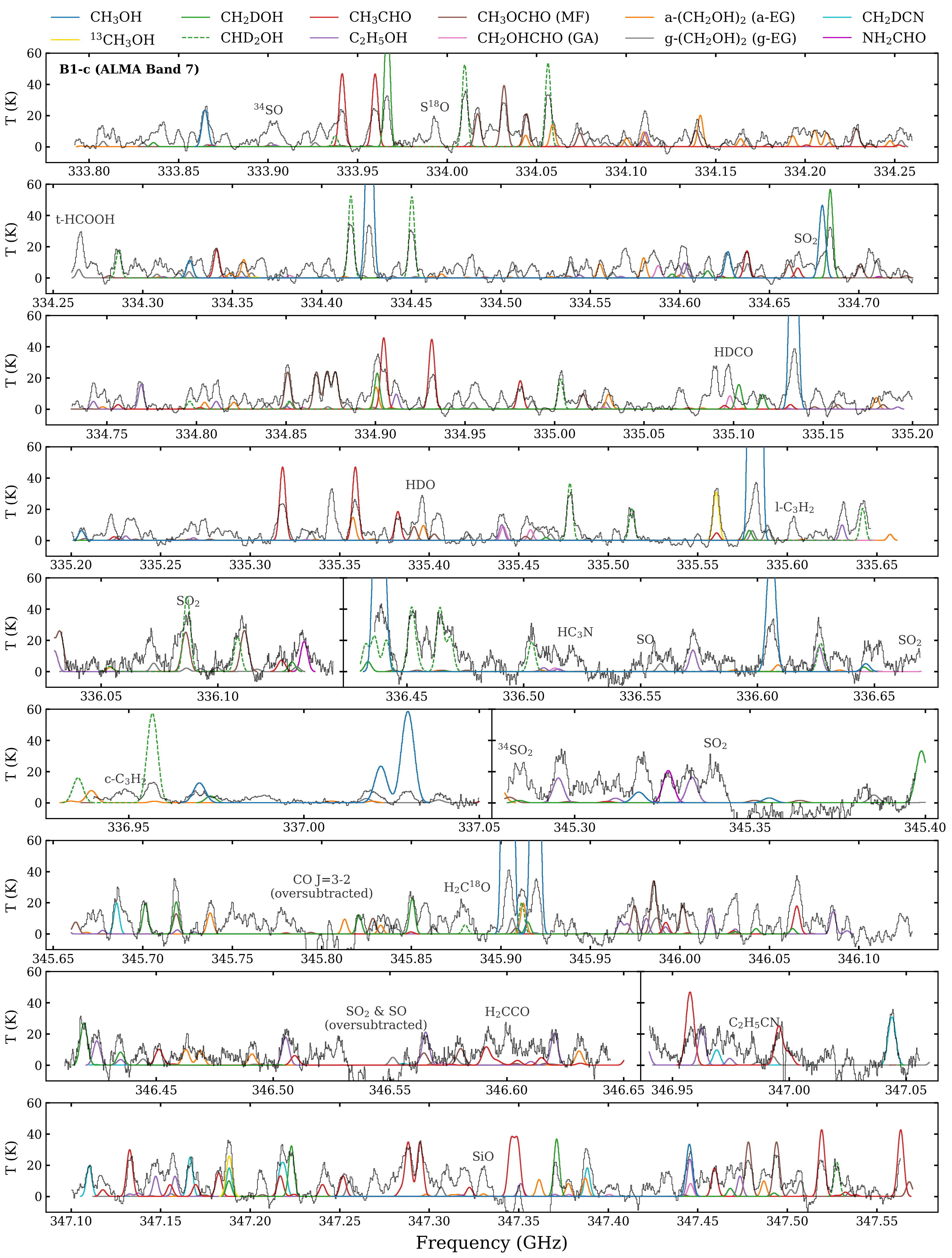}
    \caption{Same as Fig.~\ref{fig:ALMA_fit_IRAS2A} but for B1-c.}
    \label{fig:ALMA_fit_B1-c}
\end{figure*}

Figures \ref{fig:ALMA_fit_IRAS2A}--\ref{fig:ALMA_fit_B1-c} show the full ALMA spectra of IRAS~2A and B1-c and the best-fit LTE models of the detected COMs, respectively. Important strong lines of simple molecules are labelled in gray text. 

\onecolumn

\clearpage
\twocolumn
\section{\ce{H2CO} bands in the \ce{CH3OCHO} mixture}\label{appendix:H2CO}
\begin{figure}[!h]
    \centering
    \includegraphics[width=0.5\textwidth]{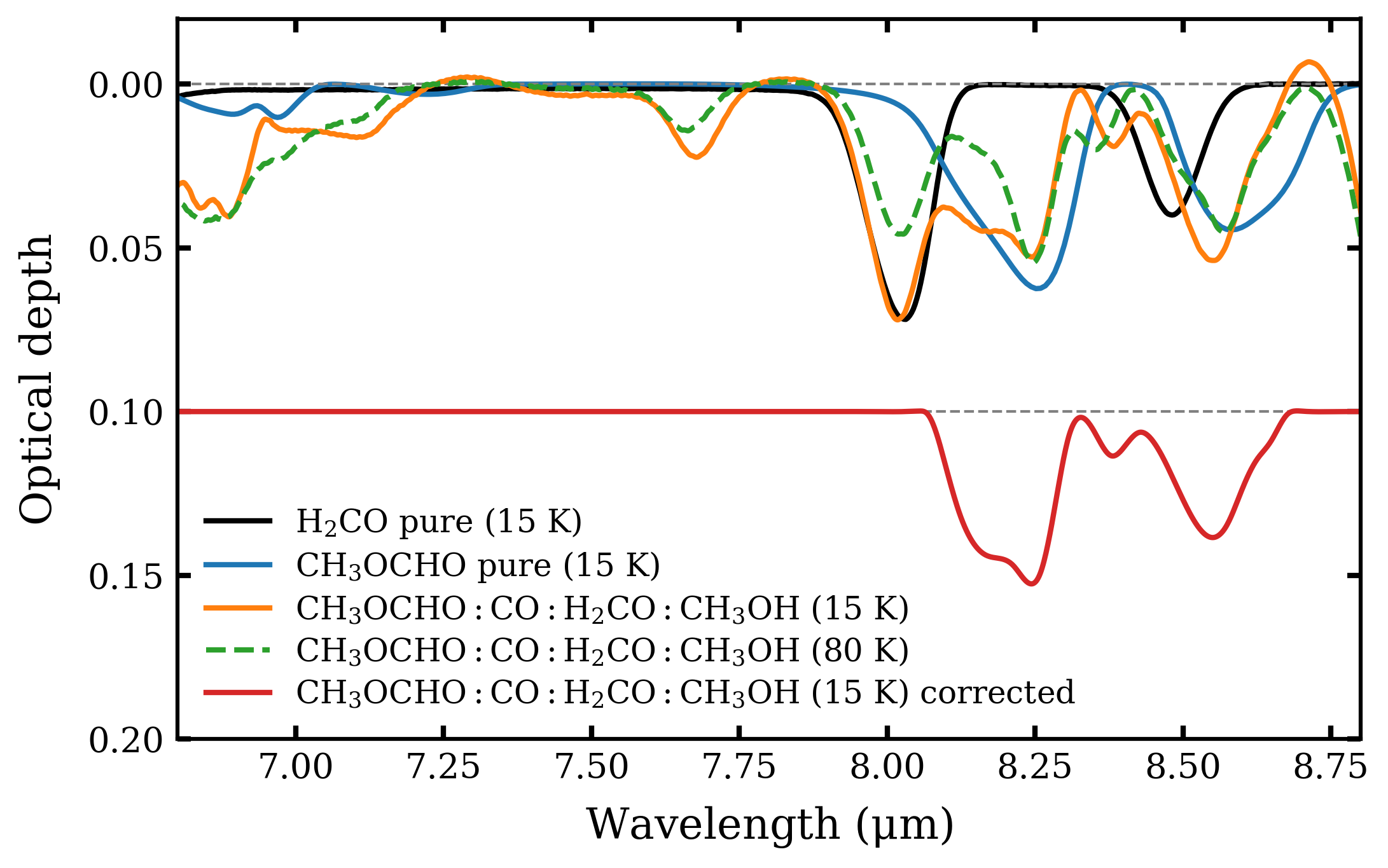}
    \caption{Comparison among the lab spectra of pure \ce{H2CO} ice (black), pure \ce{CH3OCHO} ice (blue), and \ce{CH3OCHO}:CO:\ce{H2CO}:\ce{CH3OH} mixture under 15 K (orange) and 80 K (dashed green). The corrected spectrum for the \ce{CH3OCHO} mixture under 15 K (red) is plotted at an offset place.}
    \label{fig:MF_apolar_corr}
\end{figure}

As mentioned in Sect.~\ref{sect:results_JWST_mixing_consituent} and shown in Fig.~\ref{fig:obs_vs_lab_mix}h, the lab spectrum of \ce{CH3OCHO} mixed with CO, \ce{H2CO}, and \ce{CH3OH} in a mixing ratio of 1:20:20:20 (hereafter CO-rich mixture) fits best with the observations. However, the band at 8$\mu$m does not appear in the pure ice and other ice mixtures with \ce{H2O} and \ce{CH3OH}, suggesting that this absorption feature comes from neither \ce{CH3OCHO} nor \ce{CH3OH}. In fact, this 8$\mu$m band belongs to \ce{H2CO}. Figure~\ref{fig:MF_apolar_corr} compares the lab spectra of pure \ce{H2CO}, pure \ce{CH3OCHO}, and the CO-rich mixture of \ce{CH3OCHO} under 15 K and 80 K. It is shown that the 8 $\mu$m band in the CO-rich mixture remains almost the same as in the pure \ce{H2CO}, which implies that this band is not affected by ice mixing, and therefore can be treated separately from the \ce{CH3OCHO} band at 8.25 $\mu$m.

The problem of using the original lab spectrum of the \ce{CH3OCHO} CO-rich mixture is that the mixing ratio between \ce{CH3OCHO} and \ce{H2CO} is fixed at 1:20, and therefore the intensity ratio between the \ce{CH3OCHO} band at 8.25 $\mu$m and the \ce{H2CO} band at 8.02 $\mu$m is fixed as well. The lab spectra of the \ce{CH3OCHO} CO-rich mixture at 80 K also shows that the intensity ratio between the 8.25 $\mu$m and 8.02 $\mu$m bands would increase as part of \ce{H2CO} has sublimated from the ice mixture at 80 K. However, the 1:20 ratio between \ce{H2CO} and \ce{CH3OCHO} may not be realistic; it was found in test fittings that the 8.02 $\mu$m band would be overfit and the 8.25 $\mu$m band would be underfit the observations if we used the original lab spectrum. As a result, we wanted to break the correlation between \ce{H2O} and \ce{CH3OCHO} in the CO-rich mixture to achieve a better fit at both bands.

To address this problem, we fit a Gaussian to the \ce{H2CO} band at 8.02 $\mu$m and subtracted it from the lab spectrum of the \ce{CH3OCHO} CO-rich mixture, in order to isolate the \ce{CH3OCHO} band from the \ce{H2CO} band. There is also a weaker band of \ce{H2CO} at 8.48 $\mu$m, which splits the 8.59 $\mu$m band in pure \ce{CH3OCHO} into two peaks at 8.38 and 8.55 $\mu$m in the CO-rich mixture. The presence of \ce{H2CO} also increase the relative intensity between the bands at 8.55 $\mu$m and 8.25 $\mu$m from 0.73 (in pure \ce{CH3OCHO}) to 1 (in CO-rich mixture). To further separate the influence by \ce{H2CO}, we scaled down the relatively intensity of the 8.55 $\mu$m band back to 0.73. Finally, we removed other weak feature shortward of 7.8 $\mu$m that are either absent in the pure \ce{CH3OCHO} spectrum or blended with \ce{CH3OH} features at $\sim$6.9 $\mu$m (considering that the ratio between \ce{CH3OCHO} and \ce{CH3OH} is 1:20, the contribution from \ce{CH3OCHO} is negligible). The corrected lab spectrum of \ce{CH3OCHO} CO-rich mixture is shown in Fig.~\ref{fig:MF_apolar_corr}, and was the one that used in the overall fitting.

\begin{figure}[!h]
    \centering
    \includegraphics[width=0.5\textwidth]{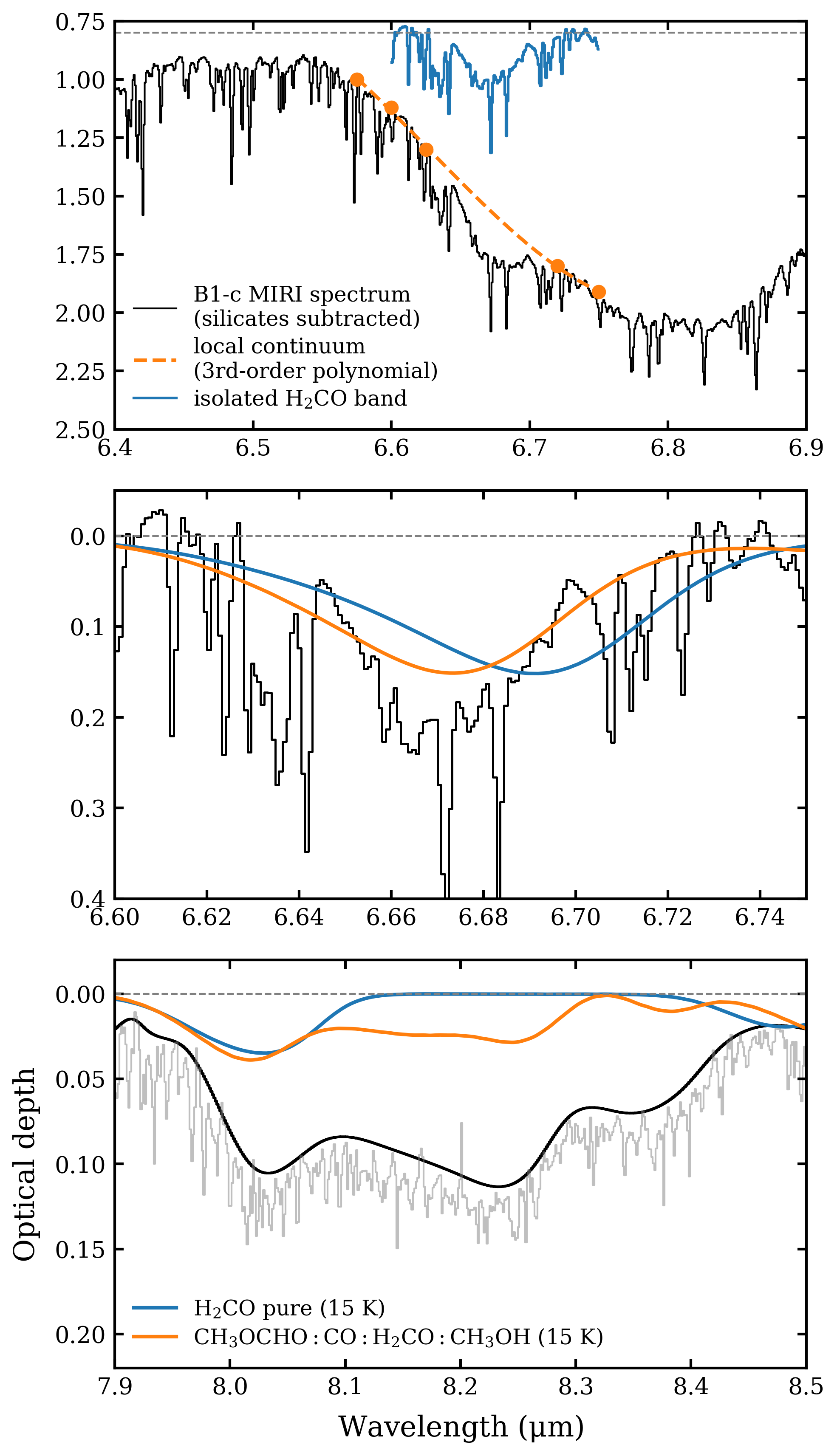}
    \caption{Top: the \ce{H2CO} band at 6.67 $\mu$m (blue) is isolated by tracing a local continuum (orange) to the observed spectrum (black); superposed gas-phase absorption lines of \ce{H2O} are not removed here. Middle: lab spectra of pure \ce{H2CO} ice (blue) and \ce{CH3OCHO}:CO:\ce{H2CO}:\ce{CH3OH} mixture (orange) are manually scaled to the isolated {H2CO} band at 6.67 $\mu$m in observations (black). Bottom: same as the middle panel but for a different wavelength range (7.9--8.5 $\mu$m).}
    \label{fig:H2CO_bands}
\end{figure}

\clearpage
\onecolumn
On the other hand, the 8.03 $\mu$m band observed in the B1-c spectrum has contribution from not only \ce{H2CO}, but also \ce{H2O}-rich mixtures of \ce{CH3COCH3} (Sect.~\ref{sect:results_JWST_mixing_consituent} and panels i--j in Fig.~\ref{fig:obs_vs_lab_mix}). To determine how much \ce{H2CO} accounts for this band, we turned into another characteristic band of \ce{H2CO} at 6.67 $\mu$m, which is outside of the COM fingerprint range. We first fit a local continuum following a similar routine to Sect.~\ref{sect:method_local_cont} to isolated this band (but without removing the gas-phase absorption lines), and then manually scaled the lab spectra of pure \ce{H2CO} and \ce{CH3OCHO} CO-rich mixture to the observations (see Fig~\ref{fig:H2CO_bands}). The observed \ce{H2CO} band at 6.67 $\mu$m is about 0.02 $\mu$m blueshifted from pure \ce{H2CO}, but matches well with the \ce{CH3OCHO} CO-rich mixture (where CO:\ce{H2CO}:\ce{CH3OH} = 1:1:1), which implies that \ce{H2CO} is likely to have surroundings rich in CO and \ce{CH3OH} in interstellar ices. The two \ce{H2CO}-relevant lab spectra were scaled to have the same intensities at 6.67 $\mu$m. The scaling factor of pure \ce{H2CO} was then determined by visual inspection, and was fixed in the overall fitting. The ice column density of \ce{H2CO} was calculated from the 8.02 $\mu$m band.

\section{Temperature analysis of COM ices}
Table~\ref{tab:T_crystal} summarizes the temperature-relevant information of the lab spectra of the five O-COMs that are considered detected in the JWST spectrum of B1-c. 

\begin{table*}[!h]
    \setlength{\tabcolsep}{0.4cm}  
    \centering
    \caption{Inferred range of crystallization temperature of COM ices under laboratory conditions ($T_\mathrm{crystal,\,lab}$). All the considered lab spectra of the five COM ices are obtained from the LIDA database \citep{Rocha2022_LIDA}.}
    \begin{tabular}{lccccc}
        \toprule
         & \multicolumn{2}{c}{pure ices} & \multicolumn{3}{c}{mixed ices$^a$} \\
        \cmidrule(lr{0.1em}){2-3}\cmidrule(lr{0.1em}){4-6}
        Species & measured $T_\mathrm{lab}$ (K)$^b$ & $T_\mathrm{crystal,\,lab}$ (K) & measured $T_\mathrm{lab}$ (K)$^b$ & $T_\mathrm{crystal,\,lab}$ (K) & fit B1-c (K)$^c$ \\
        \midrule
        \ce{CH3CHO} & \makecell{15, 30, 70, 90,\\110, 120} & 70--90 & \makecell{15, 30, 70, 90,\\110, 120, 140, 160} & 110--140 & all$^d$ \\
        \midrule
        \ce{C2H5OH} & \makecell{15, 30, 70, 100,\\120, 130, 140, 150} & 100--120 & \makecell{15, 30, 70, 100, 120,\\130, 140, 150, 160} & 100--160 & all$^d$ \\
        \midrule
        \ce{CH3OCH3} & 15, 30, 70, 90, 100 & 30$^e$--70 & \makecell{15, 30, 70, 90,\\100, 120, 140, 160} & 100--140 & < 100--120 \\
        \midrule
        \ce{CH3OCHO} & \makecell{15, 30, 50, 80,\\100, 120} & 80--100 & \makecell{15, 30, 50, 80,\\100, 120} & 50--80$^f$ & < 100\\
        \midrule
        \ce{CH3COCH3} & \makecell{15, 30, 70, 90,\\110, 120, 130, 140} & 90--110 & \makecell{15, 30, 70, 90, 100,\\110, 120, 140, 160} & 100--140 & < 90\\ 
        \bottomrule
    \end{tabular}
    \begin{minipage}{0.95\textwidth}
    $^a$ The mixing constituents are provided in the legends of Figs.~\ref{fig:obs_vs_lab_T_1} and \ref{fig:obs_vs_lab_T_2}.\\
    $^b$ In experiments, temperature is usually sampled at small intervals (e.g., 1 K), but only a small part of the spectra measured at certain temperatures are provided on databases.\\
    $^c$ The range of $T_\mathrm{lab}$ in which the lab spectra can fit into the B1-c spectrum, as long as there is no anti-correlation. \\
    $^d$ The changes in band profiles around crystallization are small, therefore it is hard to exclude the crystalline features when compared with observations. \\
    $^e$ The lower limit of $T_\mathrm{crystal,\,lab}$ of pure \ce{CH3OCH3} ice is likely higher than 30 K. \\
    $^f$ For \ce{CH3OCHO}, the $T_\mathrm{crystal,\,lab}$ of mixed ice is lower than that of pure ice. This is likely due to the crystallization or desorption of CO and \ce{H2CO}, not the crystallization of \ce{CH3OCHO} itself. \\
    \end{minipage}
    \label{tab:T_crystal}
\end{table*}

\clearpage
\section{Band strengths}\label{appendix:band_strength}
The band strength $A$ describes how strong is the absorbance of a specific band given a unit of ice column density (cm$^{-2}$). In Table~\ref{tab:JWST_results}, Col.~3 lists the characteristic absorption bands of the species (Col.~1), and Cols. 3--4 provide values and references of the corresponding band strengths. Most of these values and references are the same as those given in \cite{Rocha2024}, except for the following species:
\begin{itemize}
    \item \ce{CH3OH}: for the 9.74 $\mu$m band, we adopted $A=1.56\times10^{-17}$ cm$^{-2}$ from \cite{Luna2018}, instead of $1.8\times10^{-17}$ cm$^{-2}$ from \cite{Bouilloud2015}, considering that the former one is more updated.
    \item \ce{CH3OCH3}: the band strength of the 8.59 $\mu$m band is adopted from Table~1 in \cite{TvS2018}, and $A$ is corrected for the \ce{H2O} mixture by a factor of $\sim$0.5. This factor, known as the relative band strength, is not directly provided in \cite{TvS2018} but can only be read from the lower right panel of Fig.~C.11 therein. \cite{Rocha2024} considered it as 0.5, and the corrected $A$ is $4.9\times10^{-17}$ cm$^{-2}$. However, a more accurate estimation of this factor would be 0.56--0.57, yielding a slightly larger $A$ of  $5.55\times10^{-17}$  cm$^{-2}$.
    \item \ce{CH3COCH3}: the band strength of the 7.33 $\mu$m band is originally measured as $1.39\times10^{-17}$ cm$^{-2}$ by \cite{Hudson2018}. The relative $A$ of the \ce{H2O} mixture is not directly stated, but plotted in the lower right panel of Fig.~A.11 in \cite{Rachid2020}. The estimated relative $A$ is 0.73, and the corrected $A$ is $1.02\times10^{-17}$ cm$^{-2}$, a bit smaller than 1.2$\times10^{-17}$ cm$^{-2}$ used by \cite{Rocha2024}\footnote{In Table 1 of \cite{Rocha2024}, the identification of the 7.33 $\mu$m band of \ce{CH3COCH3} is mistakenly noted as the CCC asymmetric stretching mode, which is actually the band at 8.03 $\mu$m. The 7.33 $\mu$m band corresponds to the \ce{CH3} symmetric stretching mode.}
\end{itemize}

\section{Statistics of fitting the JWST/MIRI-MRS spectrum of B1-c}\label{appendix:JWST_fitting}
Table~\ref{tab:JWST_B1c_fit_statistics} lists the statistics of fitting the JWST/MIRI-MRS spectrum of B1-c, including the best-fit values and uncertainties of the scaling factors of the selected lab spectra.

\begin{table*}[!hp]
    \setlength{\tabcolsep}{0.15cm}
    \caption{Fitting statistics of the scaling factors $^a$ of lab spectra. The best-fit values, standard errors ($\sigma_a$) and relative errors ($a/\sigma_a$) are listed for least-squares and MCMC fittings. All the 12 candidate species were considered in the least-squares fitting, and nine of them were included in the MCMC fitting.}
    \centering
    \begin{tabular}{cccccccccc}
    \toprule
    \multicolumn{4}{c}{Ice mixture} & \multicolumn{3}{c}{Least-squares} & \multicolumn{3}{c}{MCMC}\\
    \cmidrule(lr{0.3em}){5-7}\cmidrule(lr{0.3em}){8-10}
    Constituents & Ratio & $T$ (K) & Fit range$^b$ & Best-fit $a$ & $\sigma_a$ & $a/\sigma_a$ (\%) & Best-fit $a$ & $\sigma$ & $a/\sigma_a$ (\%) \\
    \midrule
    \ce{CH4}:\ce{H2O} & 1:100 & 13 & 0--5 & 3.61 & 0.03 & 0.8 & 3.66 & 0.27 & 7.3 \\
    \ce{SO2}:\ce{CH3OH} & 1:1 & 10 & 0--1.0 & 4.2$\times10^{-7}$ & 0.56 & 1.3$\times10^{8}$ & -- & -- & -- \\
    \ce{OCN^-} & -- & 15 & 0--8.0 & 6.58 & 0.34 & 5.2 & 6.49 & 0.45 & 6.9 \\
    \ce{HCOO^-} & -- & 14 & 0-0.08 & 0.04 & 0.001 & 2.4 & 0.037 & 0.002 & 4.3 \\
    \ce{HCOOH}:\ce{H2O}:\ce{CH3OH} & 6:68:26 & 15 & 0--1.0 & 0.64 & 0.03 & 4.0 & 0.61 & 0.05 & 7.9 \\
    \ce{CH3CHO}:\ce{H2O} & 1:20 & 15 & 0--2.5 & 1.65 & 0.04 & 2.2 & 1.66 & 0.13 & 7.6 \\
    \ce{C2H5OH}:\ce{H2O} & 1:20 & 15 & 0--5.0 & 1.35 & 0.10 & 7.7 & 1.98 & 0.14 & 7.5 \\
    \ce{CH3OCH3}:\ce{H2O} & 1:20 & 15 & 0--0.8 & 0.40 & 0.05 & 12.9 & 0.38 & 0.07 & 19.3 \\
    \ce{CH3OCHO}:\ce{CO}:\ce{H2CO}:\ce{CH3OH} & 1:20:20:20 & 15 & 0--2.0 & 0.94 & 0.06 & 6.0 & 0.98 & 0.09 & 9.2 \\
    \ce{CH3COOH}:\ce{H2O} & 1:20 & 10 & 0--0.2 & 0.08 & 0.02 & 30.7 & -- & -- & -- \\
    \ce{CH3COCH3}:\ce{H2O} & 1:20 & 15 & 0--2.0 & 0.99 & 0.03 & 3.4 & 1.13 & 0.08 & 6.8 \\
    \ce{CH3CN}:\ce{H2O}:\ce{CO2} & 1:5:2 & 15 & 0--1.5 & 0.70 & 0.08 & 10.9 & -- & -- & -- \\
    \ce{H2CO} & -- & 10 & 0.8$^c$ & 0.8$^c$ & -- & -- & 0.8$^c$ & -- & -- \\
    \bottomrule
    \end{tabular}
    \begin{minipage}{0.98\textwidth}
        $^a$ The scaling factors refer to $a_i$ in Eq.~\ref{eq:linear_fitting}, which serve as the variables in the least-squares and MCMC fittings. They are proportional to the ice column densities (Eq.~\ref{eq:N_ice}).\\
        $^b$ The fit range of scaling factors were determined by manually comparing the lab spectra to the observations, as shown in Figs.~\ref{fig:obs_vs_lab_mix}, \ref{fig:obs_vs_lab_T_1}, and \ref{fig:obs_vs_lab_T_2}.\\
        $^c$ The scaling factor of the lab spectrum of \ce{H2CO} was determined by fitting to an \ce{H2CO} band at 6.67 $\mu$m, outside of the 6.8--8.8 $\mu$m range. The \ce{H2CO} band at 8.02 $\mu$m was scaled with the same factor of 0.8, which was fixed in both the least-squares and MCMC fittings.
    \end{minipage}
    \label{tab:JWST_B1c_fit_statistics}
\end{table*}

\clearpage
\twocolumn
\section{The \ce{CH3OH} band at 3.9 $\mu$m}\label{appendix:NIRSpec_CH3OH}
To better quantify the ice column density of \ce{CH3OH}, we checked the 3.9 $\mu$m \ce{CH3OH} band falling in the wavelength coverage of NIRSpec. The NIRSpec integral field unit (IFU) observations of B1-c were taken as part of GTO 1290 in a 4-point dither pattern. Details on the reduction of NIRSpec data and the spectra extraction will are not the focus of this work and will be introduced in future publication.

\begin{figure}[!htbp]
     \centering
     \begin{subfigure}[b]{0.48\textwidth}
         \centering
         \includegraphics[width=1.05\textwidth]
         {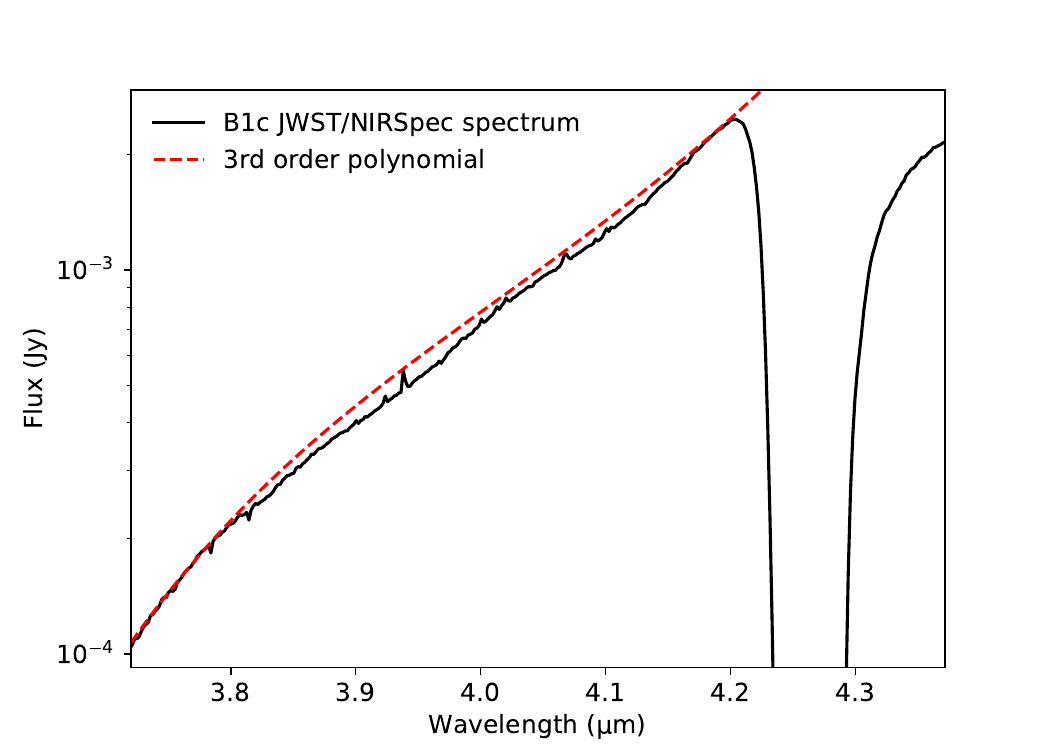}
     \end{subfigure}
     \hfill
     \begin{subfigure}[b]{0.48\textwidth}
         \centering
         \includegraphics[width=\textwidth]{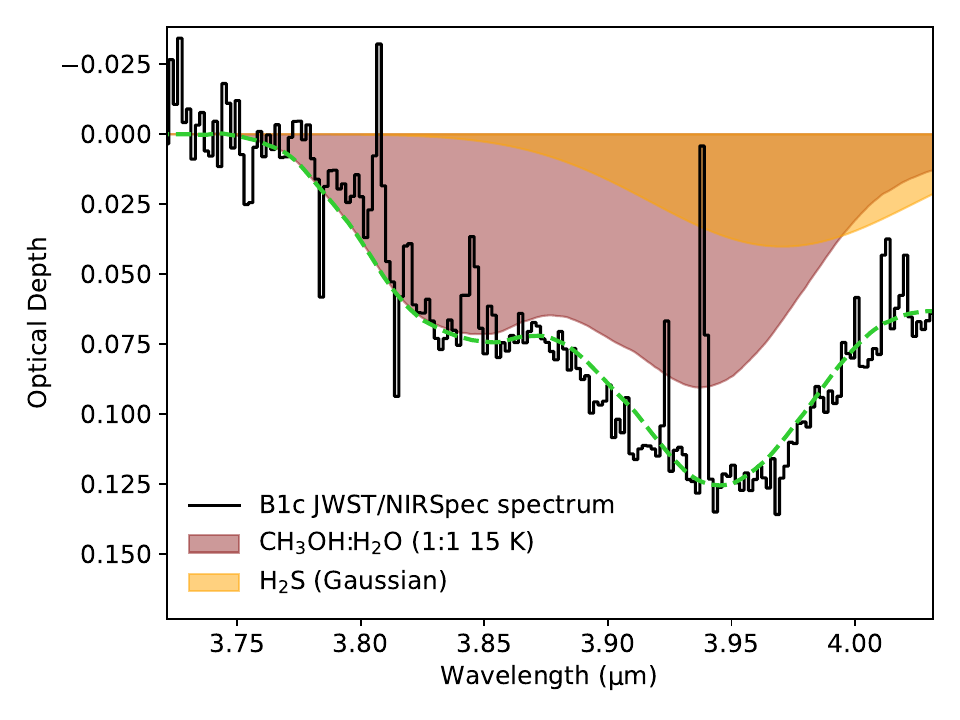}
     \end{subfigure}
     \hfill
     \begin{subfigure}[b]{0.48\textwidth}
         \centering
         \includegraphics[width=\textwidth]{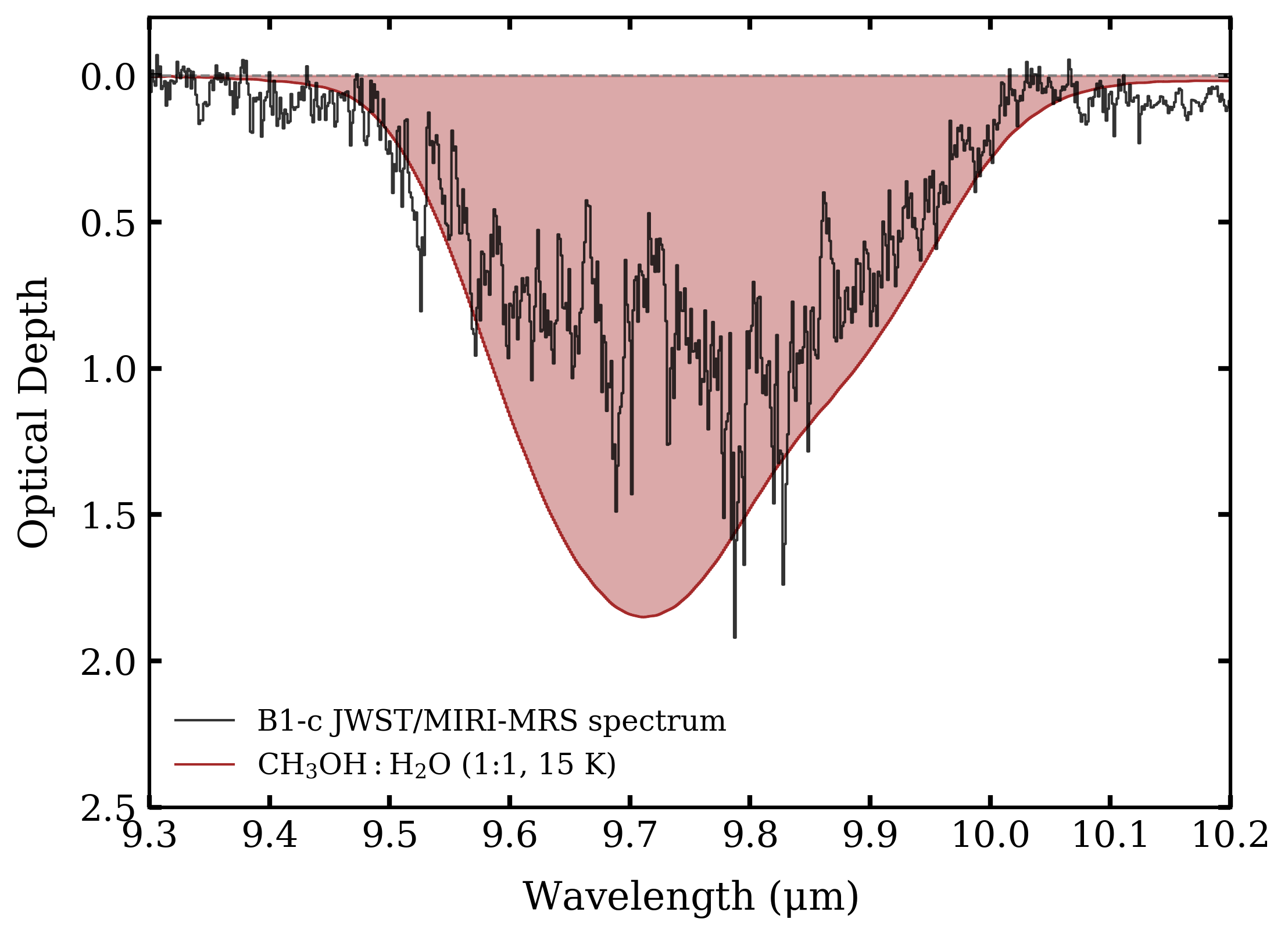}
     \end{subfigure}
    \caption{Top: the NIRSpec IFU spectrum of B1-c (solid black) and the local continuum of the 3.94$\mu$m \ce{CH3OH} band fit by a 3rd-order polynomial (dashed red). Middle: best-fit results of the observed absorption complex between 3.75 and 4.03 $\mu$m (black, the same as in the top panel but in optical depth scale). The shaded regions in red and yellow correspond to the \ce{CH3OH} and \ce{H2S} components, respectively. The overall best-fit spectrum is shown in dashed green. Bottom: the observed 9.74 $\mu$m band in the MIRI-MRS spectrum (black), overlaid with the \ce{CH3OH}:\ce{H2O} lab spectrum scaling with the same factor as in the middle panel (red).}
    \label{fig:CH3OH_NIRSpec}
\end{figure}

As shown in the top panel of Fig.~\ref{fig:CH3OH_NIRSpec}, the 3.9 $\mu$m band is next to the strong \ce{CO2} band at 4.27 $\mu$m, of which the blue wing tends to be warped by grain scattering and may need grain size correction \citep{Dartois2022, Dartois2024}. We tried fitting a 3rd-order polynomial to the observed NIRSpec spectrum in 3.72--3.8 $\mu$m and 4.17--4.2 $\mu$m. This local continuum is only considered as a preliminary analysis without a proper grain shape correction, and it is deviating from the observations longward of 4.2 $\mu$m. The observed spectrum between 3.72 and 4.2 $\mu$m was converted to optical depth scale by taking a logarithm with the local continuum.

The absorption features in this range were attributed to \ce{CH3OH}, \ce{H2S}, and HDO \citep[][the HDO feature will be discussed in a future paper]{Slavicinska2024_HDO}. The \ce{CH3OH} band was fit by a lab spectrum of \ce{CH3OH}:\ce{H2O} (1:1) ice mixture, recently measured by \cite{Slavicinska2024_HDO}. The contribution of \ce{H2S} is represented by a Gaussian function. We performed a least-squares fitting on the scaling factor of the \ce{CH3OH}:\ce{H2O} spectrum and the coefficients of the Gaussian function (best-fit results displayed in the middle panel of Fig.~\ref{fig:CH3OH_NIRSpec}). The ice column density of \ce{CH3OH} was then calculated using Eq.~\ref{eq:N_ice}, where a band strength of 1.56$\times10^{17}$ cm$^{-2}$ \citep{Luna2018} was taken. The best-fit scaling factor of the 3.9 $\mu$m \ce{CH3OH} band in the observed NIRSpec spectrum yields an ice column density of 4.0$\times10^{18}$ cm$^{-2}$ for \ce{CH3OH}. We scaled the \ce{CH3OH}:\ce{H2O} lab spectrum using the same factor as in the middle panel and compared it with the observed MIRI-MRS spectrum at 9.74 $\mu$m (the bottom panel of Fig.~\ref{fig:CH3OH_NIRSpec}). The lab spectrum overestimates the observed 9.74 $\mu$m band, but within a reasonable extent considering the extinction by the silicate band at 9.8 $\mu$m. 

As mentioned in Sect.~\ref{sect:result_CH3OH_H2O}, we finally adopted a \ce{CH3OH} ice column density of 3.0$\times10^{18}$ cm$^{-2}$, corresponding to 1.5 times the best-fit Gaussian to the observed 9.74 $\mu$m band (top panel in Fig.~\ref{fig:CH3OH_H2O_fit}). This result is based on the Gaussian fitting to the 9.74 $\mu$m band, with silicate extinction taken into account. The difference between the ice column densities derived from fitting the 3.9 $\mu$m and the 9.74 $\mu$m bands is small ($\sim$30\%), suggesting that both the estimations are reasonable.
However, considering that the analysis of the NIRSpec spectrum was only preliminary, the result of 4.0$\times10^{18}$ cm$^{-2}$ was not adopted, instead it was used as a reference to constrain the uncertainty of our fitting results to the 9.74 $\mu$m band ($\sigma$=1.0$\times10^{18}$ cm$^{-2}$).

\clearpage
\onecolumn
\section{Additional figures for the JWST fitting results}\label{appendix:JWST_fitting}
\begin{figure*}[!h]
    \centering
    \includegraphics[width=0.9\textwidth]{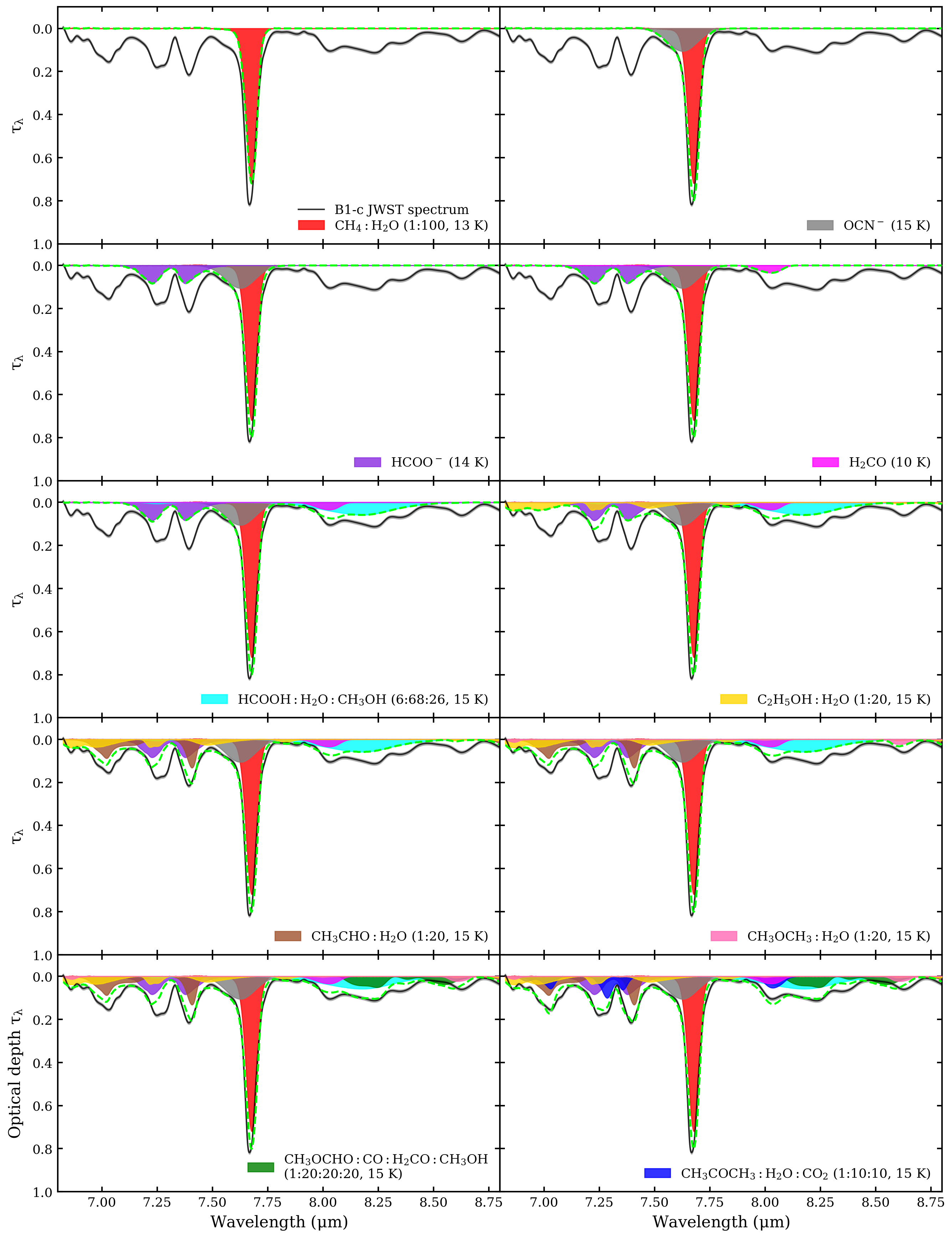}
    \caption{Same as Fig.~\ref{fig:JWST_fit_B1c_COMs} but displays the best fits species by species.}
    \label{fig:JWST_fit_B1c_steps}
\end{figure*}

\begin{figure*}[!h]
    \centering
    \includegraphics[width=1\textwidth]{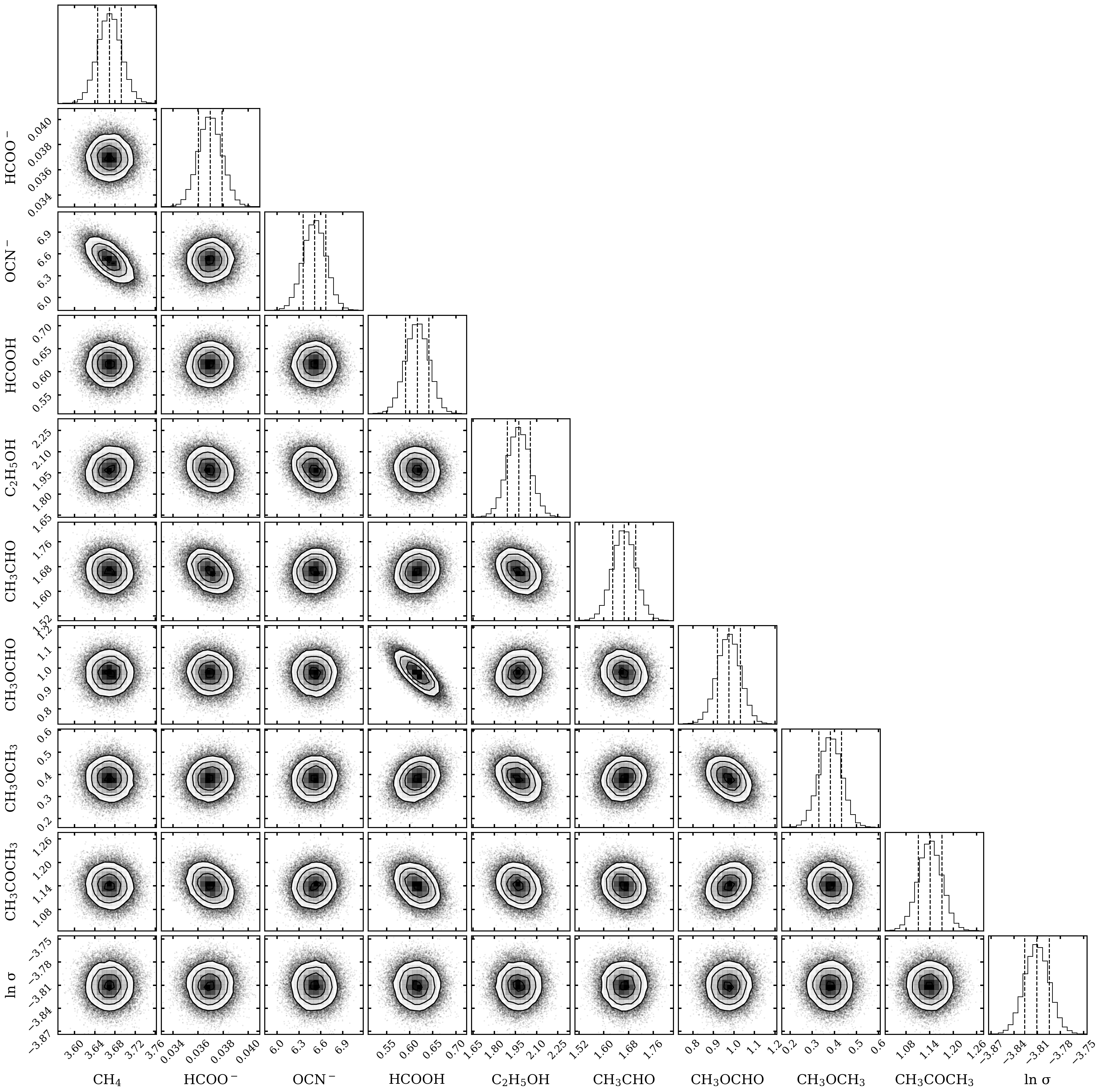}
    \caption{Corner plot of the MCMC fitting. Nine out of 12 candidate species are considered. The dashed lines in the histograms indicate the 16\%, 50\%, and 84\% quantiles from left to right.}
    \label{fig:B1c_mcmc_corner_plot}
\end{figure*}

Here we provide supplementary figures for the overall fittings of the COM fingerprint range (6.8--8.8 $\mu$m) in the JWST/MIRI-MRS spectrum of B1-c. Figure~\ref{fig:JWST_fit_B1c_steps} shows the best-fit lab spectra as in Fig.~\ref{fig:JWST_fit_B1c_COMs} but in a step-by-step form. Figure~\ref{fig:B1c_mcmc_corner_plot} shows the corner plot of the MCMC fitting introduced in Sect.~\ref{sect:results_fit_stats}.

\clearpage
\section{Additional table}\label{appendix:table}
Table~\ref{table:transition} lists the rotational transitions of the detected species reported in Table~\ref{tab:ALMA_results} covered in our ALMA Band 7 data (Sect.~\ref{sect:obs_ALMA}). Only those transitions that are used in the LTE fittings are listed.

\begin{longtable}[h]{lrclclcc}
\caption{Rotational transitions of the detected gas-phase molecules in the ALMA spectra.\label{table:transition}}\\
\toprule 
Species & \multicolumn{3}{c}{Transition} & Frequency (MHz) & $E_\mathrm{up}$ (K) & $A_\mathrm{ij}$ (s$^{-1}$) & Note \\
(Database) & & & & & & $a(b)$ = $a\times10^b$ & \\
\hline
\endfirsthead
\caption{continued.}\\
\toprule Species & \multicolumn{3}{c}{Transition} & Frequency (MHz) & $E_\mathrm{up}$ (K) & $A_\mathrm{ij}$ (s$^{-1}$) & Note\\
(Database) & & & & & & $a(b)$ = $a\times10^b$\\
\hline
\endhead
\bottomrule 
\endfoot
\bottomrule 
\endlastfoot

\textbf{\ce{CH3OH}}, vt=0--2 & 9 1 8 1 & -- & 8 2 6 2 & 333864.722 & 125.52 & 8.04(-7) & $\star$ \\
(CDMS) & 21 5 17 4 & -- & 22 4 19 4 & 334326.177 & 964.39 & 1.99(-5) & $\star$ \\
 & 3 0 3 4 & -- & 2 1 2 4 & 334426.571 & 314.47 & 5.55(-5) & $\dagger$\\
 & 22 3 19 4 & -- & 22 2 20 4 & 334626.849 & 1001.32 & 3.55(-5) & $\star$ \\
 & 25 3 23 5 & -- & 24 2 23 5 & 334679.524 & 1073.85 & 1.46(-4) & $\dagger$ \\
 & 2 2 1 0 & -- & 3 1 2 0 & 335133.570 & 44.67 & 2.69(-5) & $\dagger$ \\
 & 28 8 21 4 & -- & 28 7 21 4 & 335205.794 & 1534.70 & 2.14(-4) & $\star$ \\
 & 7 1 7 0 & -- & 6 1 6 0 & 335582.017 & 78.97 & 1.63(-4) & $\dagger$ \\
 & 14 7 7 0 & -- & 15 6 9 0 & 336438.224 & 488.22 & 3.61(-5) & $\dagger$ \\
 & 14 7 8 0 & -- & 15 6 10 0 & 336438.224 & 488.22 & 3.61(-5) & $\dagger$ \\
 & 7 1 7 6 & -- & 6 1 6 6 & 336605.889 & 747.41 & 1.64(-4) & $\dagger$ \\
 & 27 9 18 0 & -- & 26 6 21 3 & 336646.204 & 1293.65 & 2.29(-5) & \\
 & 27 9 19 0 & -- & 26 6 20 3 & 336646.533 & 1293.65 & 2.29(-5) & \\
 & 7 6 1 6 & -- & 6 6 0 6 & 336970.183 & 1022.71 & 4.57(-5) & \\
 & 7 6 2 6 & -- & 6 6 1 6 & 336970.183 & 1022.71 & 4.57(-5) & \\
 & 7 3 4 8 & -- & 6 3 3 8 & 337021.917 & 979.71 & 1.36(-4) & $\dagger$ \\
 & 7 2 6 6 & -- & 6 2 5 6 & 337029.573 & 941.38 & 1.55(-4) & $\dagger$ \\
 & 7 2 5 6 & -- & 6 2 4 6 & 337029.662 & 941.38 & 1.55(-4) & \\
 & 27 9 19 2 & -- & 28 8 21 2 & 345318.303 & 1278.18 & 5.97(-5) & \\
 & 16 1 15 0 & -- & 15 2 14 0 & 345903.916 & 332.65 & 1.04(-4) & $\dagger$ \\
 & 18 3 15 2 & -- & 17 4 14 2 & 345919.260 & 459.43 & 7.29(-5) & $\dagger$ \\
 & 19 3 16 4 & -- & 19 2 17 4 & 347445.285 & 856.25 & 4.04(-5) & \\
\hline
\textbf{\ce{^{13}CH3OH}}, vt=0--1 & 12 1 11 -0 & -- & 12 0 12 +0 & 335560.207 & 192.66 & 4.04(-4) & $\dagger$ \\
(CDMS) & 14 1 13 -0 & -- & 14 0 14 +0 & 347188.283 & 254.25 & 4.36(-4) & $\dagger$ \\
\hline
\textbf{\ce{CH3^{18}OH}}, v=0--2 & 4 0 4 1 & -- & 3 1 3 1 & 336100.324 & 35.09 & 7.42(-5) & $\ast$ \\
(CDMS) & 11 6 6 2 & -- & 12 5 8 2 & 345858.184 & 326.94 & 3.54(-5) & \\
\hline
\textbf{\ce{CH2DOH}} & 19 8 11 2 & -- & 20 7 14 2 & 333835.968 & 672.13 & 1.06(-5) & \\
(JPL) & 19 8 12 2 & -- & 20 7 13 2 & 333835.968 & 672.13 & 1.06(-5) & \\
 & 16 2 15 0 & -- & 16 1 16 0 & 333966.537 & 306.68 & 1.59(-4) & $\dagger$ \\
 & 20 3 17 1 & -- & 20 2 18 1 & 334615.482 & 494.74 & 1.40(-5) & \\
 & 16 2 15 2 & -- & 16 1 15 1 & 334683.954 & 326.58 & 1.09(-4) & $\dagger$ \\
 & 13 0 13 1 & -- & 12 1 11 2 & 334900.931 & 207.26 & 2.69(-5) & \\
 & 23 0 23 0 & -- & 22 1 21 2 & 335102.203 & 583.84 & 3.71(-5) & \\
 & 18 5 13 0 & -- & 18 4 15 2 & 335103.631 & 464.26 & 2.80(-5) & \\
 & 18 5 14 0 & -- & 18 4 14 2 & 335116.209 & 464.26 & 2.80(-5) & \\
 & 19 2 17 0 & -- & 18 3 16 1 & 335578.757 & 427.23 & 1.50(-5) & \\
 & 20 3 18 2 & -- & 19 4 15 2 & 336131.672 & 504.96 & 2.04(-5) & \\
 & 21 8 13 2 & -- & 22 7 15 1 & 336973.378 & 759.76 & 2.43(-5) & \\
 & 21 8 14 2 & -- & 22 7 16 1 & 336973.383 & 759.76 & 2.43(-5) & \\
 & 16 2 14 0 & -- & 15 3 13 0 & 345398.904 & 310.21 & 5.83(-5) & \\
 & 23 4 19 1 & -- & 23 3 21 2 & 345701.510 & 662.81 & 1.34(-4) & $\star$ \\
 & 3 2 1 1 & -- & 2 1 2 1 & 345718.718 & 39.44 & 4.23(-5) & \\
 & 19 1 19 1 & -- & 18 2 17 2 & 345820.793 & 418.03 & 2.94(-5) & $\star$ \\
 & 22 4 19 1 & -- & 22 3 19 2 & 345850.485 & 613.62 & 1.29(-4) & $\star$ \\
 & 8 2 6 2 & -- & 7 3 4 0 & 346031.073 & 112.62 & 1.76(-6) & \\
 & 20 4 16 1 & -- & 19 5 15 0 & 346042.562 & 521.63 & 1.32(-5) & \\
 & 20 4 17 1 & -- & 19 5 14 0 & 346062.966 & 521.63 & 1.32(-5) & \\
 & 21 4 18 1 & -- & 21 3 18 2 & 346419.064 & 566.56 & 1.30(-4) & $\star$ \\
 & 7 5 2 2 & -- & 8 4 5 2 & 346434.692 & 175.80 & 7.74(-6) & \\
 & 7 5 3 2 & -- & 8 4 4 2 & 346434.725 & 175.80 & 7.74(-6) & \\
 & 20 4 16 1 & -- & 20 3 18 2 & 347222.992 & 521.63 & 1.32(-4) & \\
 & 19 4 16 1 & -- & 19 3 16 2 & 347371.167 & 478.84 & 1.30(-4) & \\
 & 20 7 14 1 & -- & 21 6 15 0 & 347467.621 & 651.60 & 2.02(-5) & \\
 & 20 7 13 1 & -- & 21 6 16 0 & 347468.666 & 651.60 & 2.02(-5) & \\
\hline
\textbf{\ce{CH3CHO}} & 18 1 18 1 & -- & 17 1 17 1 & 333941.356 & 155.22 & 1.28(-3) & B1-c$\dagger$ \\
(JPL) & 18 1 18 0 & -- & 17 1 17 0 & 333959.754 & 155.15 & 1.28(-3) & B1-c$\dagger$ \\
 & 18 1 18 3 & -- & 17 1 17 3 & 334341.281 & 361.72 & 1.30(-3) & $\star$\\
 & 17 2 15 3 & -- & 16 2 14 3 & 334637.404 & 357.95 & 1.28(-3) & \\
 & 17 4 14 7 & -- & 16 4 13 7 & 334665.820 & 548.85 & 1.18(-3) & \\
 & 25 4 22 1 & -- & 25 3 23 1 & 334756.255 & 337.33 & 1.30(-4) & \\
 & 17 2 15 2 & -- & 16 2 14 2 & 334904.591 & 152.63 & 1.28(-3) & B1-c$\dagger$ \\
 & 17 2 15 0 & -- & 16 2 14 0 & 334931.415 & 152.61 & 1.28(-3) & B1-c$\dagger$ \\
 & 18 1 18 4 & -- & 17 1 17 4 & 334980.853 & 359.89 & 1.28(-3) & $\star$\\
 & 18 0 18 2 & -- & 17 0 17 2 & 335318.109 & 154.93 & 1.30(-3) & B1-c$\dagger$ \\
 & 18 0 18 0 & -- & 17 0 17 0 & 335358.722 & 154.85 & 1.29(-3) & B1-c$\dagger$ \\
 & 18 0 18 3 & -- & 17 0 17 3 & 335382.461 & 361.48 & 1.31(-3) & $\star$\\
 & 17 6 12 7 & -- & 16 6 11 7 & 335560.509 & 591.60 & 1.10(-3) & \\
 & 17 2 15 6 & -- & 16 2 14 6 & 336127.216 & 529.83 & 1.25(-3) & \\
 & 18 1 17 8 & -- & 17 1 16 8 & 345992.130 & 545.09 & 1.38(-3) & \\
 & 18 2 17 4 & -- & 17 2 16 4 & 346065.350 & 371.35 & 1.40(-3) & \\
 & 18 7 11 5 & -- & 17 7 10 5 & 346451.075 & 474.02 & 1.22(-3) & $\star$ \\
 & 18 6 12 5 & -- & 17 6 11 5 & 346591.092 & 445.29 & 1.28(-3) & \\
 & 18 7 12 0 & -- & 17 7 11 0 & 346957.556 & 268.61 & 1.22(-3) & B1-c$\dagger$ \\
 & 18 7 11 0 & -- & 17 7 10 0 & 346957.558 & 268.61 & 1.22(-3) & B1-c$\dagger$ \\
 & 18 7 12 1 & -- & 17 7 11 1 & 346995.532 & 268.57 & 1.22(-3) & \\
 & 18 7 11 6 & -- & 17 7 10 6 & 346999.913 & 646.35 & 1.22(-3) & $\star$ \\
 & 18 7 12 6 & -- & 17 7 11 6 & 346999.941 & 646.35 & 1.22(-3) & $\star$ \\
 & 18 6 13 1 & -- & 17 6 12 1 & 347132.686 & 239.32 & 1.28(-3) & \\
 & 18 4 14 8 & -- & 17 4 13 8 & 347155.125 & 573.93 & 1.37(-3) & \\
 & 9 4 6 4 & -- & 9 3 7 4 & 347155.775 & 282.78 & 1.20(-4) & \\
 & 19 0 19 0 & -- & 18 1 18 0 & 347169.520 & 171.81 & 2.09(-4) & \\
 & 18 4 14 5 & -- & 17 4 13 5 & 347182.413 & 400.38 & 1.37(-3) & $\star$ \\
 & 18 5 13 5 & -- & 17 5 12 5 & 347216.798 & 420.44 & 1.33(-3) & \\
 & 19 0 19 2 & -- & 18 1 18 1 & 347240.396 & 171.89 & 2.08(-4) & \\
 & 18 5 14 4 & -- & 17 5 13 4 & 347251.822 & 419.67 & 1.33(-3) & \\
 & 18 5 14 0 & -- & 17 5 13 0 & 347288.264 & 214.70 & 1.33(-3) & \\
 & 18 5 13 0 & -- & 17 5 12 0 & 347294.873 & 214.70 & 1.33(-3) & \\
 & 18 5 13 2 & -- & 17 5 12 2 & 347345.710 & 214.64 & 1.33(-3) & \\
 & 18 5 14 1 & -- & 17 5 13 1 & 347349.278 & 214.61 & 1.33(-3) & \\
 & 18 7 12 3 & -- & 17 7 11 3 & 347459.353 & 473.15 & 1.22(-3) & $\star$ \\
 & 18 7 11 3 & -- & 17 7 10 3 & 347459.355 & 473.15 & 1.22(-3) & $\star$ \\
 & 18 3 16 0 & -- & 17 3 15 0 & 347519.185 & 178.75 & 1.41(-3) & B1-c$\dagger$ \\
 & 18 3 16 1 & -- & 17 3 15 1 & 347563.334 & 178.71 & 1.41(-3) & B1-c$\dagger$ \\
\hline
\textbf{\ce{C2H5OH}}, v=0 & 25 7 19 2 & -- & 25 6 20 2 & 334110.890 & 334.91 & 2.39(-4) & \\
(CDMS) & 44 7 38 2 & -- & 44 6 39 2 & 334543.992 & 897.54 & 2.64(-4) & \\
 & 24 7 17 2 & -- & 24 6 18 2 & 334602.657 & 313.84 & 2.37(-4) & \\
 & 12 5 7 0 & -- & 11 4 7 1 & 334742.442 & 152.19 & 1.27(-4) & \\
 & 20 1 19 2 & -- & 19 2 18 2 & 334769.072 & 178.68 & 2.75(-4) & $\star$ \\
 & 12 5 8 0 & -- & 11 4 8 1 & 334811.021 & 152.19 & 1.27(-4) & \\
 & 24 7 18 2 & -- & 24 6 19 2 & 334911.577 & 313.83 & 2.38(-4) & \\
 & 23 7 16 2 & -- & 23 6 17 2 & 335440.539 & 293.61 & 2.36(-4) & \\
 & 23 7 17 2 & -- & 23 6 18 2 & 335630.619 & 293.61 & 2.36(-4) & \\
 & 13 3 11 2 & -- & 12 2 10 2 & 335949.651 & 87.87 & 1.43(-4) & \\
 & 19 3 16 1 & -- & 18 3 15 1 & 336572.411 & 232.34 & 3.14(-4) & \\
 & 19 2 18 2 & -- & 18 1 17 2 & 336626.401 & 162.61 & 2.71(-4) & \\
 & 21 1 21 1 & -- & 20 1 20 1 & 345295.355 & 246.22 & 3.67(-4) & \\
 & 21 0 21 0 & -- & 20 0 20 0 & 345333.442 & 241.54 & 3.72(-4) & \\
 & 20 13 7 1 & -- & 19 13 6 1 & 345966.686 & 443.28 & 2.10(-4) & \\
 & 20 13 8 1 & -- & 19 13 7 1 & 345966.686 & 443.28 & 2.10(-4) & \\
 & 20 14 6 1 & -- & 19 14 5 1 & 345970.743 & 476.27 & 1.85(-4) & \\
 & 20 14 7 1 & -- & 19 14 6 1 & 345970.743 & 476.27 & 1.85(-4) & \\
 & 20 12 8 1 & -- & 19 12 7 1 & 345981.237 & 412.70 & 2.33(-4) & \\
 & 20 12 9 1 & -- & 19 12 8 1 & 345981.237 & 412.70 & 2.33(-4) & \\
 & 20 11 9 1 & -- & 19 11 8 1 & 346017.212 & 384.56 & 2.55(-4) & $\star$ \\
 & 20 11 10 1 & -- & 19 11 9 1 & 346017.212 & 384.56 & 2.55(-4) & $\star$ \\
 & 20 10 10 1 & -- & 19 10 9 1 & 346085.563 & 358.85 & 2.74(-4) & $\star$ \\
 & 20 10 11 1 & -- & 19 10 10 1 & 346085.563 & 358.85 & 2.74(-4) & $\star$ \\
 & 20 10 10 0 & -- & 19 10 9 0 & 346424.583 & 353.45 & 2.91(-4) & $\star$ \\
 & 20 10 11 0 & -- & 19 10 10 0 & 346424.583 & 353.45 & 2.91(-4) & $\star$ \\
 & 20 9 12 0 & -- & 19 9 11 0 & 346505.347 & 330.25 & 3.09(-4) & \\
 & 20 9 11 0 & -- & 19 9 10 0 & 346505.347 & 330.25 & 3.09(-4) & \\
 & 20 7 14 1 & -- & 19 7 13 1 & 346565.082 & 296.37 & 3.22(-4) & \\
 & 20 7 13 1 & -- & 19 7 12 1 & 346565.398 & 296.37 & 3.22(-4) & \\
 & 20 17 3 0 & -- & 19 17 2 0 & 346610.251& 584.43 & 1.12(-4) & \\
 & 20 17 4 0 & -- & 19 17 3 0 & 346610.251 & 584.43 & 1.12(-4) & \\
 & 20 8 13 0 & -- & 19 8 12 0 & 346620.325 & 309.51 & 3.26(-4) & $\star$ \\
 & 20 8 12 0 & -- & 19 8 11 0 & 346620.333 & 309.51 & 3.26(-4) & $\star$ \\
 & 20 6 14 1 & -- & 19 6 13 1 & 346938.874 & 280.48 & 3.34(-4) & \\
 & 21 0 21 2 & -- & 20 1 20 2 & 346962.603 & 185.84 & 4.45(-4) & $\star$ \\
 & 20 6 15 0 & -- & 19 6 14 0 & 347147.209 & 275.46 & 3.55(-4) & \\
 & 20 6 14 0 & -- & 19 6 13 0 & 347157.997 & 275.46 & 3.55(-4) & \\
 & 14 3 12 2 & -- & 13 2 11 2 & 347351.089 & 99.66 & 1.56(-4) & \\
 & 21 1 21 2 & -- & 20 0 20 2 & 347445.568 & 185.85 & 4.47(-4) & \\
 & 20 5 16 1 & -- & 19 5 15 1 & 347473.561 & 267.09 & 3.45(-4) & \\
\hline
\textbf{\ce{CH3OCH3}}, v=0 & 37 7 30 0 & -- & 37 6 31 0 & 347340.892 & 710.39 & 2.38(-4) & $\ast$ \\
(CDMS) & 37 7 30 1 & -- & 37 6 31 1 & 347342.977 & 710.39 & 2.38(-4) & $\ast$ \\
 & 37 7 30 5 & -- & 37 6 31 5 & 347345.059 & 710.39 & 2.38(-4) & $\ast$ \\
 & 37 7 30 3 & -- & 37 6 31 3 & 347345.066 & 710.39 & 2.38(-4) & $\ast$ \\
\hline
\textbf{\ce{CH3OCHO}} & 15 6 10 1 & -- & 14 5 9 2 & 333972.824 & 94.90 & 2.42(-5) & \\
(JPL) & 27 11 16 2 & -- & 26 11 15 2 & 334017.031 & 303.76 & 4.74(-4) & $\star$ \\
 & 27 11 17 0 & -- & 26 11 16 0 & 334031.781 & 303.76 & 4.74(-4) & $\dagger$ \\
 & 27 11 16 0 & -- & 26 11 15 0 & 334031.781 & 303.76 & 4.74(-4) & \\
 & 27 11 17 1 & -- & 26 11 16 1 & 334044.362 & 303.76 & 4.74(-4) & \\
 & 27 8 20 3 & -- & 26 8 19 3 & 334074.073 & 453.00 & 5.18(-4) & \\
 & 15 6 10 0 & -- & 14 5 9 0 & 334109.114 & 94.90 & 3.79(-5) & \\
 & 27 6 22 4 & -- & 26 6 21 4 & 334138.972 & 435.17 & 5.39(-4) & \\
 & 29 5 24 2 & -- & 28 6 23 1 & 334179.438 & 282.10 & 2.99(-5) & \\
 & 26 5 21 5 & -- & 25 5 20 5 & 334228.598 & 416.22 & 5.48(-4) & $\star$ \\
 & 29 5 24 0 & -- & 28 6 23 0 & 334235.866 & 282.10 & 2.99(-5) & $\star$ \\
 & 27 8 19 5 & -- & 26 8 18 5 & 334660.939 & 453.20 & 5.11(-4) & \\
 & 27 8 19 3 & -- & 26 8 18 3 & 334700.891 & 453.07 & 5.21(-4) & $\star$ \\
 & 27 10 17 2 & -- & 26 10 16 2 & 334850.947 & 290.09 & 4.93(-4) & \\
 & 27 10 18 0 & -- & 26 10 17 0 & 334867.014 & 290.09 & 4.94(-4) & $\star$ \\
 & 27 10 17 0 & -- & 26 10 16 0 & 334872.791 & 290.09 & 4.94(-4) & $\star$ \\
 & 27 10 18 1 & -- & 26 10 17 1 & 334877.566 & 290.08 & 4.93(-4) & $\star$ \\
 & 27 7 21 3 & -- & 26 7 20 3 & 335015.896 & 443.53 & 5.33(-4) & $\star$ \\
 & 27 8 20 4 & -- & 26 8 19 4 & 335391.536 & 453.05 & 5.14(-4) & \\
 & 27 9 19 0 & -- & 26 9 18 0 & 336028.165 & 277.85 & 5.14(-4) & \\
 & 27 9 19 1 & -- & 26 9 18 1 & 336032.357 & 277.85 & 4.90(-4) & \\
 & 27 9 18 2 & -- & 26 9 17 2 & 336086.182 & 277.86 & 4.90(-4) & \\
 & 27 9 18 0 & -- & 26 9 17 0 & 336111.324 & 277.86 & 5.15(-4) & \\
 & 9 9 0 0 & -- & 8 8 1 0 & 345718.662 & 80.32 & 9.86(-5) & \\
 & 9 9 1 0 & -- & 8 8 0 0 & 345718.662 & 80.32 & 9.86(-5) & \\
 & 28 6 23 4 & -- & 27 6 22 4 & 345828.557 & 451.76 & 5.99(-4) & \\
 & 28 12 16 2 & -- & 27 12 15 2 & 345974.664 & 335.44 & 5.16(-4) & \\
 & 28 12 17 0 & -- & 27 12 16 0 & 345985.381 & 335.44 & 5.16(-4) & \\
 & 28 12 16 0 & -- & 27 12 15 0 & 345985.381 & 335.44 & 5.16(-4) & \\
 & 28 12 17 1 & -- & 27 12 16 1 & 346001.616 & 335.43 & 5.16(-4) & $\star$ \\
 & 28 9 20 4 & -- & 27 9 19 4 & 346564.463 & 480.56 & 5.69(-4) & \\
 & 27 6 21 3 & -- & 26 6 20 3 & 346580.244 & 437.64 & 6.04(-4) & \\
 & 27 5 22 2 & -- & 26 5 21 2 & 347478.251 & 247.25 & 6.14(-4) & $\dagger$ \\
 & 27 5 22 0 & -- & 26 5 21 0 & 347493.965 & 247.26 & 6.14(-4) & $\dagger$ \\
\hline
\textbf{\ce{CH2OHCHO}}, v=0 & 37 8 29 0 & -- & 36 9 28 0 & 334382.061 & 439.41 & 1.83(-4) & $\ast$ \\
(CDMS) & 35 14 21 0 & -- & 35 13 22 0 & 334587.394 & 469.56 & 5.85(-4) & \\
 & 35 14 22 0 & -- & 35 13 23 0 & 334588.479 & 469.56 & 5.85(-4) & \\
 & 50 7 43 0 & -- & 50 6 44 0 & 334859.254 & 750.32 & 4.85(-4) & \\
 & 17 6 11 0 & -- & 16 5 12 0 & 335097.924 & 107.27 & 3.89(-4) & \\
 & 34 14 20 0 & -- & 34 13 21 0 & 335441.062 & 450.00 & 5.81(-4) & \\
 & 34 14 21 0 & -- & 34 13 22 0 & 335441.575 & 450.00 & 5.81(-4) & \\
 & 30 5 25 0 & -- & 29 6 24 0 & 335456.548 & 282.11 & 4.31(-4) & \\
 & 47 15 32 0 & -- & 47 14 33 0 & 346572.448 & 765.89 & 7.26(-4) & $\ast$ \\
 & 18 6 12 0 & -- & 17 5 13 0 & 347445.973 & 117.44 & 4.02(-4) & \\
\hline
\textbf{$a$-\ce{(CH2OH)2}} & 32 18 14 1 & -- & 31 18 13 0 & 334043.836 & 418.76 & 6.24(-4) & \\
(CDMS) & 32 18 15 1 & -- & 31 18 14 0 & 334043.836 & 418.76 & 6.24(-4) & \\
 & 32 17 15 1 & -- & 31 17 14 0 & 334058.781 & 401.61 & 6.55(-4) & \\
 & 32 17 16 1 & -- & 31 17 15 0 & 334058.781 & 401.61 & 6.56(-4) & \\
 & 32 19 13 1 & -- & 31 19 12 0 & 334059.121 & 436.89 & 5.91(-4) & \\
 & 32 19 14 1 & -- & 31 19 13 0 & 334059.121 & 436.89 & 5.91(-4) & \\
 & 32 20 12 1 & -- & 31 20 11 0 & 334100.228 & 456.00 & 5.57(-4) & \\
 & 32 20 13 1 & -- & 31 20 12 0 & 334100.228 & 456.00 & 5.57(-4) & \\
 & 32 16 16 1 & -- & 31 16 15 0 & 334109.885 & 385.45 & 6.85(-4) & \\
 & 32 16 17 1 & -- & 31 16 16 0 & 334109.885 & 385.45 & 6.85(-4) & \\
 & 36 2 35 0 & -- & 35 2 34 1 & 334141.377 & 310.88 & 9.15(-4) & $\dagger$ \\
 & 36 1 35 0 & -- & 35 1 34 1 & 334141.397 & 310.88 & 9.15(-4) & $\dagger$ \\
 & 32 21 11 1 & -- & 31 21 10 0 & 334163.829 & 476.08 & 5.20(-4) & \\
 & 32 21 12 1 & -- & 31 21 11 0 & 334163.829 & 476.08 & 5.21(-4) & \\
 & 33 10 24 0 & -- & 32 10 23 1 & 334193.102 & 325.62 & 8.29(-4) & \\
 & 32 15 17 1 & -- & 31 15 16 0 & 334205.280 & 370.28 & 7.13(-4) & $\star$ \\
 & 32 15 18 1 & -- & 31 15 17 0 & 334205.280 & 370.28 & 7.14(-4) & $\star$ \\
 & 33 10 23 0 & -- & 32 10 22 1 & 334211.959 & 325.62 & 8.30(-4) & \\
 & 32 22 10 1 & -- & 31 22 9 0 & 334247.377 & 497.14 & 4.82(-4) & \\
 & 32 22 11 1 & -- & 31 22 10 0 & 334247.377 & 497.14 & 4.82(-4) & \\
 & 32 23 9 1 & -- & 31 23 8 0 & 334348.904 & 519.17 & 4.43(-4) & \\
 & 32 23 10 1 & -- & 31 23 9 0 & 334348.904 & 519.17 & 4.43(-4) & \\
 & 32 14 18 1 & -- & 31 14 17 0 & 334356.392 & 356.11 & 7.40(-4) & $\star$ \\
 & 32 14 19 1 & -- & 31 14 18 0 & 334356.392 & 356.11 & 7.40(-4) & $\star$ \\
 & 32 24 8 1 & -- & 31 24 7 0 & 334466.874 & 542.17 & 4.01(-4) & \\
 & 32 24 9 1 & -- & 31 24 8 0 & 334466.874 & 542.17 & 4.01(-4) & \\
 & 32 6 26 0 & -- & 31 6 25 1 & 334555.253 & 282.34 & 8.95(-4) & \\
 & 32 13 20 1 & -- & 31 13 19 0 & 334579.759 & 342.94 & 7.66(-4) & \\
 & 32 13 19 1 & -- & 31 13 18 0 & 334579.760 & 342.94 & 7.66(-4) & \\
 & 36 0 36 0 & -- & 35 1 35 0 & 334804.512 & 300.16 & 1.85(-4) & \\
 & 36 1 36 0 & -- & 35 0 35 0 & 334804.513 & 300.16 & 1.85(-4) & \\
 & 36 0 36 1 & -- & 35 1 35 1 & 334821.007 & 300.51 & 1.85(-4) & \\
 & 36 1 36 1 & -- & 35 0 35 1 & 334821.008 & 300.51 & 1.85(-4) & \\
 & 32 12 21 1 & -- & 31 12 20 0 & 334900.249 & 330.78 & 7.91(-4) & \\
 & 32 12 20 1 & -- & 31 12 19 0 & 334900.284 & 330.78 & 7.91(-4) & \\
 & 32 6 27 1 & -- & 31 6 26 0 & 335030.116 & 279.36 & 9.34(-4) & \\
 & 33 9 25 0 & -- & 32 9 24 1 & 335179.801 & 316.68 & 8.51(-4) & \\
 & 32 11 22 1 & -- & 31 11 21 0 & 335356.975 & 319.65 & 8.15(-4) & \\
 & 32 11 21 1 & -- & 31 11 20 0 & 335357.726 & 319.65 & 8.15(-4) & \\
 & 33 9 24 0 & -- & 32 9 23 1 & 335396.713 & 316.71 & 8.53(-4) & \\
 & 34 4 31 1 & -- & 33 3 30 1 & 336053.816 & 297.13 & 1.33(-4) & \\
 & 34 4 31 0 & -- & 33 3 30 0 & 336054.127 & 296.80 & 1.12(-4) & \\
 & 34 5 29 1 & -- & 33 6 28 1 & 336608.832 & 311.78 & 3.62(-4) & \\
 & 33 11 23 1 & -- & 32 11 22 0 & 345737.011 & 335.95 & 9.00(-4) & \\
 & 33 11 22 1 & -- & 32 11 21 0 & 345738.443 & 335.95 & 9.01(-4) & \\
 & 34 9 26 0 & -- & 33 9 25 1 & 345812.719 & 333.57 & 9.36(-4) & \\
 & 34 8 27 0 & -- & 33 8 26 1 & 345832.886 & 325.79 & 5.43(-4) & \\
 & 38 1 38 0 & -- & 37 1 37 1 & 345912.524 & 333.62 & 1.02(-3) & \\
 & 38 0 38 0 & -- & 37 0 37 1 & 345912.524 & 333.62 & 1.02(-3) & \\
 & 34 7 28 0 & -- & 33 7 27 1 & 346462.657 & 318.83 & 9.43(-4) & $\star$ \\
 & 33 10 24 1 & -- & 32 10 23 0 & 346468.981 & 325.91 & 9.26(-4) & $\star$ \\
 & 33 10 23 1 & -- & 32 10 22 0 & 346490.976 & 325.91 & 9.26(-4) & \\
 & 34 5 29 0 & -- & 33 5 28 1 & 346630.908 & 311.52 & 1.07(-3) & \\
 & 36 4 33 0 & -- & 35 4 32 1 & 347361.377 & 330.36 & 1.02(-3) & \\
 & 36 3 33 0 & -- & 35 3 32 1 & 347377.705 & 330.36 & 1.01(-3) & $\star$ \\
 & 32 6 26 1 & -- & 31 6 25 0 & 347387.171 & 282.65 & 9.81(-4) & \\
 & 33 9 25 1 & -- & 32 9 24 0 & 347486.902 & 316.98 & 9.51(-4) & \\
 & 33 8 25 1 & -- & 32 8 25 1 & 347596.647 & 309.41 & 1.23(-4) & \\
\hline
\textbf{$g$-\ce{(CH2OH)2}}  & 23 5 19 0 & -- & 22 4 19 1 & 333807.874 & 147.69 & 1.88(-4) & \\
(CDMS) & 31 17 14 1 & -- & 31 16 16 0 & 333830.012 & 382.39 & 1.47(-4) & \\
 & 31 17 15 1 & -- & 31 16 15 0 & 333830.012 & 382.39 & 1.47(-4) & \\
 & 30 17 13 1 & -- & 30 16 15 0 & 333926.319 & 367.26 & 1.43(-4) & \\
 & 30 17 14 1 & -- & 30 16 14 0 & 333926.319 & 367.26 & 1.43(-4) & \\
 & 32 5 27 1 & -- & 31 5 26 0 & 334079.683 & 276.07 & 3.19(-4) & \\
 & 28 17 11 1 & -- & 28 16 13 0 & 334088.087 & 338.47 & 1.35(-4) & \\
 & 28 17 12 1 & -- & 28 16 12 0 & 334088.087 & 338.47 & 1.35(-4) & \\
 & 32 7 25 0 & -- & 31 7 24 1 & 334098.077 & 284.34 & 3.31(-4) & \\
 & 34 3 31 1 & -- & 33 4 30 1 & 334112.202 & 295.02 & 1.78(-4) & \\
 & 34 3 31 0 & -- & 33 4 30 0 & 334120.029 & 294.96 & 3.08(-4) & \\
 & 27 17 10 1 & -- & 27 16 12 0 & 334155.000 & 324.80 & 1.30(-4) & \\
 & 27 17 11 1 & -- & 27 16 11 0 & 334155.000 & 324.80 & 1.30(-4) & \\
 & 20 7 13 1 & -- & 19 6 13 0 & 334167.497 & 126.59 & 2.43(-4) & \\
 & 26 17 9 1 & -- & 26 16 11 0 & 334213.515 & 311.63 & 1.25(-4) & \\
 & 26 17 10 1 & -- & 26 16 10 0 & 334213.515 & 311.63 & 1.25(-4) & \\
 & 34 4 31 1 & -- & 33 3 30 1 & 334253.644 & 295.03 & 3.07(-4) & \\
 & 34 4 31 0 & -- & 33 3 30 0 & 334264.186 & 294.97 & 1.79(-4) & \\
 & 25 17 8 1 & -- & 25 16 10 0 & 334264.285 & 298.94 & 1.19(-4) & \\
 & 25 17 9 1 & -- & 25 16 9 0 & 334264.285 & 298.94 & 1.19(-4) & \\
 & 24 17 7 1 & -- & 24 16 9 0 & 334307.936 & 286.74 & 1.12(-4) & \\
 & 24 17 8 1 & -- & 24 16 8 0 & 334307.936 & 286.74 & 1.12(-4) & \\
 & 18 8 11 1 & -- & 17 7 10 1 & 334320.081 & 114.67 & 1.86(-4) & \\
 & 18 8 10 1 & -- & 17 7 11 1 & 334325.869 & 114.67 & 1.86(-4) & \\
 & 23 17 6 1 & -- & 23 16 8 0 & 334345.068 & 275.03 & 1.04(-4) & \\
 & 23 17 7 1 & -- & 23 16 7 0 & 334345.068 & 275.03 & 1.04(-4) & \\
 & 18 8 11 0 & -- & 17 7 10 0 & 334354.965 & 114.62 & 1.88(-4) & \\
 & 18 8 10 0 & -- & 17 7 11 0 & 334360.695 & 114.62 & 1.89(-4) & \\
 & 12 11 1 1 & -- & 11 10 2 1 & 334506.538 & 96.91 & 3.56(-4) & $\star$ \\
 & 12 11 2 1 & -- & 11 10 1 1 & 334506.538 & 96.91 & 3.56(-4) & $\star$ \\
 & 12 11 1 0 & -- & 11 10 2 0 & 334538.884 & 96.86 & 3.56(-4) & $\star$ \\
 & 12 11 2 0 & -- & 11 10 1 0 & 334538.884 & 96.86 & 3.56(-4) & $\star$ \\
 & 20 7 14 1 & -- & 19 6 14 0 & 334562.201 & 126.58 & 2.05(-4) & \\
 & 16 9 8 1 & -- & 15 8 7 1 & 334599.304 & 105.79 & 2.28(-4) & \\
 & 16 9 7 1 & -- & 15 8 8 1 & 334599.324 & 105.79 & 2.28(-4) & \\
 & 14 10 4 1 & -- & 13 9 5 1 & 334604.473 & 99.88 & 2.81(-4) & \\
 & 14 10 5 1 & -- & 13 9 4 1 & 334604.473 & 99.88 & 2.81(-4) & \\
 & 16 9 8 0 & -- & 15 8 7 0 & 334633.280 & 105.74 & 2.30(-4) & \\
 & 16 9 7 0 & -- & 15 8 8 0 & 334633.300 & 105.74 & 2.30(-4) & \\
 & 14 10 4 0 & -- & 13 9 5 0 & 334637.793 & 99.83 & 2.82(-4) & \\
 & 14 10 5 0 & -- & 13 9 4 0 & 334637.793 & 99.83 & 2.82(-4) & \\
 & 36 0 36 1 & -- & 35 0 35 0 & 334709.849 & 298.92 & 3.61(-4) & $\star$ \\
 & 36 1 36 1 & -- & 35 1 35 0 & 334709.849 & 298.92 & 3.61(-4) & $\star$ \\
 & 33 20 13 0 & -- & 32 20 12 1 & 334838.561 & 467.86 & 2.18(-4) & \\
 & 33 20 14 0 & -- & 32 20 13 1 & 334838.561 & 467.86 & 2.18(-4) & \\
 & 33 21 12 0 & -- & 32 21 11 1 & 334840.150 & 487.72 & 2.05(-4) & \\
 & 33 21 13 0 & -- & 32 21 12 1 & 334840.150 & 487.72 & 2.05(-4) & \\
 & 33 19 14 0 & -- & 32 19 13 1 & 334851.819 & 448.96 & 2.31(-4) & \\
 & 33 19 15 0 & -- & 32 19 14 1 & 334851.819 & 448.96 & 2.31(-4) & \\
 & 33 22 11 0 & -- & 32 22 10 1 & 334852.879 & 508.55 & 1.91(-4) & \\
 & 33 22 12 0 & -- & 32 22 11 1 & 334852.879 & 508.55 & 1.91(-4) & \\
 & 33 18 15 0 & -- & 32 18 14 1 & 334884.208 & 431.04 & 2.43(-4) & \\
 & 33 18 16 0 & -- & 32 18 15 1 & 334884.208 & 431.04 & 2.43(-4) & \\
 & 33 17 16 0 & -- & 32 17 15 1 & 334940.915 & 414.08 & 2.55(-4) & \\
 & 33 17 17 0 & -- & 32 17 16 1 & 334940.915 & 414.08 & 2.55(-4) & \\
 & 22 6 17 1 & -- & 21 5 17 0 & 334954.517 & 141.61 & 2.37(-4) & \\
 & 33 5 29 0 & -- & 32 4 28 0 & 334987.470 & 286.45 & 3.98(-4) & \\
 & 33 16 17 0 & -- & 32 16 16 1 & 335028.496 & 398.10 & 2.66(-4) & \\
 & 33 16 18 0 & -- & 32 16 17 1 & 335028.496 & 398.10 & 2.66(-4) & \\
 & 33 15 18 0 & -- & 32 15 17 1 & 335155.577 & 383.11 & 2.76(-4) & \\
 & 33 15 19 0 & -- & 32 15 18 1 & 335155.577 & 383.11 & 2.76(-4) & \\
 & 23 4 20 0 & -- & 22 3 20 1 & 335226.462 & 142.78 & 1.46(-4) & \\
 & 33 14 20 0 & -- & 32 14 19 1 & 335333.954 & 369.09 & 2.86(-4) & \\
 & 33 14 19 0 & -- & 32 14 18 1 & 335333.954 & 369.09 & 2.86(-4) & \\
 & 34 4 31 1 & -- & 33 4 30 0 & 335422.139 & 295.03 & 2.84(-4) & $\star$ \\
 & 34 3 31 1 & -- & 33 3 30 0 & 335459.569 & 295.02 & 4.14(-4) & \\
 & 18 8 10 1 & -- & 17 7 10 0 & 335462.789 & 114.67 & 2.38(-4) & \\
 & 18 8 11 1 & -- & 17 7 11 0 & 335467.703 & 114.67 & 2.37(-4) & \\
 & 36 6 30 1 & -- & 35 7 29 1 & 335488.573 & 349.86 & 1.07(-4) & \\
 & 33 13 21 0 & -- & 32 13 20 1 & 335580.432 & 356.08 & 2.96(-4) & \\
 & 33 13 20 0 & -- & 32 13 19 1 & 335580.435 & 356.08 & 2.96(-4) & \\
 & 35 3 33 0 & -- & 34 3 32 1 & 335588.334 & 302.57 & 3.60(-4) & \\
 & 35 2 33 0 & -- & 34 2 32 1 & 335589.204 & 302.57 & 3.55(-4) & \\
 & 32 7 25 1 & -- & 31 7 24 0 & 336072.528 & 284.39 & 3.32(-4) & \\
 & 33 22 11 1 & -- & 32 22 10 0 & 336092.641 & 508.57 & 1.93(-4) & \\
 & 33 22 12 1 & -- & 32 22 11 0 & 336092.641 & 508.57 & 1.93(-4) & \\
 & 33 20 13 1 & -- & 32 20 12 0 & 336099.583 & 467.89 & 2.20(-4) & \\
 & 33 20 14 1 & -- & 32 20 13 0 & 336099.583 & 467.89 & 2.20(-4) & \\
 & 33 23 10 1 & -- & 32 23 9 0 & 336116.773 & 530.36 & 1.78(-4) & \\
 & 33 23 11 1 & -- & 32 23 10 0 & 336116.773 & 530.36 & 1.78(-4) & \\
 & 33 15 18 1 & -- & 32 15 17 0 & 336558.515 & 383.14 & 2.79(-4) & \\
 & 33 15 19 1 & -- & 32 15 18 0 & 336558.515 & 383.14 & 2.79(-4) & \\
 & 23 4 20 1 & -- & 22 2 20 1 & 337017.819 & 142.83 & 1.10(-4) & \\
 & 25 6 20 0 & -- & 24 5 19 0 & 337038.379 & 177.09 & 1.09(-4) & \\
 & 34 15 19 0 & -- & 33 15 18 1 & 345385.355 & 399.71 & 3.07(-4) & \\
 & 34 15 20 0 & -- & 33 15 19 1 & 345385.355 & 399.71 & 3.07(-4) & \\
 & 13 11 2 1 & -- & 12 10 2 0 & 345836.731 & 103.25 & 3.87(-4) & $\star$ \\
 & 13 11 3 1 & -- & 12 10 3 0 & 345836.731 & 103.25 & 3.87(-4) & $\star$ \\
 & 17 9 8 1 & -- & 16 8 8 0 & 345841.998 & 114.10 & 2.83(-4) & $\star$ \\
 & 17 9 9 1 & -- & 16 8 9 0 & 345841.998 & 114.10 & 2.83(-4) & $\star$ \\
 & 34 13 22 0 & -- & 33 13 21 1 & 345862.205 & 372.71 & 3.27(-4) & $\star$ \\
 & 34 13 21 0 & -- & 33 13 20 1 & 345862.211 & 372.71 & 3.27(-4) & $\star$ \\
 & 15 10 5 1 & -- & 14 9 5 0 & 345906.318 & 107.20 & 3.25(-4) & \\
 & 15 10 6 1 & -- & 14 9 6 0 & 345906.318 & 107.20 & 3.25(-4) & \\
 & 36 2 34 1 & -- & 35 3 33 1 & 345985.032 & 319.24 & 3.20(-4) & \\
 & 36 3 34 1 & -- & 35 2 33 1 & 345986.658 & 319.24 & 3.23(-4) & \\
 & 36 2 34 0 & -- & 35 3 33 0 & 345987.306 & 319.18 & 3.24(-4) & \\
 & 36 3 34 0 & -- & 35 2 33 0 & 345988.992 & 319.18 & 3.21(-4) & \\
 & 34 17 17 1 & -- & 33 17 16 0 & 346444.360 & 430.71 & 2.87(-4) & \\
 & 34 17 18 1 & -- & 33 17 17 0 & 346444.360 & 430.71 & 2.87(-4) & \\
 & 24 3 21 1 & -- & 23 2 21 0 & 346551.132 & 154.12 & 1.52(-4) & \\
 & 35 5 30 0 & -- & 34 6 29 0 & 346551.443 & 326.48 & 1.29(-4) & \\
 & 35 5 30 1 & -- & 34 6 29 1 & 346568.765 & 326.51 & 2.15(-4) & \\
 & 34 16 18 1 & -- & 33 16 17 0 & 346577.753 & 414.74 & 2.98(-4) & \\
 & 34 16 19 1 & -- & 33 16 18 0 & 346577.753 & 414.74 & 2.98(-4) & \\
 & 29 6 24 1 & -- & 28 5 23 1 & 346986.538 & 231.40 & 1.13(-4) & \\
 & 34 14 21 1 & -- & 33 14 20 0 & 346993.104 & 385.75 & 3.20(-4) & \\
 & 34 14 20 1 & -- & 33 14 19 0 & 346993.104 & 385.75 & 3.20(-4) & \\
 & 36 3 34 1 & -- & 35 3 33 0 & 347254.922 & 319.24 & 3.93(-4) & \\
 & 36 2 34 1 & -- & 35 2 33 0 & 347255.462 & 319.24 & 3.97(-4) & \\
 & 34 13 22 1 & -- & 33 13 21 0 & 347306.553 & 372.75 & 3.30(-4) & \\
 & 34 13 21 1 & -- & 33 13 20 0 & 347306.559 & 372.75 & 3.30(-4) & \\
 & 34 10 25 0 & -- & 33 10 24 1 & 347502.019 & 339.86 & 3.56(-4) & \\
 & 37 2 36 0 & -- & 36 2 35 1 & 347507.368 & 326.07 & 4.01(-4) & \\
 & 37 1 36 0 & -- & 36 1 35 1 & 347507.368 & 326.07 & 4.01(-4) & \\
 & 34 10 24 0 & -- & 33 10 23 1 & 347535.163 & 339.86 & 3.56(-4) & \\
\hline
\textbf{t-HCOOH} & 35 4 31 & -- & 35 3 32 & 334247.840 & 739.34 & 7.49(-6) & \\
(CDMS) & 15 2 14 & -- & 14 2 13 & 334265.833 & 141.58 & 4.17(-4) & $\dagger$ \\
\hline
\textbf{\ce{H2CCO}} (CDMS) & 17 1 16 & -- & 16 1 15 & 346600.451 & 162.79 & 4.74(-4) & \\
\hline
\textbf{\ce{CH2DCN}} & 20 1 20 & -- & 19 1 19 & 345685.375 & 179.63 & 3.60(-3) & \\
(CDMS) & 20 7 13 & -- & 19 7 12 & 346968.947 & 438.92 & 3.20(-3) & \\
 & 20 7 14 & -- & 19 7 13 & 346968.947 & 438.92 & 3.20(-3) & \\
 & 20 0 20 & -- & 19 0 19 & 347043.458 & 174.96 & 3.65(-3) & $\star$ \\
 & 20 6 14 & -- & 19 6 13 & 347044.649 & 368.95 & 3.32(-3) & $\star$ \\
 & 20 6 15 & -- & 19 6 14 & 347044.649 & 368.95 & 3.32(-3) & $\star$ \\
 & 20 5 15 & -- & 19 5 14 & 347110.032 & 309.72 & 3.42(-3) & $\star$ \\
 & 20 5 16 & -- & 19 5 15 & 347110.032 & 309.72 & 3.42(-3) & $\star$ \\
 & 20 4 17 & -- & 19 4 16 & 347166.409 & 261.24 & 3.50(-3) & $\star$ \\
 & 20 4 16 & -- & 19 4 15 & 347166.421 & 261.24 & 3.50(-3) & $\star$ \\
 & 20 2 19 & -- & 19 2 18 & 347188.253 & 196.56 & 3.61(-3) & \\
 & 20 3 18 & -- & 19 3 17 & 347216.880 & 223.53 & 3.57(-3) & \\
 & 20 3 17 & -- & 19 3 16 & 347219.365 & 223.53 & 3.57(-3) & \\
 & 20 2 18 & -- & 19 2 17 & 347388.249 & 196.62 & 3.62(-3) & \\
\hline
\textbf{\ce{C2H5CN}}, v=0 & 37 5 32 & -- & 36 5 31 & 333921.554 & 330.94 & 3.11(-3) & \\
(CDMS) & 39 3 37 & -- & 38 3 36 & 345921.198 & 344.46 & 3.50(-3) & \\
 & 38 3 35 & -- & 37 3 34 & 346983.834 & 333.42 & 3.53(-3) & \\
\hline
\textbf{\ce{NH2CHO}} & 16 2 15 & -- & 15 2 14 & 336136.877 & 149.67 & 2.76(-3) & \\
(CDMS) & 16 1 15 & -- & 15 1 14 & 345326.688 & 145.16 & 3.02(-3) & \\
\end{longtable}
\begin{minipage}{1.0\textwidth}
    \textbf{Note}:\\
    $\star$: for the robustly detected species, key transitions that are used to constrain the column densities and excitation temperatures; must be optically thin and unblended. \\
    $\ast$: for the tentatively detected species, key transitions that are used to estimate upper limits of column densities. \\
    $\dagger$: transitions that are (expected to be) optically thick and not considered in the fitting. \\
    \end{minipage}
\end{appendix}

\end{document}